\documentclass{aa}  
\usepackage{graphicx}
\usepackage{xcolor}
\usepackage{txfonts}
\usepackage{booktabs}
\usepackage{mathrsfs}
\usepackage{comment}
\newcommand{\dd}[1]{\,\text{d}#1}

\begin{document} 

    \title{Cosmology with the submillimetre galaxies magnification bias} \subtitle{Tomographic analysis\thanks{The cleaned MCMC chains for the $\Lambda$CDM, $\omega_0$CDM, and $\omega_0\omega_a$CDM cases are only available in electronic form at the CDS via anonymous ftp to cdsarc.u-strasbg.fr (130.79.128.5) or via http://cdsweb.u-strasbg.fr/cgi-bin/qcat?J/A+A/}}
    \titlerunning{Tomography with Magnification Bias}
    \authorrunning{Bonavera L. et al.}

   \author{Bonavera L.\inst{1,2}, Cueli M. M.\inst{1,2}, Gonz{\'a}lez-Nuevo J.\inst{1,2}, Ronconi T.\inst{3,4,5}, Migliaccio M.\inst{6,7}, Lapi A.\inst{3,5}, Casas J. M.\inst{1,2}, Crespo D.\inst{1,2}
   }

   \institute{$^1$Departamento de Fisica, Universidad de Oviedo, C. Federico Garcia Lorca 18, 33007 Oviedo, Spain\\
              $^2$Instituto Universitario de Ciencias y Tecnologías Espaciales de Asturias (ICTEA), C. Independencia 13, 33004 Oviedo, Spain\\
              $^3$International School for Advanced Studies (SISSA), via Bonomea 265, I-34136 Trieste, Italy\\
              $^4$INFN Sezione di Trieste, via Valerio 2, I-34127 Trieste, Italy\\
              $^5$Institute for Fundamental Physics of the Universe (IFPU), Via Beirut 2, I-34014 Trieste, Italy\\
              $^6$Dipartimento di Fisica, Universit\`a di Roma Tor Vergata, Via della Ricerca Scientifica, 1, Roma, Italy\\
              $^7$INFN, Sezione di Roma 2, Universit\`a di Roma Tor Vergata, Via della Ricerca Scientifica, 1, Roma, Italy\\
              }

   \date{Received xxx, xxxx; accepted xxx, xxxx}

% \abstract{}{}{}{}{} 
% 5 {} token are mandatory
 
  \abstract
   {High-z submillimetre galaxies (SMGs) can be used as a background sample for gravitational lensing studies thanks to their magnification bias. In particular,  the magnification bias can be exploited in order to constrain the free parameters of a halo occupation distribution (HOD) model and some of the main cosmological parameters. A pseudo-tomographic analysis shows that the tomographic approach should improve the parameter estimation.}
   {In this work the magnification bias has been evaluated as cosmological tool in a tomographic set-up. The cross-correlation function (CCF) data have been used to jointly constrain the astrophysical parameters $M_{min}$, $M_{1}$, and $\alpha$ in each of the selected redshift bins as well as the cosmological parameters $\Omega_{M}$, $\sigma_{8}$, and $H_0$ for the lambda cold dark matter ($\Lambda$CDM) model. Moreover, we explore the possible time evolution of the dark energy density by also introducing  the $\omega_0, \omega_a$ parameters in the joint analysis ($\omega_0$CDM and $\omega_0\omega_a$CDM).}
   {The CCF was measured between a foreground spectroscopic sample of Galaxy And Mass Assembly (GAMA) galaxies and a background sample of \textit{Herschel} Astrophysical Terahertz Large Area Survey (H-ATLAS) galaxies. The foreground sample was divided into four redshift bins (0.1-0.2, 0.2-0.3, 0.3-0.5, and 0.5-0.8) and the sample of H-ATLAS galaxies has photometric redshifts >1.2. The CCF was modelled using a halo model description that depends on HOD and cosmological parameters. Then a Markov chain Monte Carlo method was used to estimate the parameters for different cases.}
   {For the $\Lambda$CDM model the analysis yields a maximum posterior value at 0.26 with  $[0.17,0.41]$ 68\%  C.I.  for $\Omega_M$ and at 0.87 with $[0.75,1]$ 68\% C.I. for $\sigma_8$. With our current results $H_0$ is not yet constrained. With a more general $\omega_0$CDM model, the constraints on $\Omega_M$ and $\sigma_8$ are similar, but we found a maximum posterior value for $\omega_0$ at -1 with $[-1.56, -0.47]$ 68\% C.I. In the $\omega_0\omega_a$CDM model, the results are -1.09 with $[-1.72, -0.66]$ 68\% C.I. for $\omega_0$ and -0.19 with $[-1.88, 1.48]$ 68\% C.I. for $\omega_a$.}
  {The results on $M_{min}$ show a trend towards higher values at higher redshift confirming recent findings. The tomographic analysis presented in this work improves the constraints in the $\sigma_8-\Omega_M$ plane with respect to previous findings exploiting the magnification bias and it confirms that magnification bias results do not show the degeneracy found with cosmic shear measurements. Moreover, related to dark energy, we found a trend of higher $\omega_0$ values for lower $H_0$ values.}


   \maketitle
%
%-------------------------------------------------------------------

\section{Introduction}

The gravitational lensing effects of area stretching and amplification cause in general an apparent modification of the number of high-redshift sources observed in the proximity of low-redshift structures \citep[see e.g.][]{SCH92}. This effect, called magnification bias, thus increases the probability of such sources being included in a flux-limited sample \citep[see e.g.][]{ARE11}. As a result, a non-negligible signal is expected when computing the cross-correlation function (CCF) between two source samples with non-overlapping redshift distributions, for example  galaxies and quasars \citep{SCR05, MEN10}, Herschel sources and Lyman-break galaxies \citep{HIL13}, and the cosmic microwave background \citep[CMB,][]{BIA15, BIA16}. 

Throughout this work we use  submillimetre galaxies (SMGs) as the background sample. Given their steep luminosity function, very faint emission in the optical band, and typical redshifts of $ z > 1-1.5 $, they are an optimal sample \citep[see e.g. discussion on their boost in fluxes  in][]{ARE11} for the CCF studies when used as background sample, as confirmed for example in \citet{BLA96, NEG07, NEG10, GON12,BUS12,BUS13,FU12, WAR13, CAL14, NAY16, NEG17, GON21}, and \citet{BAK20}.
In addition, \cite{DUN20} report a serendipitous direct observation of high-redshift SMGs around a statistically complete sample of 12 250 $\mu$m selected galaxies at $z = 0.35$, which were targeted by ALMA in a study of gas tracers.

Moreover, the magnification bias signal by SMGs was observed by \citet{WAN11} and measured at $> 10\sigma$ by \citet{GON14}. Such measurements were improved and used in studies within the halo model formalism in \citet{GON17} to conclude that the lenses are massive galaxies or even galaxy groups and/or clusters, with a minimum mass of $M_{\text{lens}}\,$$\sim$$\,10^{13}M_{\odot}$. Along the same line, \citet{BON19} used the magnification bias signal to study the mass properties of a sample of quasi-stellar objects (QSOs) at $0.2<z<1.0$ acting as a signpost for a lens, yielding a minimum mass of $M_{\text{min}}\,$=\,$ 10^{13.6_{-0.4}^{+0.9}} M_{\odot}$ and suggesting that the lensing effect is produced by a cluster-size halo near the QSOs positions. Furthermore, \citet{CUE20} assessed the potential of magnification bias CCF measurements as a way of constraining the halo mass function according to two common parametrisations (the Sheth \&\ Tormen and the Tinker fits), finding general agreement with traditional values along with a hint at a slightly higher number of halos at intermediate and high masses for the Sheth \&\ Tormen fit.

Since the gravitational deflection of light travelling close to the lenses depends on cosmological distances and galaxy halo properties, magnification bias can be used as an independent cosmological probe to address the estimation of the parameters in the standard cosmological model. In \citet{BON20}, the measurements of this observable are used in a proof of concept with the purpose of constraining some cosmological parameters, finding interesting limits for $\Omega_M$ $(>\,$$0.24$ as a 95\%  C.I.) and $\sigma_8$ ($<\,$$1.0$ as a 95\% C.I.). Finally, in \citet{GON21}, different biases in the source samples are analysed and corrected in order to obtain the optimal strategy to precisely measure and analyse unbiased CCFs at cosmological scales.
 Even if measurements based on different observables \citep[CMB anisotropies, Big Bang nucleosynthesis, SNIa observations, BAOs;][]{HIN13, PLA16_XIII, PLA18_VI, FIE06, BET14, ROS15} are in general agreement with the current lambda cold dark matter ($\Lambda$CDM) model, the need for independent cosmological probes is driven by the fact that recently there have been some hints that suggest a modification of this framework, such as the Hubble constant's differing measurements ($H_0=74.03 \pm 1.42$ km/s/Mpc and $67.4 \pm 0.5$ km/s/Mpc by \citet{RIE19} and \citet{PLA18_VI}, respectively) and the degenerate $\Omega_M$ and $\sigma_8$ relationship \citep[e.g.][]{HEY13,PLA16_XXIV,HIL17, PLA18_VI}.

A plethora of works have addressed the issue of probing the distribution of mass in the Universe via measurements of cosmic shear, the gravitational lensing effect causing a coherent distortion of galaxy images. Performed for the first time at the beginning of the millennium \citep{BA00,VW00,WI00,RHO01}, cosmic shear measurements have benefited from a rapid and considerable refinement of its methodology in order to analyse the results from recent weak lensing surveys such as the Dark Energy Survey \citep[DES; ][]{DES16}, the Kilo-Degree Survey \citep[KiDS;][]{dJO13}, and the Hyper Suprime-Cam survey \citep[HSC;][]{AI18}. Constraints on $\Omega_M$, $\sigma_8$, and the dark energy equation-of-state parameter ($\omega$) have thus been obtained in recent works \citep[e.g.][]{HIL17,TRO18,HI19} by means of a tomographic analysis, that is, by splitting the redshift distribution of galaxies in several bins and measuring the shear CCF between galaxies in different bins \citep{HU99}.

Analogously  to what is done in shear studies, a tomographic analysis might improve the results on the cosmological parameters, particularly by allowing a better estimation of the halo occupation distribution (HOD) parameters.  Slightly different values for these parameters are expected when probed in separate redshift bins. This is demonstrated in \citet{GON17}, where tomographic studies on the HOD parameters are also performed by splitting the foreground sample in four bins of redshift, namely $0.1 < z < 0.2, 0.2 < z < 0.3, 0.3 < z < 0.5$ and $0.5 < z < 0.8 $. It has been found that while $M_{1}^{lens}$ is almost redshift independent, there is a clear evolution in $M_{min}^{lens}$, which increases with redshift in agreement with theoretical estimations. Moreover, \citet{GON21} use a background sample of the \textit{Herschel} Astrophysical Terahertz Large Area Survey (H-ATLAS)  galaxies with photometric redshifts $z > 1.2$ and two independent foreground samples, Galaxy And Mass Assembly (GAMA) galaxies with spectroscopic redshifts and Sloan Digital Sky Survey (SDSS) galaxies with photometric redshifts, with 0.2 < z < 0.8 in order to perform a simplified tomographic analysis, obtaining constraints on the cosmological parameters $\Omega_M$ ($0.50_{-0.20}^{+0.14}$) and $\sigma_8$ ($0.75_{-0.10}^{+0.07}$) as a 68\% C.I.

This work constitutes a step forward from \citet{BON20}. We explore the possibility of improving the constraints on $\Omega_M$, $\sigma_8$, and $H_0$ by adopting the unbiased samples of \citet{GON21} and introducing a tomographic analysis by jointly estimating the HOD parameters in four different bins of redshift along with the aforementioned cosmological parameters. We also add the possibility of allowing different values for the dark energy equation of state parameter, $\omega$, to study the possible redshift evolution of the predominant energy contribution of the Universe.

For a flat cosmology with a constant dark energy equation of state, \cite{ALL08} found $\omega=-1.14 \pm 0.31$, using Chandra measurements of the X-ray gas mass fraction in 42 galaxy clusters. When combining their data with independent constraints from cosmic microwave background and SNIa studies, they obtain improved value of $\omega = -0.98 \pm 0.07$. In \cite{SUZ12}, for the same constant-$\omega$ model, SNe Ia alone give $\omega=-1.001^{+0.348}_{-0.398}$ and SNe combined with the constraints from BAO, CMB, and $H_0$ measure  $\omega=-1.013^{+0.068}_{-0.073}$ (68\% C.I.). 
\cite{PLA18_VI} set the tightest constraints from the combination of \textit{Planck} TT,TE,EE+lowE+lensing+SNe+BAO to $\omega_0=-1.03 \pm 0.03$, but when using \textit{Planck} TT,TE,EE+lowE+lensing alone is considerably less constraining \citep[see Fig.30 in][]{PLA18_VI}; for example, they obtain $\omega_0=-1.56^{+0.19}_{-0.39}$ (68\% C.I.) with the baseline model base\_w\_plikHM\_TT\_lowl\_lowE.
When the $\omega_a$ parameter is not fixed to $\omega_a=0$, they find the tightest constraints for the data combination \textit{Planck} TT,TE,EE+lowE+lensing+BAO+SNe to $\omega_0=-0.957 \pm 0.080 $ and $\omega_a=-0.29_{-0.26}^{+0.32}$ (68\% C.I.), but $\omega_0=-0.76 \pm 0.20$ and $\omega_a=-0.72^{+0.62}_{-0.54}$ (68\% C.I.) \textit{Planck}+BAO/RSD+WL \citep[see Table 6 in][]{PLA18_VI} and $\omega_0=-0.59_{-0.26}^{+0.29}$ and $\omega_a=-1.33 \pm 0.79$ (68\% C.I.) for the baseline base\_w\_wa\_plikHM\_TT\_lowl\_lowE\_BAO (parameter grid table in the PLA). 

For $\omega_a$ \cite{ALA17} find a value of $-0.39 \pm 0.34$ with a cosmological analysis of BOSS Galaxies combining \textit{Planck}+BAO+FS+SN and a value of $-1.16 \pm 0.55$ with \textit{Planck}+BAO. They also find a strong $\omega_0-\omega_a$ correlation.

Results on the $\omega$ parameter have also been  provided by the Dark Energy Survey  \citep[DES][]{ABB18} Year1. From a starting point using only the Y1 $\xi_{\pm}(\theta)$ information, a value  was found of $\omega=-0.99^{+0.33}_{-0.39}$  at 68\%\ C.I. By combining  different observations, the best finding is $\omega=-1.00^{+0.05}_{-0.04}$ for the DES Y1+\textit{Planck}+JLA+BAO combination at 68\%\ C.I. An improvement of 12\% is reported by \cite{POR21} from the combination of galaxy clustering and galaxy-galaxy lensing by optimising the lens galaxy sample selection using information from  DES Year3 data and assuming the DES Year1 METACALIBRATION sample for the sources. Recently, \citet{ABB21} have found a value of $\omega=-0.98^{+0.32}_{-0.20}$ with the DES Y3 3$\times$2pt analysis.

This paper is  organised as follows. Section \ref{sec:data} provides a description of the data used in this work. Section \ref{sec:methodology} presents the methodology we  followed in the measurement, modelling, and study of our observable. The results we obtained on the astrophysical constraints with a fixed $\Lambda$CDM cosmology are described in Section \ref{sec:astro_results}, and those on the cosmological constraints in Section \ref{sec:cosmo_results}. Finally, Section \ref{sec:conclusion} summarises the conclusions that can be drawn from our findings.

%--------------------------------------------------------------------
\section{Data}
\label{sec:data}
In this work we use the same background and foreground samples as in \cite{BON20}, and the CCF data are computed according to the mini-Tiles scheme detailed in \citet{GON21}. We prefer the mini-Tiles to the Tiles scheme because of the minimum surface area corrections required and the better statistics (about 96 mini tiles of $\approx 2\times 2 \,\text{deg}^2$, to be compared with about 24 different tiles of $\approx 4\times 4 \,\text{deg}^2$  each).

\begin{figure}[ht]
\centering
\includegraphics[width=0.49\textwidth]{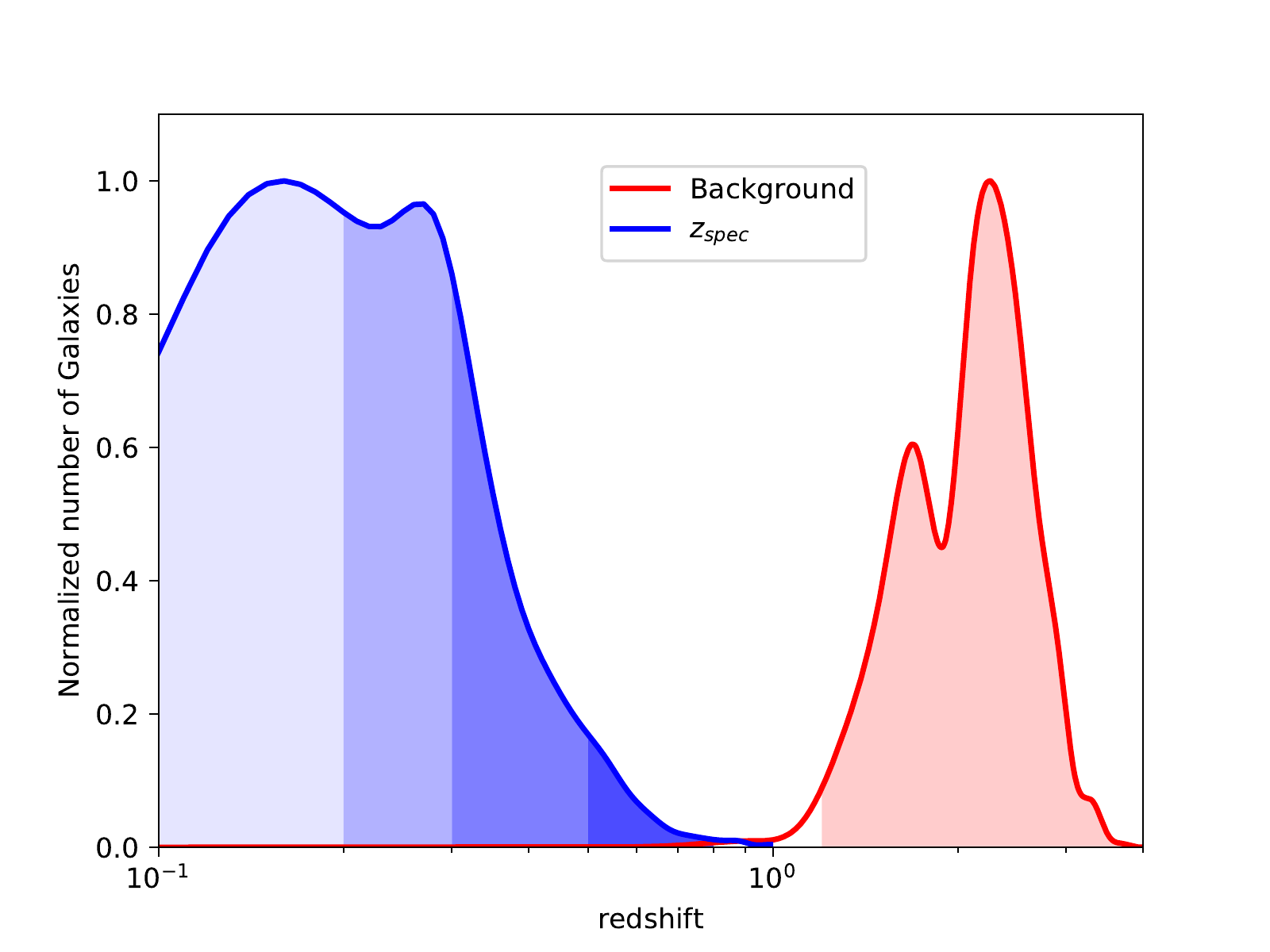}
    \caption{Normalised redshift distribution of the background (red line) and foreground (blue line) samples. The coloured areas represents the selected ranges of redshift: in red the $1.2<z<4.0$ interval for the background SMGs sample and in different shades of blue the $0.1<z<0.2$, $0.2<z<0.3$, $0.3<z<0.5$, and $0.5<z<0.8$ bins for the foreground GAMA galaxies.}
    \label{Fig:z_distro}
\end{figure}

The foreground sources are selected in the GAMA II spectroscopic survey \citep{DRI11, BAL10,BAL14,LIS15}. They consist of $\sim 225000$ galaxies with spectroscopic redshifts of $0.1 < z < 0.8$ and a median value of $z_{\text{med}} = 0.28$. The same sample was used in \citet{GON17}, \citet{BON20}, and \citet{GON21} and is referred to as the $z_{spec}$ sample in the last of the three. In this work we  divided the foreground sample into four redshift bins: $0.1-0.2$, $0.2-0.3$, $0.3-0.5$, and $0.5-0.8$. These redshift ranges were  chosen in order to obtain a similar number of sources in each bin. Moreover, with respect to our previous works, we  included sources at lower redshifts (0.1-0.2) in order to add some  potentially useful information for the tomographic analysis. Figure \ref{Fig:z_distro} shows  the redshift distribution of the background sources and the regions that represent the four selected bins of redshift. Since the spectroscopic redshifts for the foreground sources have negligible errors, there is no overlap in the redshift distribution among bins and no source is misplaced in the wrong bin.

The background sample consists of Herschel \citep{PIL10, EAL10} sources detected in the three GAMA fields (total area of $\sim 147 \,\text{deg}^2$) and the part of the south galactic region (SGP) that overlaps with the foreground sample ($\sim 60 \,\text{deg}^2$).
Details concerning the generation of the H-ATLAS catalogue can be found in \citet{IBA10, PAS11, RIG11, VAL16, BOU16}, and \citet{MAD20}.
A photometric redshift selection of 1.2 < z < 4.0 has been applied to ensure no overlap in the redshift distributions of lenses and background sources (red area in Figure \ref{Fig:z_distro}), thus being left with 57930 galaxies ($\sim 24\%$  of the initial sample). The process of redshift estimation is  described in detail in \cite{GON17} and \citet{BON19}. 
It consists in a minimum $\chi^2$ fit to the SPIRE data, and to PACS data when available, of the SMM J2135-0102 template SED \citep[the Cosmic Eyelash at z = 2.3;][]{IVI10, SWI10}. It has been found to be the best overall template having $\Delta z/(1+z)=-0.07$ with a 0.153 dispersion \citep{IVI16,LAP11, GON12}.

The background redshift distribution used during the analysis (red line) is the estimated $p(z|W)$ of the galaxies selected by our window function (i.e. a top hat for 1.2 < z < 4.0), taking into account the effect of random errors in photometric redshifts on the redshift distribution, as done in \citet{GON17}. In the same work it is concluded that the possible cross-contamination (sources at lower redshift, $z$ < 0.8, with photometric redshifts > 1.2) can be considered statistically negligible even when considering the photometric redshift uncertainties \citep[see][and references therein]{GON17}.

%--------------------------------------------------------------------
\section{Methodology}
\label{sec:methodology}

\subsection{Foreground angular auto-correlation function}
First of all, we performed an auto-correlation analysis on the GAMA sample to obtain a first estimate of the HOD parameters for the foreground sample. As in \citet{BON20}, we relied just on the H-ATLAS fields G09, G12, and G15 and selected the normalising redshift quality (column \texttt{NQ}) greater than $2$ and the spectroscopic redshift (column \texttt{Z}) in the range $0.05 < z_\text{spec} < 0.6$, resulting in a sample with $z_\text{med} = 0.23$. 

In each field the number density of galaxies is estimated as $V_\text{field} = A_\text{field} \cdot \Delta d_C$, where $\Delta d_C$ is the comoving distance difference between the chosen maximum and minimum redshifts and $A_\text{field}$ is the area of the field, computed as in \cite{BON20}. The angular auto-correlation function $w(\theta)$ is estimated with the Landy-Szalay estimator \citep [Eq. \ref{eq:LS_est},][]{LAN93}, 
\begin{equation}
    \label{eq:LS_est}
    \tilde{w}(\theta) = \dfrac{DD(\theta) - 2 DR(\theta) + RR(\theta)}{RR(\theta)},
\end{equation}
where $DD$, $DR,$ and $RR$ are the normalised data-data, data-random, and random-random pair counts for a given angular separation $\theta$. Its errors are obtained by extracting 128 bootstrap samples from each field. 

According to the halo model formalism \citep[see e.g.][]{COO02}, the average number density of galaxies at a given redshift can be expressed as
\begin{equation}
    \label{eq:ng01}
    n_g(z) = \int_{M_\text{min}}^{M_\text{max}} \langle N_g \rangle (M) n(M, z) \dd{M}
,\end{equation}
where $n(M, z)$ is the halo mass function, which we assume to match the Sheth \& Tormen fit in this work \citep{SHE99}, and $N_g(M) = N_\text{cen}(M) + N_\text{sat}(M)$ is the mean number of galaxies in a halo of mass $M$, which we split into contributions from central and satellite galaxies, respectively. As in \cite{BON20}, we adopt the three-parameter $(\alpha,M_{min},M_1)$ HOD model by \citet{ZHE05}, where $M_{min}$ is the minimum halo mass required to host a (central) galaxy, and $M_1$ and $\alpha$ describe the power-law behaviour of satellite galaxies:  $N_{\text{sat}}(M)=(M/M_1)^{\alpha}$. 
Moreover, we assume the projected two-point correlation function (obtained using the standard Limber approximation), $w(r_p, z)$, approximately constant in the given redshift bins $[z_1, z_2]$. This is a good approximation considering that the auto-correlation varies less than 10\% between different bins, and thus even less inside a single bin (see Fig. \ref{fig:ACF}). In any case, the variation is smaller than the error bars. Therefore, the angular auto-correlation function can be approximated as
\begin{equation}
    \label{eq:wt02}
    w ( \theta, z ) \approx  \biggl[ \int_{z_1}^{z_2} \dd{z} \dfrac{\dd{V}(z)}{\dd{z}} n_f^2(z) \biggr] \cdot w[ r_p( \theta ), \overline{z} ) ] = A \cdot w[ r_p( \theta ), \overline{z} ) ]
,\end{equation}
where $dV/dz$ is the comoving volume unit, $n_f(z)$ is the normalised redshift distribution of the foreground population, and $\bar{z}$ denotes the mean redshift of the interval. In this  way the normalisation factor, $A$, can be taken into account as a free parameter along with $M_{min},M_1$, and $\alpha$, which simplifies  and speeds up the calculations.

\subsection{Foreground-background angular CCF}
Having performed a preliminary analysis of the auto-correlation within the foreground sample, we now turn to the corresponding analysis with the CCF, the actual main observable we mean to probe. As described in detail in \cite{GON17} and \citet{BON20}, its measurement is performed via a modified version of the \cite{LAN93} estimator \citep{HER01},
\begin{equation}
\tilde{w}_{fb}(\theta)=\frac{\rm{D}_f\rm{D}_b(\theta)-\rm{D}_f\rm{R}_b(\theta)-\rm{D}_b\rm{R}_f(\theta)+\rm{R}_f\rm{R}_b(\theta)}{\rm{R}_f\rm{R}_b(\theta)},
\end{equation}
where $\rm{D}_f\rm{D}_b$, $\rm{D}_f\rm{R}_b$, $\rm{D}_b\rm{R}_f$, and $\rm{R}_f\rm{R}_b$ are the normalised foreground-background, foreground-random, background-random, and random-random pair counts for a given angular separation $\theta$.

\begin{figure*}[ht]
\centering
\includegraphics[width=0.49\textwidth]{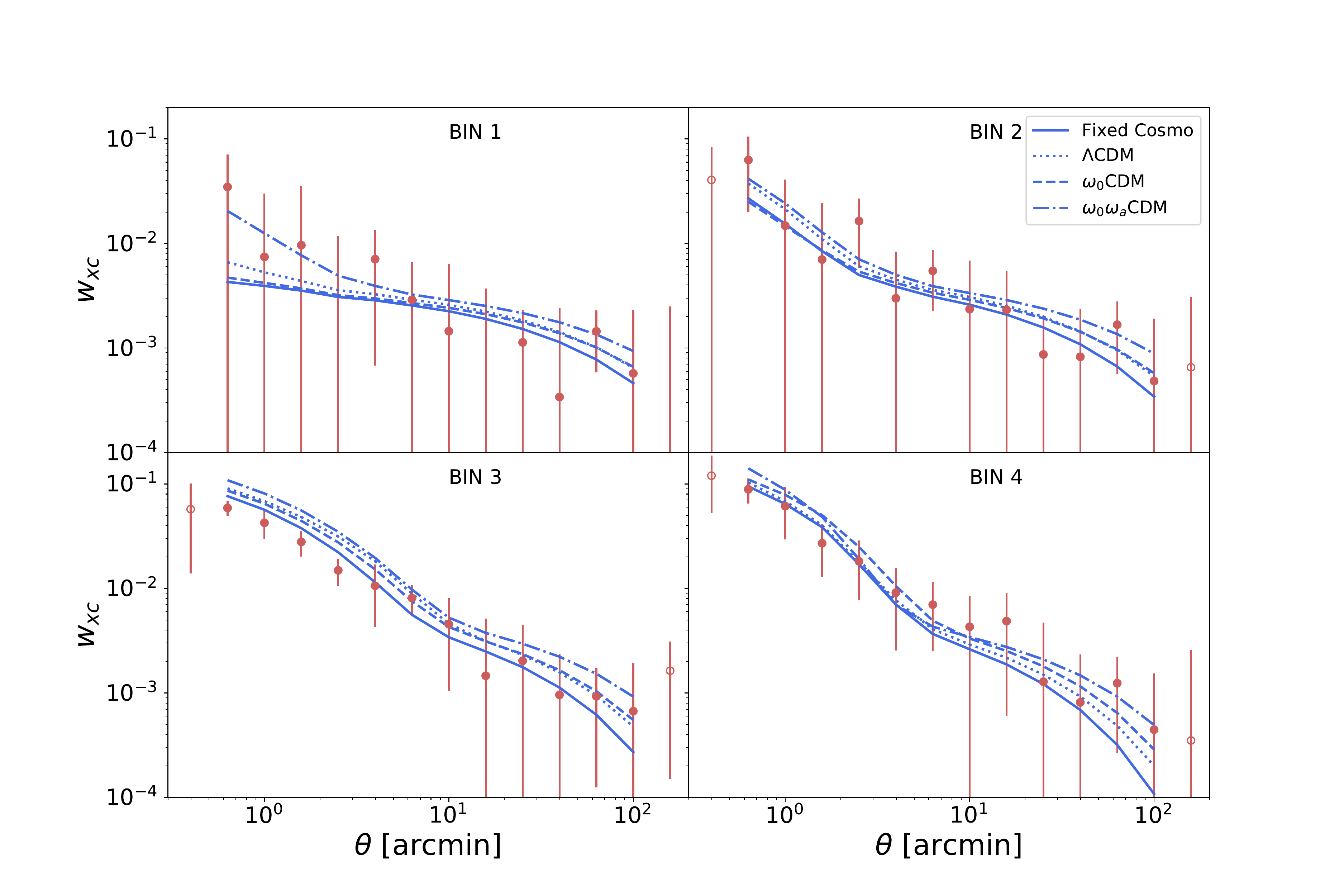}
\includegraphics[width=0.49\textwidth]{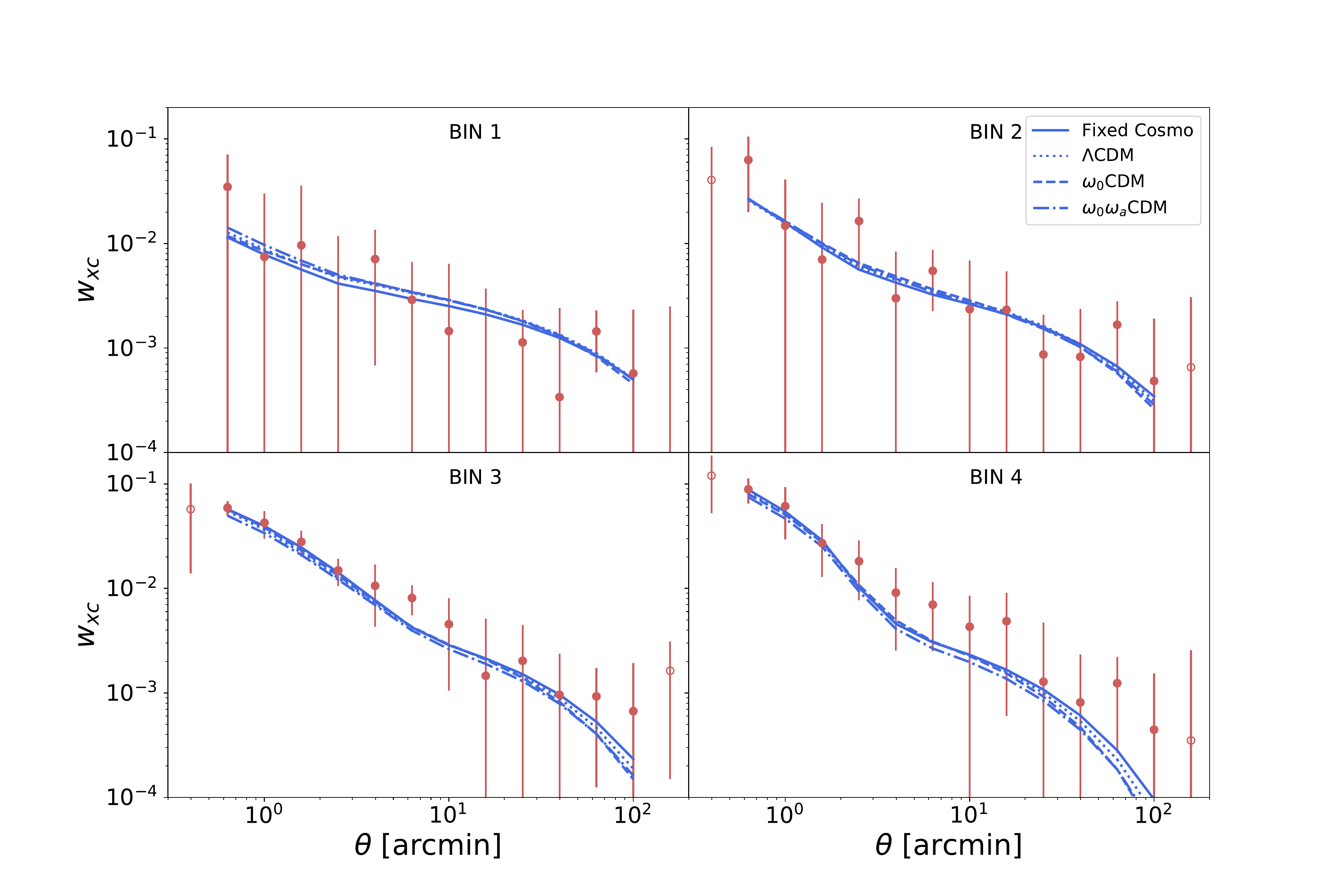}
    \caption{CCF data computed on Herschel SMGs (as background sample) and GAMA galaxies (as lenses sample) in the mini-Tiles data sets (red points) for the different redshift bins (from left to right, top to bottom): 0.1--0.2 (Bin1), 0.2--0.3 (Bin2), 0.3--0.5 (Bin3), and 0.5--0.8 (Bin4). The empty circles are the data discarded from the MCMC analysis, as explained in the text. The best fits performed in this work are also shown: the runs allowing for astrophysical parameters only (solid line) and for astrophysical and cosmological parameters:  $\sigma_8$, $\Omega_M$, and $h$ (dotted line); $\sigma_8$, $\Omega_M$, $h$, and $\omega_0$ (dashed line); and $\sigma_8$, $\Omega_M$, $h$, $\omega_0$, and $\omega_a$ (dash-dotted line). In the left panel the peaks of the marginalised 1D posterior distributions are used, and in the right panel the mean is used.}
    \label{Fig:xcorr_data}
\end{figure*}

The tiles scheme is the one described in \cite{GON21} and referred to as mini-Tiles therein, shown with red points in Fig. \ref{Fig:xcorr_data}. The selected redshift bins are $0.1-0.2$ (top left), $0.2-0.3$ (top right), $0.3-0.5$ (bottom left), and $0.5-0.8$ (bottom right). The measurement was  computed as in \citet{BON20}, the only difference being that median values were preferred because splitting the data into bins of redshifts reduces the statistics.
For each redshift bin, we jointly estimate the corresponding HOD parameters together with the cosmological parameters. The only difference with respect to the auto-correlation analysis is that, in the CCF case, the HOD describes only those galaxies and satellites acting as lenses, not the statistical properties of the full sample.

For the theoretical description, we adopt once again the halo model formalism by \citet{COO02}, as in \citet{GON17} and \citet{BON19,BON20}, and compute the foreground-background source CCF adopting the standard Limber \citep{LIM53} and flat-sky approximations \citep[see e.g.][and references therein]{KIL17} as
\begin{equation}
    \begin{split}
    w_{fb}(\theta)=2(\beta -1)\int^{\infty}_0 \frac{dz}{\chi^2(z)}n_f(z)W^{\text{lens}}(z) \\
    \int_{0}^{\infty}dl\frac{l}{2\pi}P_{\text{gal-dm}}(l/\chi^2(z),z)J_0(l\theta),
    \end{split}\label{crossth}
\end{equation}
where
\begin{equation}
    W^{\text{lens}}(z)=\frac{3}{2}\frac{H_0^2}{c^2}\bigg[\frac{E(z)}{1+z}\bigg]^2\int_z^{\infty} dz' \frac{\chi(z)\chi(z'-z)}{\chi(z')}n_b(z'),
\end{equation}
$n_b(z)$ ($n_f(z))$ is the unit-normalised background (foreground) redshift distribution, $\chi(z)$ is the comoving distance to redshift z, $J_0$ is the zeroth-order Bessel function of the first kind, and $\beta$ is the logarithmic slope of the background sources number counts $N(S) = N_0 S^{-\beta}$, which is assumed to be $\beta=3$ \citep[as in][]{LAP11, LAP12, CAI13, BIA15, BIA16, GON17, BON19}. The quantity 
\begin{equation*}
    E(z)\equiv \sqrt{\Omega_M(1+z)^3+\Omega_{\text{DE}}f(z)},
\end{equation*}
which neglects all contributions to the energy density from curvature or radiation, includes a redshift-dependent function $f(z)$ that considers the potential time evolution of the dark energy density. By assuming that the barotropic index of dark energy is given by
\begin{equation}
    \label{eq:w}
    \omega(z)=\omega_0+\omega_a\frac{z}{1+z},
\end{equation}
one easily obtains that the dark energy density evolves with redshift as $\rho(z)=\rho_0f(z)$, where
\begin{equation}
    f(z)=(1+z)^{3(1+\omega_0+\omega_a)}e^{-3\omega_a\frac{z}{1+z}}
\end{equation}
and $\rho_0$ is the dark energy density at $z=0$. Equivalently, the redshift evolution of the dark energy density parameter is given by
\begin{equation}
    \Omega_{\text{DE}}(z)=\frac{\Omega_{\text{DE}}f(z)}{\Omega_M(1+z)^3+\Omega_{\text{DE}}f(z)}.
\end{equation}
It should be noted that the cosmological constant is recovered by setting $\omega_0=-1$ and $\omega_a=0$.
 
Some comments should be made regarding the halo modelling of the galaxy-dark matter power spectrum in \eqref{crossth}, $P_{\text{gal-dm}}(k,z)$. As in the auto-correlation analysis, we use the HOD model of \citet{ZHE05} to describe how galaxies populate dark matter halos. Additionally, halos are identified as spherical regions with an overdensity equal to its virial value, which is estimated at every redshift using the approximation of \cite{WEI03}. We adopt the usual NFW halo density profile \citep{NAV96} and concentration parameter given in \cite{BUL01}.

Lastly, the cross-correlation is computed between each individual bin and the background sample, and not among the different bins. Thus, there are only two possible ways to introduce a cross-contamination between different foreground bins. The first, only important at small angular scales, would be the unlikely double lensing event or the alignment along the line of sight of at least three galaxies: two lenses in two different bins of the foreground sample and a source from the background sample. Considering the typical lensing probability, $\mathcal{O}(10^{-3})$, the double-lensing probability would be of $\mathcal{O}(10^{-6})$, thus completely negligible from the statistical point of view. 

The second possibility is related to a background source amplified by a lens belonging to a higher redshift bin that appears by chance at intermediate angular separation from a foreground galaxy belonging to a lower redshift bin. From the point of view of the lower redshift bin sample, as the lens and the foreground galaxy are not physically related, this magnified background source is indistinguishable from the other background field galaxies. In other words, this background source is taken into account both from the data and the random pairs counting, cancelling out any potential contribution to the final measured CCF for the lower redshift bin sample. For this reason, this second potential cross-contamination is also negligible. Therefore, the CCFs estimated for the different foreground redshift bins can be considered statistically independent.

\subsection{Parameter estimation}
\label{subsec:paramest}

The estimation of parameters is carried out via a Markov chain Monte Carlo (MCMC) algorithm, performed using the open source {\texttt{emcee}} software package \citep{EMCEE}, which is an MIT licensed pure-Python implementation of the Affine Invariant MCMC Ensemble sampler by \cite{GOO10}. In each MCMC run the number of walkers is set to four times the number of parameters to be estimated, $N$, and the number of iterations is chosen between 3000 and 5000. Therefore, up to $20000 N$ posterior samples are generated per run. With these posteriors, we then compute an average autocorrelation time, $\tau$, of the estimated autocorrelation times for each parameter (with the \texttt{get\_autocorr\_time()} function). Next we process the MCMC output by introducing a burn-in of $2\tau$ and a thinning of $\tau/2$ when flattening the chain\footnote{as suggested in the tutorials section of the official \texttt{emcee3.0.2} documentation at \texttt{https://emcee.readthedocs.io/en/stable/}}. The resulting chains differ in length for each case, which depends not only on the different number of walkers used in the MCMC run, but also on the $\tau$ value on each case. We lastly test the convergence of the chains using \texttt{GETDIST} \citep{GETDIST}.

As already discussed, in the most basic run the astrophysical parameters to be estimated in our analysis are $M_{min}$, $M_1$, and $\alpha$ (different for each redshift bin), while the cosmological parameters are $\Omega_M$, $\sigma_8$, and $h$ (defined via $H_0=100\,h\,\,\text{km}\,\text{s}^{-1}\text{Mpc}^{-1}$) since with the current samples we cannot constrain $\Omega_b$, $\Omega_{\text{DE}}$, and $n_s$ yet. Consequently, we assume a flat universe with $\Omega_{\text{DE}}=1-\Omega_M$ and keep the rest fixed to the  best-fit \citet{PLA18_VI} values of $\Omega_b=0.0486$ and $n_s=0.9667$. As already explained, this tomographic analysis also introduces the possibility of a redshift variation of the dark energy density and allows us to explore beyond $\Lambda$CDM by estimating the $\omega_0$ and $\omega_a$ parameters, according to equation \ref{eq:w}. Moreover, given that we are in the weak lensing approximation \citep[see][for a detailed discussion]{BON19}, only the CCF data in this regime ($\theta \ge 0.2$ arcmin) are being taken into account (red filled circles in Fig. \ref{Fig:xcorr_data}).

Within the Bayesian approach to parameter estimation,  both a prior and a likelihood distribution need to be specified. For the latter 
a traditional Gaussian likelihood function is used.
Taking into account the spectroscopic redshifts of the foreground sample (i.e. no overlap between bins) and that the different CCFs can be considered statistically independent (i.e. negligible cross-contamination), there is no need to take into account the covariance inside the likelihood function.

With regard to the selection of the astrophysical parameter priors, we perform a preliminary MCMC run on astrophysical parameters only (setting all cosmological parameters to \textit{Planck} values) with flat (uniform) priors on each redshift bin corresponding to those of \citet{BON20}. With this test we conclude that Bin 1 results indicate lower masses for the lenses: the galaxies in this redshift bin are very local, so their mass is smaller than that estimated in \citet{BON20} and we need a smaller value for $M_{min}$. On the other hand, Bin 4 needs higher priors for $M_{min}$: either the mass of the lenses is bigger at higher redshifts or we need higher mass to measure the lensing effect.
As a consequence, considering that masses are likely to increase with redshift, we decide to set the following uniform priors for the astrophysical parameters in redshift bin $i$:\\\\ $\log{M_{min}}\sim\mathcal{U}[a_i,b_i]$, where
\begin{align*}
    [a_1,b_1]&=[10.0,13.0]\quad[a_2,b_2]=[11.0,13.0],\\
    [a_3,b_3]&=[11.5,13.5]\quad[a_4,b_4]=[13.0,15.5];
\end{align*}
$\log{M_{1}}\sim\mathcal{U}[c_i,d_i]$, where
\begin{align*}
    [c_1,d_1]=[12.0,15.5]\quad[c_2,d_2]=[12.0,15.5],\\
    [c_3,d_3]=[12.5,15.5]\quad[c_4,d_4]=[13.0,15.5];
\end{align*}
and $\alpha\sim\mathcal{U}[0.5,1.5]$ for all bins. Such values are different from the flat priors ranges set for the auto-correlation case ($\log{M_{min}}\sim\mathcal{U}[9,16]$, $\log{M_{1}}\sim\mathcal{U}[9,16]$ and $\alpha\sim\mathcal{U}[0.1,2.5]$): they are still included in the wider auto-correlation priors, but knowing already the outcome of the preliminary run, they have been chosen to be smaller in order to avoid possible waste of iteration in the runs. As regards the prior distributions of the cosmological parameters to be estimated, they are set to the same ones used in \cite{BON20}:   $\Omega_M\sim\mathcal{U}[0.1-0.8]$, $\sigma_8\sim\mathcal{U}[0.6-1.2]$, and $h\sim\mathcal{U}[0.5-1.0]$. The $\omega_0$ and $\omega_a$ uniform priors were chosen to follow the $\mathcal{U}[-2.0,0.0]$ and $\mathcal{U}[-3.0,3.0]$, respectively.

%--------------------------------------------------------------------
\section{Astrophysical constraints with fixed $\Lambda$CDM cosmology}
\label{sec:astro_results}

This section is dedicated to discussing the results from runs where the astrophysical parameters were the only free parameters within a \textit{Planck} $\Lambda$CDM cosmology using the auto- and cross-correlation function measurements. The corner plots and more detailed results of all the cases described here are provided in Appendices \ref{App:auto_data} and \ref{App:corner_table}.

\subsection{Auto-correlation analysis}

\begin{figure*}
    \centering
    \includegraphics[width=0.9\textwidth]{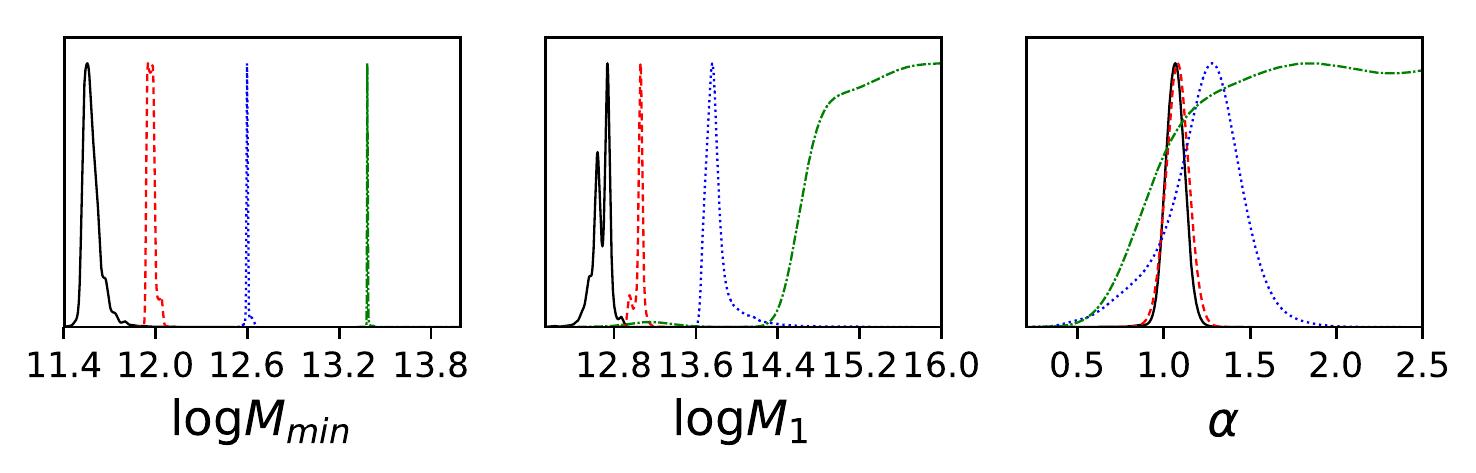}
    \caption{One-dimensional posterior distributions of the HOD parameters $\log{(M_{min}/M_{\odot})}$, $\log{(M_1/M_{\odot}),}$ and $\alpha$ (from left to right) in each redshift bin for the auto-correlation run. Results for Bins 1 to 4 are depicted as black solid, red dashed, blue dotted, and green dash-dotted lines, respectively.}
    \label{fig:auto_1D_distro}
\end{figure*}

The priors and results for the auto-correlation run are listed in Table \ref{Tab:auto-corr}: from left to right, the columns denote the redshift bin, the parameter in question, its prior distribution and the mean ($\mu$), median, $68\%$ credible interval and peak of its marginalised 1D posterior distribution. Figure \ref{fig:auto_1D_distro} depicts the one-dimensional  posterior distributions of the astrophysical parameters for this run: from left to right, $\log M_{min}$, $\log M_1$, and $\alpha$ for Bin 1 (black solid line), Bin 2 (red dashed line), Bin 3 (blue dotted line), and Bin 4 (green dash-dotted line).

An increase with redshift in $\log{M_{min}}$ is observed, the peak in the posterior distributions being at values of 11.55, 11.95, 12.60, and 13.38, from Bin 1 to Bin 4. This is due to a typical selection effect: at a higher redshift, objects with a higher luminosity are more likely to be observed \citep[see Fig. 23 in][]{DRI11}. The same increase is present in the plot for $\log{M_1}$, peaking at 12.74, 13.06, and 13.76 in Bin 1, Bin 2, and Bin 3, respectively. This parameter does not display a peak for Bin 4, as happens for $\alpha$, which cannot be constrained in this redshift interval. As regards the rest of the bins, $\alpha$ peaks at almost the same value in Bin 1 and Bin 2 (1.07 and 1.08, respectively) and at a slightly higher value (1.28) in Bin 3.

\subsection{CCF analysis}
\label{sec:results_astro_xcorr}
\begin{figure*}
    \centering
    \includegraphics[width=0.9\textwidth]{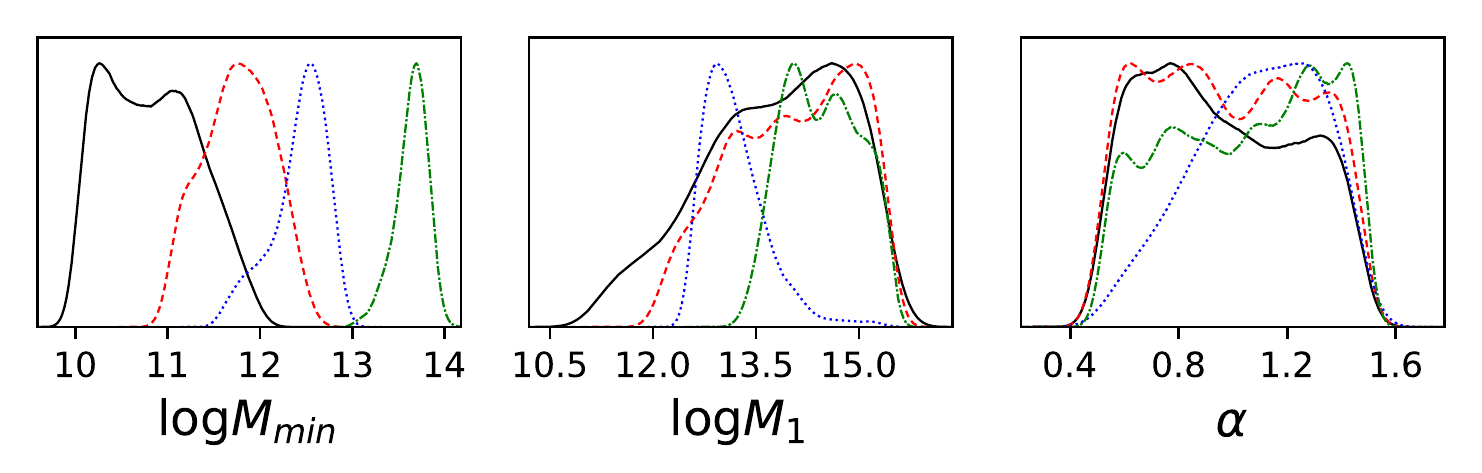}
    \caption{One-dimensional posterior distributions of the HOD parameters $\log{(M_{min}/M_{\odot})}$, $\log{(M_1/M_{\odot})}$ and $\alpha$ (from left to right) in each redshift bin for the CCF run with a fixed cosmology. Results for Bins 1 to 4 are depicted as black solid, red dashed, blue dotted and green dash-dotted lines, respectively.}
    \label{Fig:1D_FPA}
\end{figure*}

As already discussed, we first performed a MCMC run on only the astrophysical parameters in all four redshift bins, the cosmology being fixed to \textit{Planck} values.  The resulting corner plots for each bin are depicted in Fig. \ref{Fig:FPA}, and the corresponding marginalised posterior distributions are summarised in Table \ref{Tab:FPA}.
The one-dimensional posterior distributions for the $\log{M_{min}}$, $\log{M_1}$, and $\alpha$ parameters are shown in Fig. \ref{Fig:1D_FPA} (from left to right) for the four different bins (black solid, red dashed, blue dotted, and green dash-dotted lines for Bin 1, 2, 3, and 4, respectively). 

\begin{figure}
    \centering
    \includegraphics[width=0.45\textwidth]{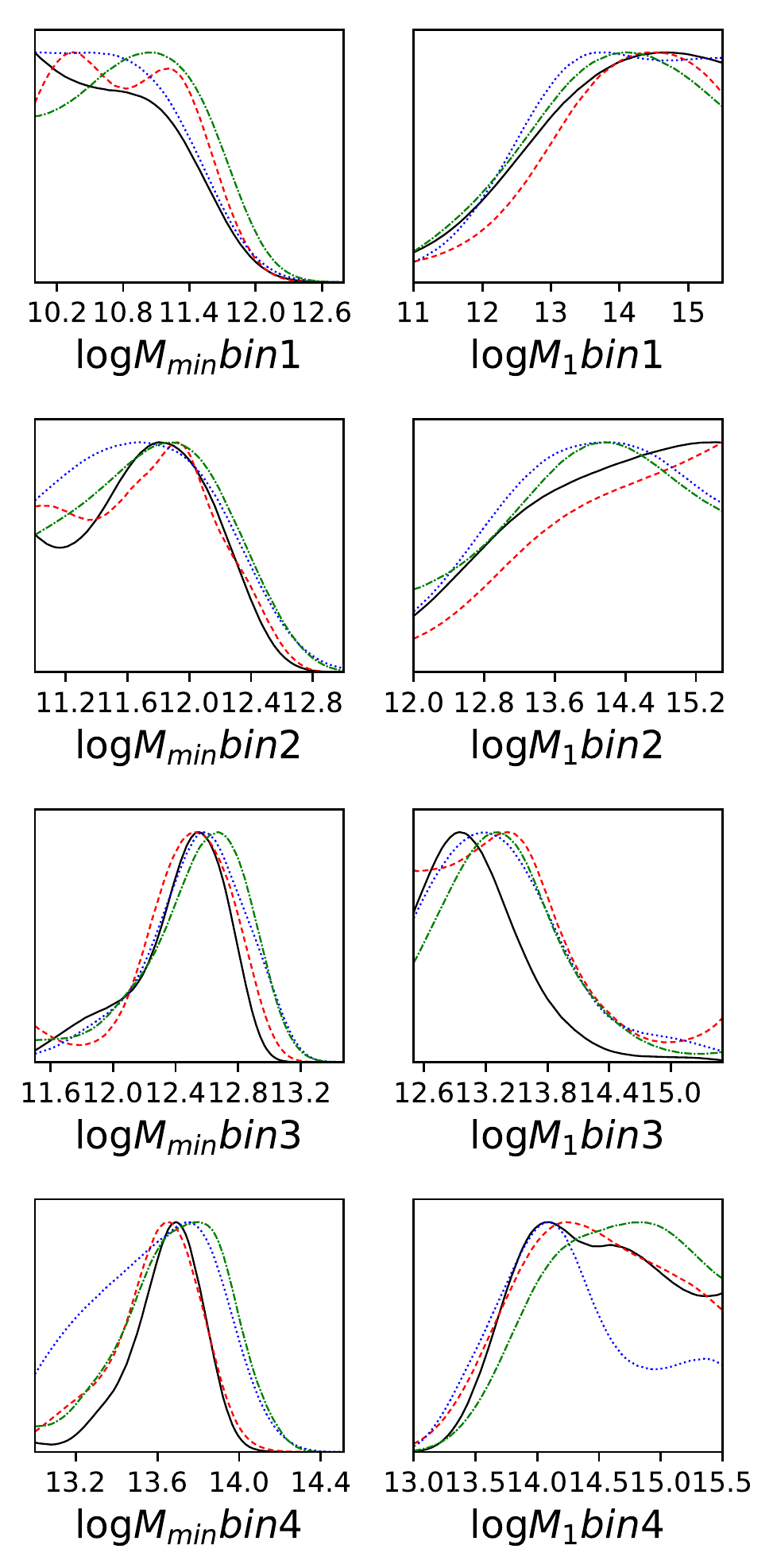}
    \caption{One-dimensional posterior distributions for $\log{(M_{min}/M_{\odot})}$ (first column) and $\log{(M_1/M_{\odot})}$ (second column) in Bins 1, 2, 3, and 4 (from top to bottom, respectively) for the run on just the HOD parameters (black solid lines) and for the three runs on the astrophysical and cosmological parameters jointly, namely   within $\Lambda$CDM (red dashed lines),   within $\omega_0$CDM (blue dotted lines), and   within $\omega_0\omega_a$CDM (green dash-dotted lines).}
    \label{Fig:Astro_all}
\end{figure}

The results regarding $M_{min}$ show a general agreement with what is found in the auto-correlation run. The wider distributions still indicates an increase with redshift, and although Bin 1 results are not informative (only providing a 68\% upper limit of 11.09), Bin 2 and Bin 3 peak at 11.80 and 12.54, respectively, and a slight disagreement can be observed for Bin 4, which peaks at 13.69. 

The results for $M_1$ are generally different with respect to the auto-correlation case, where the increase with redshift is highlighted in Fig. \ref{fig:auto_1D_distro}. In particular, in Bin 1 and Bin 2 not much can be said about the posterior distributions (lower limits at 13.28 and 13.56, respectively). Moreover, Bin 3 peaks at 12.95 and Bin 4 at 14.09, contrary to the auto-correlation results, which provide just a lower limit. Concerning the $\alpha$ parameter, Bin 3 is the only one showing a peak at 1.16 in the marginalised posterior distribution, being the results on the rest of the bins inconclusive. 
A similar trend is found by \cite{GON17}: they obtain $M_{min}=12.61^{+0.06}_{-0.09}$ and $M_1=15.01^{+0.30}_{-0.38}$ for Bin 1, $M_{min}=12.74^{+0.09}_{-0.11}$ and $M_1=14.79^{+0.52}_{-0.38}$ for Bin 2, $M_{min}=13.27^{+0.11}_{-0.12}$ and $M_1=14.97^{+0.35}_{-0.38}$ for Bin 3, and $M_{min}=14.36^{+0.14}_{-0.10}$ and $M_1=14.96^{+0.28}_{-0.19}$ for Bin 4.
The results are also in agreement with those from the corresponding non-tomographic run by \cite{BON20}: they give $M_{min}=12.41^{+0.17}_{-0.07}$ and $M_1=13.65^{+0.29}_{-0.54}$ at 68\% C.I.

Therefore, there is a clear disagreement between the CCF and the auto-correlation results. This is not unexpected since the CCF is related only to the characteristics of the lenses and not to those of the global sample. From the point of view of the CCF, those foreground galaxies not acting as lenses are invisible or non-existent, and they only contribute to the random part. 
For example, the higher $M_{min}$ value in the last bin with respect to the auto-correlation case suggests that, at a higher redshift (with respect to the other three bins), the mass of the object acting as a lens is higher than the average mass of the whole Bin 4 sample. This is due to the fact that, at a greater distance, only the most massive objects can actually act as lenses in order to compensate the lower lensing probability \citep{LAP12,GON17} . We therefore think that it is not safe to use the auto-correlation results as Gaussian priors on the astrophysical parameters for the cosmological runs, or even to perform a joint analysis, since they might be non-representative of the lens sample in the CCF data. Any use of information from the auto-correlation analysis would introduce an uncontrollable bias in the cosmological results. 
Moreover, given the tight auto-correlation HOD constraints, they will completely overrun the cosmological analysis  if used as Gaussian priors or in a joint analysis.
For these reasons we decide to focus  on the CCF measurements and to only perform runs with uniform HOD priors in this work.

\section{Cosmological constraints}
\label{sec:cosmo_results}
This section constitutes the core of this work. We study the cosmological constraints that can be derived from the measured CCFs. We performed a series of MCMC runs on the HOD and cosmological parameters jointly under the assumptions of different cosmological models ($\Lambda$CDM, $\omega_0$CDM, and $\omega_0\omega_a$CDM). In particular, we focus on the cosmological parameters $\Omega_M$, $\sigma_8$, $h$, $\omega_0$, and $\omega_a$, to whose effect the CCF is more sensitive, while keeping the rest fixed to \textit{Planck} values. The corner plots and more detailed results of all the cases described here are provided in Appendix \ref{App:corner_table}.

\begin{figure*}
    \centering
    \includegraphics[width=0.9\textwidth]{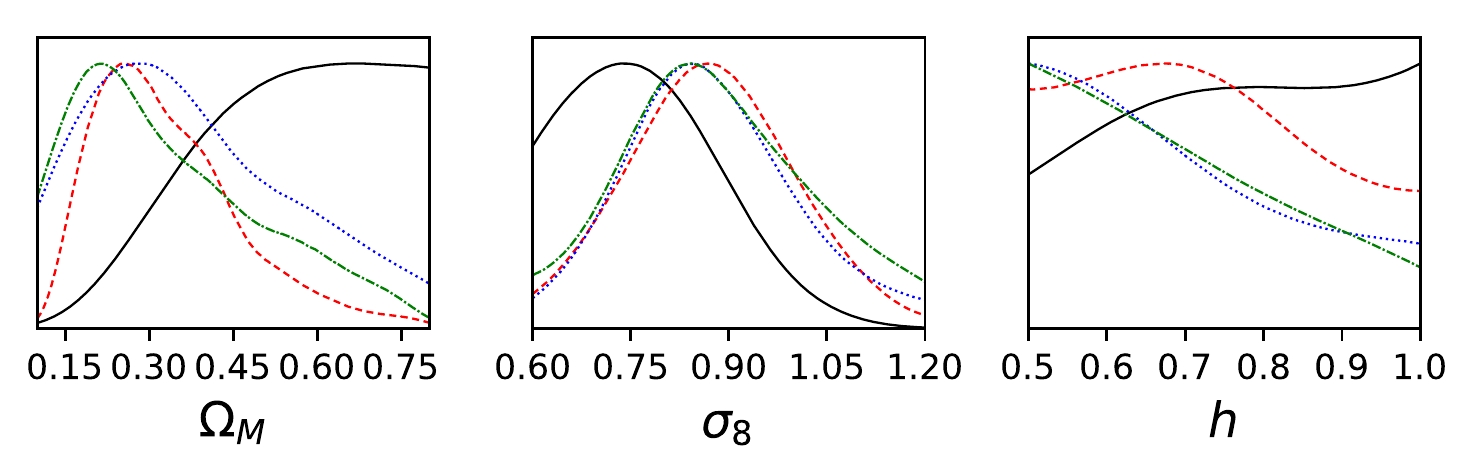}
    \caption{One-dimensional posterior distributions of $\Omega_M$, $\sigma_8$, and $h$ (from left to right) for the results in \cite{BON20} (black solid lines), and for the three MCMC runs performed jointly on the astrophysical and cosmological parameter in this work, namely   within $\Lambda$CDM (red dashed lines),   within $\omega_0$CDM (blue dotted lines), and  within $\omega_0\omega_a$CDM (green dash-dotted lines).}
    \label{Fig:Cosmo_all}
\end{figure*}

\subsection{$\Lambda$CDM model vs $\omega_0$CDM and $\omega_0\omega_a$CDM models}

We performed three different MCMC runs: the first within the $\Lambda$CDM cosmology (i.e. setting $\omega_0$ and $\omega_a$ to -1 and 0, respectively); the second  within the so-called flat $\omega_0$CDM model (where $\omega_0$ is a free parameter, but $\omega_a$ is set to 0); and the  third and   most general model, a flat $\omega_0\omega_a$CDM model (where both parameters are left free). In all three runs the astrophysical HOD parameters are also allowed to vary in the different ranges we discussed in \ref{subsec:paramest}.

The derived marginalised posterior distributions of the astrophysical $\log{M_{min}}$ (first column) and $\log{M_1}$ (second column) parameters in each redshift bin (1 to 4, from top to bottom) are compared in Fig. \ref{Fig:Astro_all} ($\alpha$ is not shown because it is unconstrained in most of the cases). We also add the results from the fixed cosmology case (see section \ref{sec:astro_results}). There is overall agreement among all cases.

On the other hand, Fig. \ref{Fig:Cosmo_all} compares the corresponding results for the cosmological $\Omega_M$, $\sigma_8$ and $h$ parameters (from left to right, respectively) of the three cosmological models assumed in this subsection (depicted as before with red dashed, blue dotted, and green dash-dotted lines) with the previous results by \cite{BON20} (shown in solid black). 

The results for the $\Lambda$CMD model, which includes only $\Omega_M$, $\sigma_8$, and $h$ in the cosmological analysis, are shown in Fig. \ref{Fig:FPA_FPC} and Table \ref{Tab:FPA_FPC}. The trend for the astrophysical parameters (see red dashed line in Fig. \ref{Fig:Astro_all}) is similar to the fixed cosmology run in section \ref{sec:astro_results}. For the cosmological parameters, there is a substantial improvement with respect to the non-tomographic analysis carried out in \cite{BON20} (black solid line in Fig. \ref{Fig:Cosmo_all}).
They obtain a lower limit of $ 0.24$ at 95\% C.I. on $\Omega_M$, only a tentative peak around 0.75 for $\sigma_8$ with an upper limit of $1$ at 95\% C.I., while we have a detection of $\Omega_M = 0.33$ [0.17, 0.41] (mean value and 68\% C.I.) and a maximum in the posterior distribution at 0.26.
We note that the preference for lower values of $\Omega_M$ in the present analysis with respect to \cite{BON20} is a consequence of a better treatment of the estimator biases \citep{GON21}.
We also report a constraint on $\sigma_8$ of $0.87$ [0.75, 1.0] at 68\% C.I. The $\Omega_M$ and $\sigma_8$ values are respectively slightly lower and higher than those found in \citet{GON21} (0.50 and 0.75, respectively).
However, when opening the dark energy related parameters, the constraints on $\sigma_8$ and $\Omega_M$ loosen, especially that on $\Omega_M$.
In the $\Lambda$CDM case the one-dimensional plots in Fig. \ref{Fig:Cosmo_all} (red dashed lines) display clear peaks except in the case of the $h$ parameter, where we just obtain a low significant peak at 0.67 for $h$.

The corresponding corner plots and summarised results from the $\omega_0$CMD model, which allows the $\omega_0$ parameter to be free while keeping $\omega_a$ fixed to 0, are shown in Fig. \ref{Fig:FPA_FPC_w0} and Table \ref{Tab:FPA_FPC_w0}. The trend for the parameter distributions in this second case (blue dotted lines in Fig. \ref{Fig:Astro_all} and \ref{Fig:Cosmo_all}) is similar to the $\Lambda$CDM case. The only exceptions are a broader posterior distribution for $M_{min}$ in Bin2 and Bin4, a clear peak for $M_1$ in Bin 2 (there is no peak in the $\Lambda$CDM case), and $h$, which is not constrained, showing a preference towards lower values. In particular, the $\Omega_M$ and $\sigma_8$ parameters peak at 0.26 and 0.85, respectively (being compatible with the findings of the $\Lambda$CDM case). 

The results from the third and most general model $\omega_0\omega_a$CDM, which gives freedom to both $\omega_0$ and $\omega_a$ are shown in Fig. \ref{Fig:FPA_FPC_wa} and table \ref{Tab:FPA_FPC_wa}. As shown in Fig. \ref{Fig:Astro_all} and Fig. \ref{Fig:Cosmo_all} (green dash-dotted lines), there is overall agreement with the other cases: we obtain peaks for $\Omega_M$ and $\sigma_8$ at 0.21 and 0.84, respectively, which are slightly lower than the other cases. Instead, $h$  is unconstrained, favouring lower values as in the other cases.

\begin{figure}
    \centering
    \includegraphics[width=0.45\textwidth]{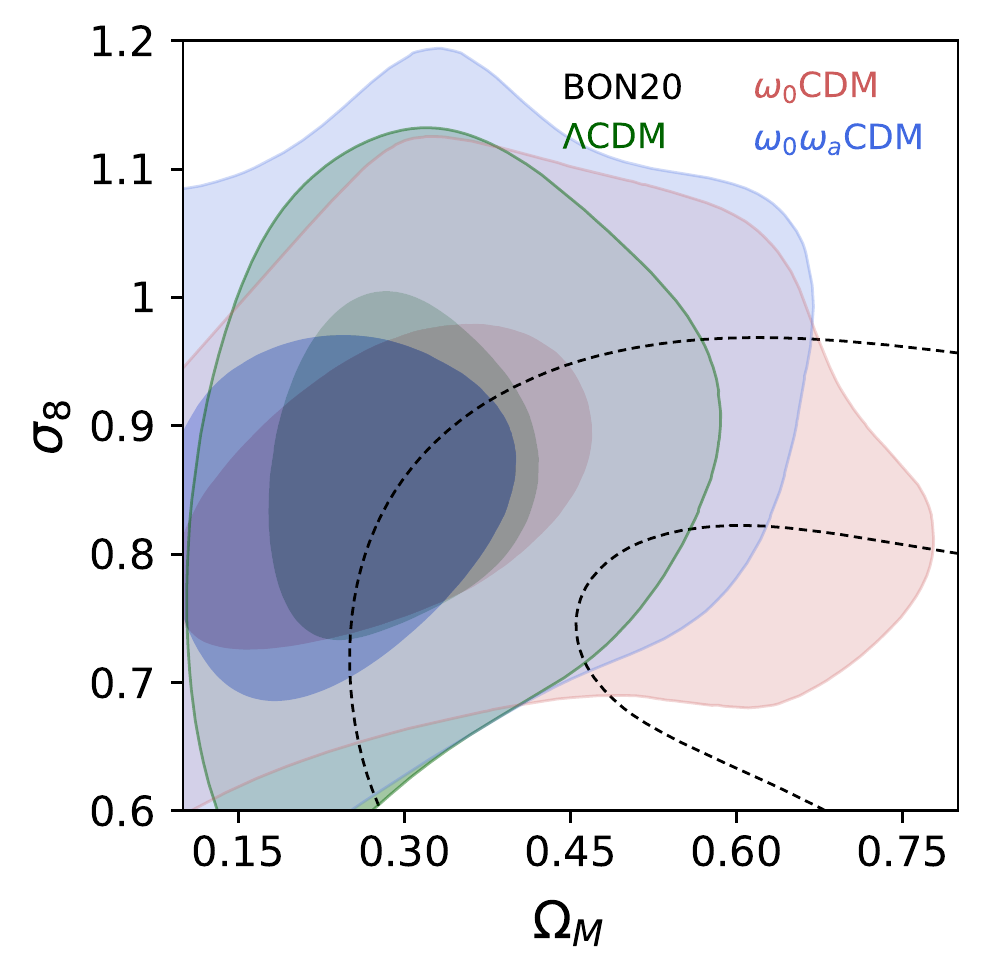}
    \caption{Contour plot of the two-dimensional posterior distribution of $\Omega_M$ and  $\sigma_8$ for the three MCMC runs where the astrophysical and cosmological parameters are jointly analysed, namely within $\Lambda$CDM (in green), within $\omega_0$CDM (in red), and within $\omega_0\omega_a$CDM (in blue). Moreover, in order to compare with previous results, those by \citet{BON20} are shown in black. The contours are set to 0.393 and 0.865.}
    \label{Fig:Comp_Om_s8}
\end{figure}

The contour plots of the two-dimensional posterior distribution for $\Omega_M$ and $\sigma_8$ in all three cases are shown in Fig. \ref{Fig:Comp_Om_s8} (in green, red, and blue, respectively)\footnote{The $1\sigma$ and $2\sigma$ levels for the 2D contours are 39.3\% and 86.5\%, not 68\% and 95\%. Otherwise, there is no direct comparison with the 1D histograms above the contours.}. For comparison purposes, the results by \citet{BON20} are also shown (in black). 
As in Fig. \ref{Fig:Cosmo_all}, the improvement with respect to previous results with magnification bias can be appreciated: the lower limit on $\Omega_M$ is still present, especially in the $\Lambda$CDM case and even if it lowers and worsens in the $\omega_0$CDM and $\omega_0\omega_a$CDM cases (see Tables \ref{Tab:FPA_FPC_w0} and \ref{Tab:FPA_FPC_wa}, respectively), they are in general more constraining in the $\Omega_M$-$\sigma_8$ plane with respect to the non-tomographic case. On the other hand, upper constraints to the distributions are achieved in the tomographic case with the three runs.

The best-fit models for the CCFs obtained in all these cases are shown in Fig. \ref{Fig:xcorr_data}. In the left panel we use the peak of each one of the marginalised 1D posterior distributions (although it is not the same as the maximum of the $n$-dimensional posterior distribution), while we use the associated mean values for the right panel. In particular, the solid line represents the results from the fixed cosmology case whereas the dotted, dashed, and dash-dotted lines are used for the three runs where cosmology was taken into account ($\Lambda$CDM, $\omega_0$CDM, and $\omega_0\omega_a$CDM, respectively). When using the maxima of the posterior distributions, the fixed cosmology results provides the lowest curves while the $\omega_0\omega_a$CDM one the highest. On the other hand, when using the mean values major differences among the different runs are not found.

The reduced $\chi^2$ values for the best fits with the mean (peaks values of the 1D posterior distributions) are 0.36 (0.54), 0.44 (1.42), 0.44 (1.14), and 0.56 (2.86) for the cases of  fixed cosmology (36 d.o.f.), $\Lambda$CDM (33 d.o.f.), $\omega_0$CDM (32 d.o.f.), and $\omega_0\omega_a$CDM (31 d.o.f.), respectively. It should be noted that when using the $n$-dimensional best fit, the reduced $\chi^2$ obviously improves (1.05 for the $\omega_0\omega_a$CDM case). The bigger values for the peaks case are mostly due to the worst best fit in Bin 3. It is interesting that the $\omega_0$CDM case has a comparable or better $\chi^2$ with respect to the $\Lambda$CDM case. It should be noted that \emph{emcee} is not an optimiser, but a sampler. As such, and in the context of Bayesian statistics, the single tuple of parameters that best fits the data (that is, the maximum of the $n$-dimensional posterior distribution) cannot be associated with credible intervals for each individual parameter. What can be done is to marginalise over the rest of the parameters in order to obtain one-dimensional distributions, the maximum of which will not, in general, coincide with the individual value of the aforementioned tuple. This explains why $\chi^2$ values for individual fits have to be interpreted with care, since the most meaningful information comes from the overall distributions, not from point estimates.
In conclusion, the different cases considered in this work show a general agreement  regarding the posterior distributions, although there are slight changes in the peak positions, well within  1$\sigma$ limits in most cases.

\begin{figure}
    \centering
    \includegraphics[width=0.45\textwidth]{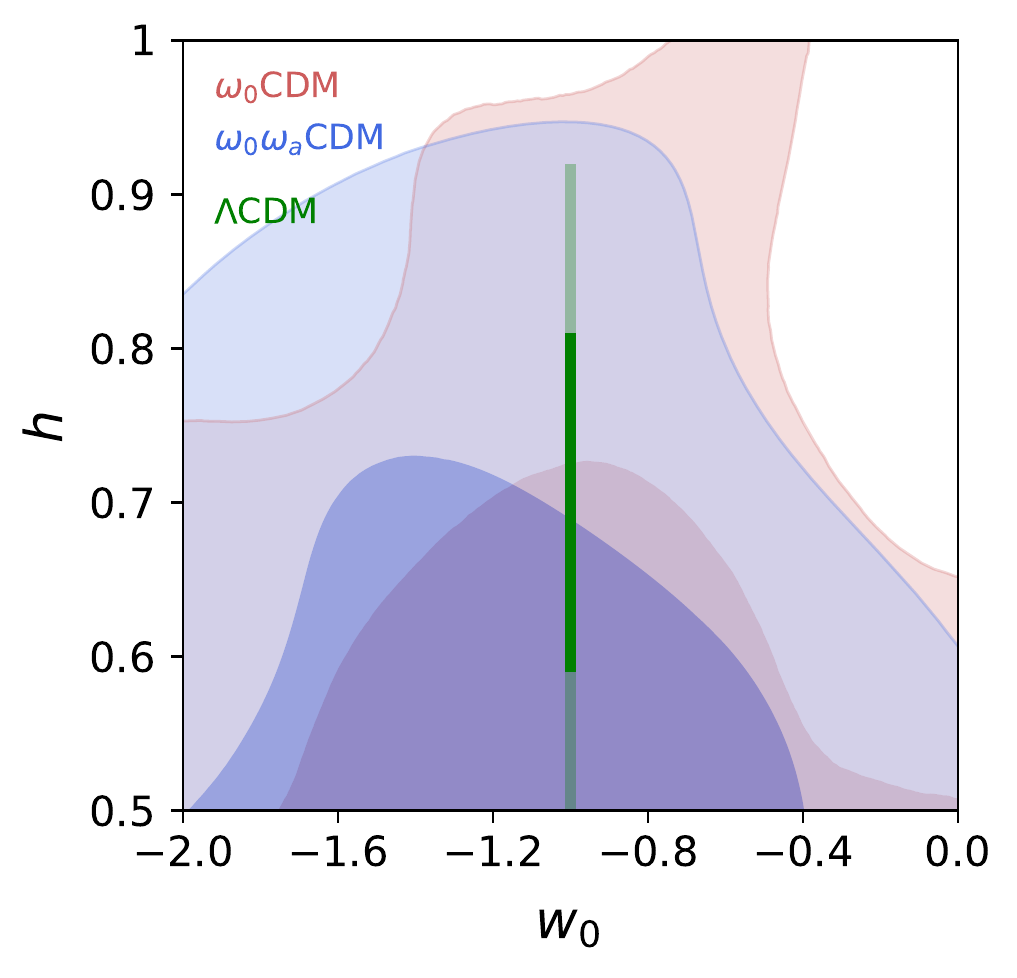}
    \caption{Contour plot of the two-dimensional posterior distribution of $h$ and  $\omega_0$ for the three MCMC runs where the astrophysical and cosmological parameters are jointly analysed, namely  within $\Lambda$CDM (in green), the  within $\omega_0$CDM (in red), and  within $\omega_0\omega_a$CDM (in blue). The contours are set to 0.393 and 0.865.}
    \label{Fig:Comp_h_w0}
\end{figure}

\begin{figure}
    \centering
    \includegraphics[width=0.45\textwidth]{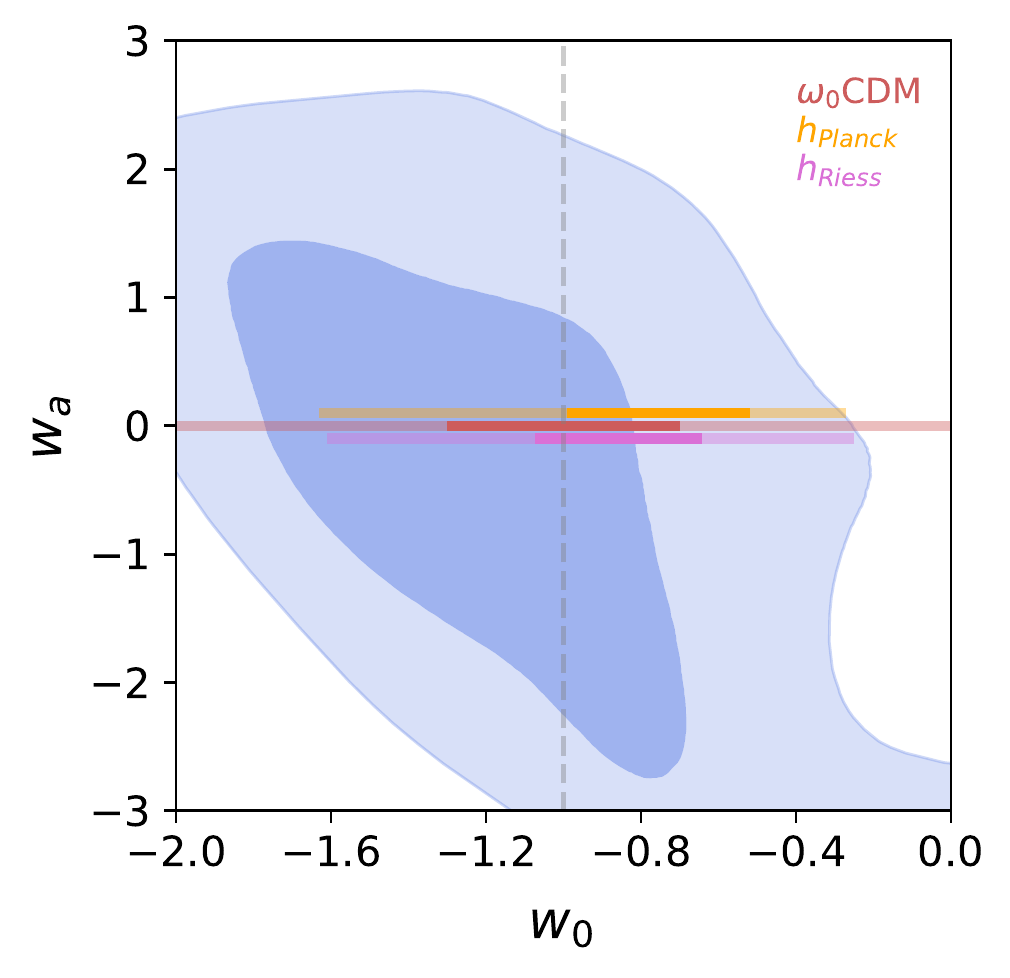}
    \caption{Contour plot of the two-dimensional posterior distribution of $\omega_0$ and  $\omega_a$ for the MCMC run within $\omega_0\omega_a$CDM. The contours are set to 0.393 and 0.865.}
    \label{Fig:Comp_wa_w0}
\end{figure}

\subsection{The Dark Energy equation of state}
In this subsection we focus on the dark energy equation of state and what we can learn about it from our results.
Figure \ref{Fig:Comp_h_w0} shows the contour plot of the two-dimensional posterior distribution for $h$ and $\omega_0$. The results from the run within $\omega_0$CDM is shown in red, while those corresponding to $\omega_0\omega_a$CDM are depicted in blue. The green bar represents the run within $\Lambda$CDM. We find a mean value of $-1.00$ and a maximum of the posterior distribution of -0.97 for $\omega_0$ ($[-1.56, -0.47]$ 68\% C.I.) for the $\omega_0$CDM model. Under the $\omega_0\omega_a$CDM model we derive -1.09 and -0.92 for $\omega_0$ $[-1.72,-0.66]$ C.I. and -0.19 and -0.20 for $\omega_a$ $[-1.88,1.48]$ 68\% C.I. for the $\omega_0\omega_a$CDM model.

Moreover, Fig. \ref{Fig:Comp_wa_w0} displays the contour plot of the two-dimensional posterior distribution of $\omega_0$ and $\omega_a$. 
There seems to be a degeneracy direction, with $\omega_a$ having higher values for lower values of $\omega_0$. This anti-correlation is at 30\%.
Taking into account the results from these alternative models ($\omega_0$CDM and $\omega_0\omega_a$CDM), our measurements are in agreement with  $\Lambda$CDM ($\omega_0=-1$ and $\omega_a$=0) at  68\% C.I.

However, since the previous cases were not able to constrain the $h$ parameter, we now investigate if $\omega_0$ is somehow affected by changes in $H_0$. In order to do so, we performed a MCMC run on the astrophysical parameters and $\omega_0$ by fixing the rest of the cosmology to \textit{Planck} values, and then comparing the results with a similar MCMC run with the exception of $H_0$ being fixed to the value from \citet[][$H_0=74.03 \pm 1.42 \,\text{km} s^{-1} \text{Mpc}^{-1}$]{RIE19}.

The results are shown in Figs. \ref{Fig:FPA_w0} and \ref{Fig:FPA_w0_hRiess} and summarised in Tables \ref{Tab:FPA_w0} and \ref{Tab:FPA_w0_hRiess} (for the cases  fixing $H_0$ to the  \textit{Planck} and \cite{RIE19} values, respectively). 
The marginalised posterior distribution for $\omega_0$ peaks at -0.69 ([-1.31, -0.38] at 68\% C.I.) and -0.86 ([-1.29, -0.45] at 68\% C.I.) for the \textit{Planck} and Riess value of $H_0$, respectively. These results show that $\omega_0$ tends to higher values when the \textit{Planck} $H_0$ value is used. 

However, we note a degeneracy in the $\omega_0-M_{min}$ plane: a higher $\omega_0$ tends to prefer a lower $M_{min}$, a phenomenon that is evident in Bin 3, but especially in Bin 4. In both bins the degeneracy is at about  50\%. 
The same results can also be seen in the previous general runs (see Figs. \ref{Fig:FPA_FPC_w0} and \ref{Fig:FPA_FPC_wa}), for both redshift bins. In Bin 4 is where we generally have the greatest difference in $M_{min}$ when compared with the auto-correlation results. This difference disappear for higher $\omega_0$ values, making $M_{min}$ in agreement with the auto-correlation findings.

Therefore, we observe that fixing $H_0$ to the $h_{Riess}$ and $h_{Planck}$ values produces shifts towards $\omega_0>-1$ values in the derived maximum of the 1D marginalised posterior distribution. However, both cases are still compatible with the $\Lambda$CDM model at the 68\% C.I.

\subsection{Comparison with other results}

\begin{figure}
    \centering
    \includegraphics[width=0.45\textwidth]{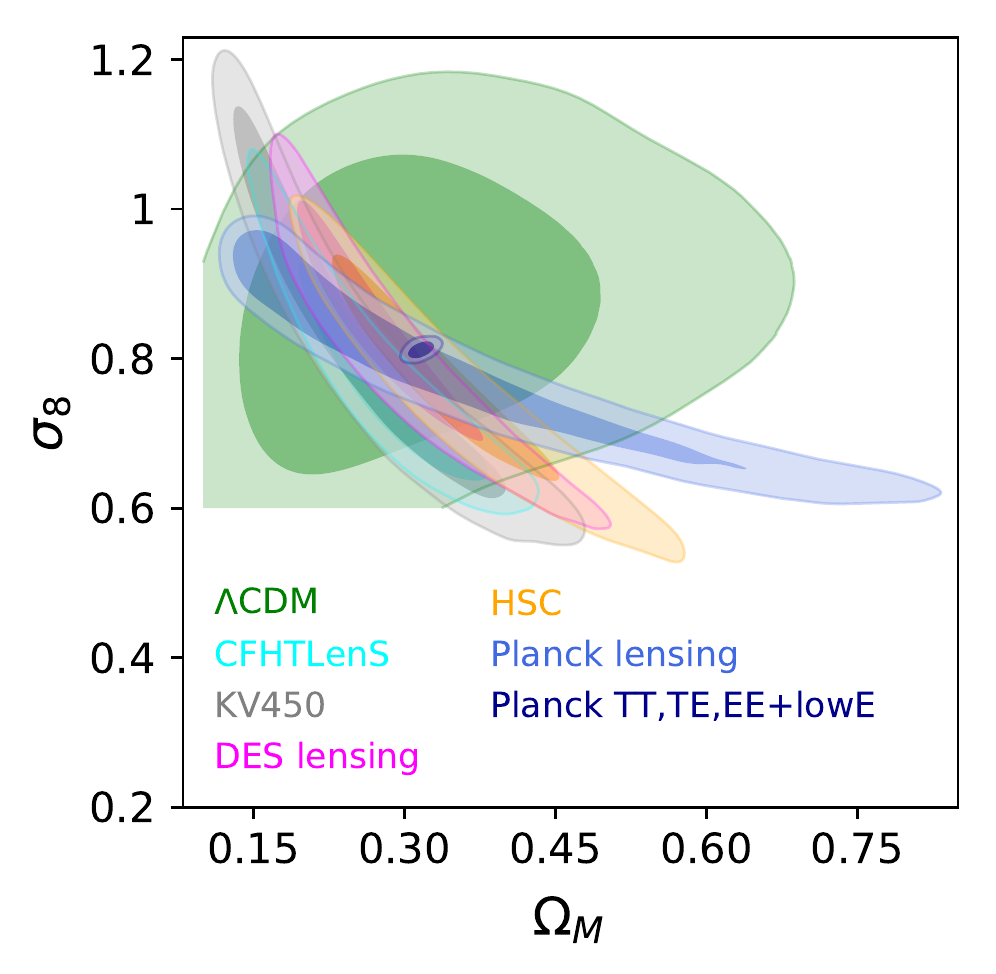}
    \caption{Comparison among the contour plots in the $\Omega_M$ -- $\sigma_8$ plane from \textit{Planck} CMB lensing (blue),CFHTLenS (cyan), KV450 (grey), DES (magenta), and HSC (yellow) and the results from this work (green). Differently from the other 2D plots in this work, the contours here are set to 0.68 and 0.95.}
    \label{Fig:Comp_ext_data}
\end{figure}

As in \cite{BON20}, we compare our cosmological constraints with those obtained using the shear of the weak gravitational lensing (some of which are known to be in tension). In particular, Fig. \ref{Fig:Comp_ext_data} shows the contour plots for the posterior distribution of $\Omega_M$ and $\sigma_8$ (68\% and 95\% C.I., for a direct comparison with the literature, which are different from the values of 39.3\% and 86.5\% used in the corner plots shown in the previous sections) from the publicly released CMB lensing from \textit{Planck} \citep[][in blue]{PLA18_VIII}, cosmic shear tomography measurements of the Canada-France-Hawaii Telescope Lensing Survey \citep[CFHTLenS,][in cyan]{JOU17}, first combined cosmological measurements of the Kilo Degree Survey and VIKING based on 450 $\text{deg}^2$ data \citep[KV450,][in grey]{HIL20}, first-year lensing data from the Dark Energy Survey \citep[DES][in magenta]{TRO18} and two-point correlation functions of the Subaru Hyper Suprime-Cam first-year data \cite[HSC,][in yellow]{HAM20}. For completeness, we also show results from \textit{Planck} CMB temperature and polarisation angular power spectra (in dark blue). The results of this work are shown in green ($\Lambda$CDM).

It should be noted that we do not attempt to adjust the different priors to our fiducial set-up as an in-depth comparison is beyond the scope of this paper. Moreover, it should be considered that these results span different redshift ranges and are affected by completely different systematic effects with respect to the magnification bias measurements (which thus represent a complementary probe to weak lensing). 

The combination of this external data set identifies degeneracy directions that are clearly visible in Fig. \ref{Fig:Comp_ext_data}.
As already anticipated in \citet{BON20}, the tomographic results confirm that the magnification bias does not show the  degeneracy  found  with  cosmic  shear  measurements.
In order to further confirm this conclusion, we perform a principal component analysis (PCA) on the $\Lambda$CDM results. The first principal component is $\sigma_8 \Omega_M^{-0.36}$, almost orthogonal to the cosmic shear degeneracy. At present this component is constrained at a level of $\sim18$\%, and therefore it cannot be considered a proper degeneracy. Better constraints are needed before deciding if magnification bias results can be further improved by analysing this component, as commonly done with $S_8$ in cosmic shear analyses.

\begin{figure}[ht]
\centering
\includegraphics[width=0.49\textwidth]{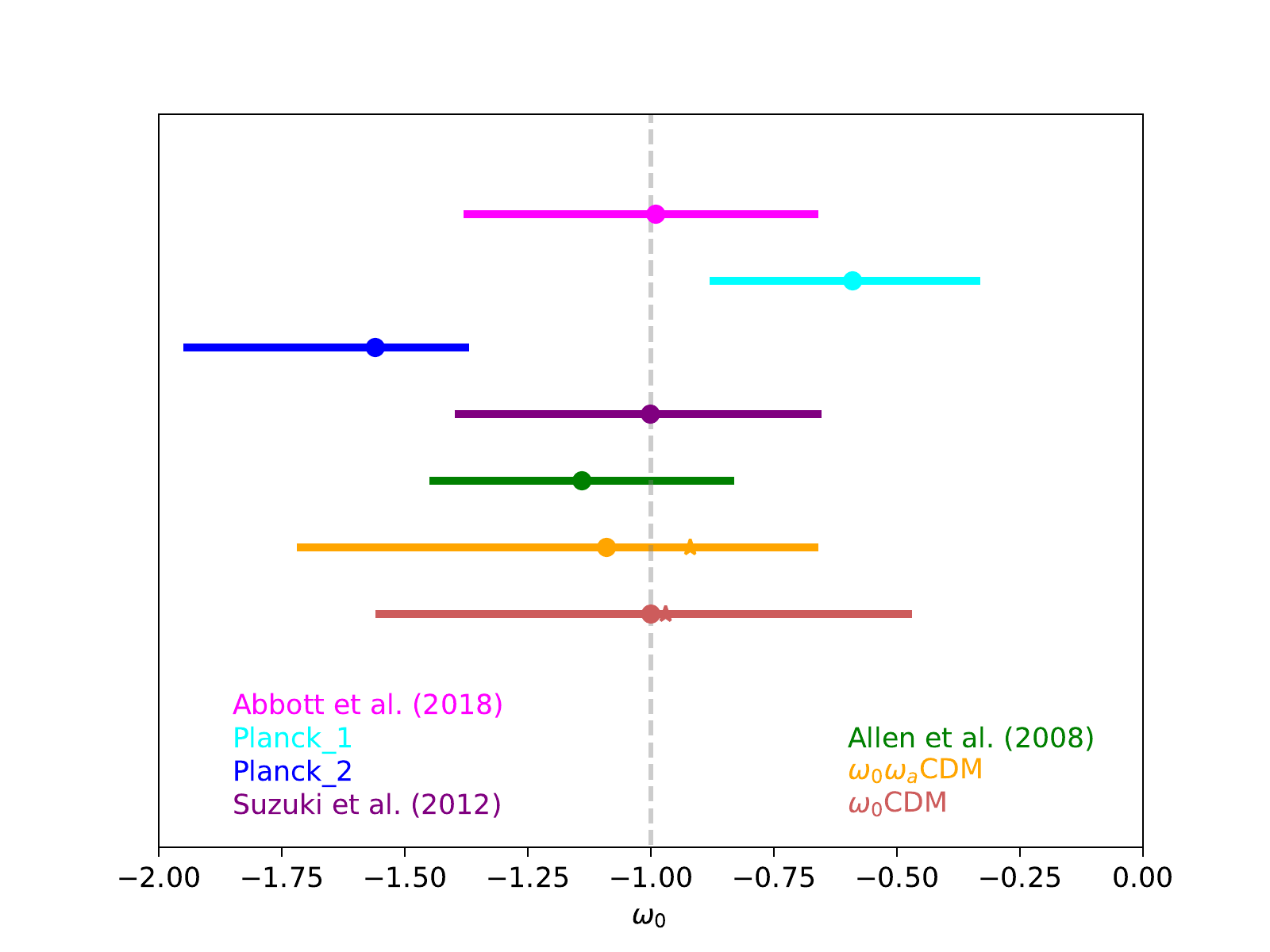}
    \caption{Comparison of $\omega_0$  results of this work at  68\% C.I. ($\omega_0$CDM model in red and $\omega_0\omega_a$CDM model in yellow) with findings from some other experiments: DES (magenta), \textit{Planck}\_1 (baseline base\_w\_wa\_plikHM\_TT\_lowl\_lowE\_BAO, cyan), \textit{Planck}\_2 (baseline base\_w\_plikHM\_TT\_lowl\_lowE, blue), supernovae (purple), x-ray measurements (green).}
    \label{Fig:w0_comp}
\end{figure}

With respect the $\omega_0$CDM and $\omega_0\omega_a$CDM models, our results are in agreement with recent findings in literature. They also have a comparable precision when referring to non-combined results from other experiments see Figure \ref{Fig:w0_comp} for the $\omega_0$ comparison of the 68\% C.I. from this work (red bar for the $\omega_0$CDM case and yellow one for the $\omega_0\omega_a$CDM case) with findings by other experiments.
For example, in the $\omega_0$CDM case, \cite{ALL08} (green bar) obtain $\omega=-1.14 \pm 0.31$, \cite{SUZ12} (purple bar) $\omega=-1.001^{+0.348}_{-0.398}$, \textit{Planck}\footnote{\label{note1}See the parameter grid tables in the PLA \textit{Planck} Collaboration. 2018, \textit{Planck} Legacy Archive, {\url{https://pla.esac.esa.int/}}} $\omega_0=-1.56^{+0.19}_{-0.39}$ (baseline base\_w\_plikHM\_TT\_lowl\_lowE; blue bar), and  \cite{ABB18} (magenta bar) $\omega=-0.99^{+0.33}_{-0.39}$.
In the $\omega_0\omega_a$CDM case, \textit{Planck}$^{\ref{note1}}$ results obtain $\omega_0=-0.59^{+0.29}_{-0.26}$ and $\omega_a = -1.33 \pm 0.79$ (baseline base\_w\_wa\_plikHM\_TT\_lowl\_lowE\_BAO; cyan bar). Moreover, the strong degeneracy between $\omega_0$ and $\omega_a$ is also found in \citet{ALA17} using models that allow the evolving equation of state (see Eq. \ref{eq:w}) in a flat universe, and constraints on $\omega_a$ are also found to be poor $\omega_a = -0.39 \pm 0.34$.

Even if our results are not yet very constraining, they put somewhat interesting limits to the other external results and do not present the typical degeneracy that characterises cosmic shear results. It is very likely that in the future, more constraining magnification bias samples can offer a complementary probe able to break the degeneracy and improve the dark energy constraints.

%--------------------------------------------------------------------
\section{Conclusions}
\label{sec:conclusion}

In this work we  exploited the magnification bias effect, similarly to \cite{BON20} and following the work by \citet{GON17} and \cite{BON19} on high-$z$ submillimetre galaxies, to perform a tomographic analysis with a view of searching for constraints on the free parameters of the HOD model ($M_{min}$, $M_1$, and $\alpha$), but also on the cosmological $\Omega_M$, $\sigma_8$, $h$, $\omega_0$, and $\omega_a$ parameters in four different cases (fixed cosmology, $\Lambda$CDM, $\omega_0$CDM, and $\omega_0\omega_a$CDM). To perform this analysis we  split the foreground sample into four bins of redshift with approximately the same number of lenses: 0.1-0.2, 0.2-0.3, 0.3-0.5, 0.5-0.8. 

For the astrophysical parameters, we could not  successfully constrain the $\alpha$ parameters in any of our redshift bins. In some cases the posterior distributions showed a broad peak (e.g. in Bin 3 with a mean value at 1.09 [0.92, 1.43] 68\% C.I. for the $\omega_0$CDM case).             
In $M_{min}$ we found upper limits in most of the cases for Bin1 ($<10$), suggesting that at lower redshifts the average minimum mass needed might be lower than in the other cases and more difficult to constrain, perhaps also due to the larger uncertainties in these data. In Bin2 the situation slightly improved; we obtained   broad peaks in all cases at $\log_{10}(M_{min}/M_\odot)\sim12$. In Bin3 and Bin4 the 1D distributions have clear maxima for most of the cases at $\log_{10}(M_{min}/M_\odot)\sim 12.5$ and at 13.7, respectively. The results on $M_{min}$ confirm the values by \citet{GON17}, showing the trend towards higher values at higher redshift. For $M_1$ the situation worsens again, setting lower limits for most of the cases with the exception of Bin3 and Bin4 (peaking at about $\log_{10}(M_{1}/M_\odot)\sim 13$ and $14$, respectively). 

For the cosmological parameters, for $\Omega_M$ we obtained somewhat lower values with respect to the \cite{GON21} findings;  the highest are   the $\Lambda$CDM and $\omega_0$CDM cases with a maximum at 0.26 and the lowest the $\omega_0\omega_a$CDM case with a maximum at 0.21. 
On the other hand, we found slightly higher values for $\sigma_8$, peaking in all cases between 0.84 and 0.87. 
Unfortunately $h$ is not constrained yet, obtaining just a low significant peak at 0.67 in the $\Lambda$CDM case.

The tomographic analysis presented in this work improves the constraints in the $\sigma_8-\Omega_M$ plane with respect to \cite{BON20} and it confirms that magnification bias results do not show the degeneracy found with cosmic shear measurements. Further progress with this approach is subject to the possibility of increasing the statistics in each redshift bin, which is necessary to reduce the uncertainties on the CCF data at large scales.

Moreover, we were able to study the barotropic index of dark energy $\omega$ and its possible evolution with redshift, obtaining results in agreement with current findings in literature and with a similar constraining level, finding a peak at -0.97 ($[-1.56, -0.47]$ 68\% C.I.) and at -0.92 ($[-1.72,-0.66]$ 68\% C.I.) for $\omega_0$ in the $\omega_0$CDM and $\omega_0\omega_a$CDM cases, respectively. For $\omega_a$ we obtain a peak at -0.20 ($[-1.88,1.48]$ 68\% C.I.) and a 30\% anti-correlation with $\omega_0$.

Finally, we explored the impact on $\omega_0$ when fixing $H_0$ to either the \cite{PLA18_VIII} or the \cite{RIE19} values. We set the cosmological parameters to the \textit{Planck} values \citep{PLA18_VIII} and ran only on the HOD parameters and $\omega_0$ with $H_0$ first fixed to the value determined by \textit{Planck} and then to the value by \cite{RIE19}. We found a trend of higher $\omega_0$ values for lower $H_0$ values. 

\begin{acknowledgements}
We deeply thank the anonymous referee for quite useful comments on the original draft.\\
LB, JGN, MMC, JMC, DC acknowledge the PGC 2018 project PGC2018-101948-B-I00 (MICINN/FEDER).
MMC acknowledges support from PAPI-20-PF-23 (Universidad de Oviedo).
MM is supported by the program for young researchers ``Rita Levi Montalcini" year 2015 and acknowledges support from INFN through the InDark initiative.
AL acknowledges partial support from PRIN MIUR 2015 Cosmology and Fundamental Physics: illuminating the Dark Universe with Euclid, by the RADIOFOREGROUNDS grant (COMPET-05-2015, agreement number 687312) of the European Union Horizon 2020 research and innovation program, and the MIUR grant 'Finanziamento annuale individuale attivita base di ricerca'.
We deeply acknowledge the CINECA award under the ISCRA initiative, for the availability of high performance computing resources and support. In particular the projects “SIS19\_lapi”, “SIS20\_lapi” in the framework “Convenzione triennale SISSA-CINECA”.\\
The Herschel-ATLAS is a project with Herschel, which is an ESA space observatory with science instruments provided by European-led Principal Investigator consortia and with important participation from NASA. The H-ATLAS web-site is http://www.h-atlas.org. GAMA is a joint European-Australasian project based around a spectroscopic campaign using the Anglo-Australian Telescope. The GAMA input catalogue is based on data taken from the Sloan Digital Sky Survey and the UKIRT Infrared Deep Sky Survey. Complementary imaging of the GAMA regions is being obtained by a number of independent survey programs including GALEX MIS, VST KIDS, VISTA VIKING, WISE, Herschel-ATLAS, GMRT and ASKAP providing UV to radio coverage. GAMA is funded by the STFC (UK), the ARC (Australia), the AAO, and the participating institutions. The GAMA website is: http://www.gama-survey.org/.\\
In this work, we made extensive use of \texttt{GetDist} \citep{GETDIST}, a Python package for analysing and plotting MC samples. In addition, this research has made use of the python packages \texttt{ipython} \citep{ipython}, \texttt{matplotlib} \citep{matplotlib} and \texttt{Scipy} \citep{scipy}
\end{acknowledgements}


\bibliographystyle{aa} % style aa.bst
\bibliography{XCORR_COSMO} % your references Yourfile.bib
\appendix

\section{Auto-correlation data}
\label{App:auto_data}

\begin{figure}[ht]
\centering
\includegraphics[width=0.4\textwidth]{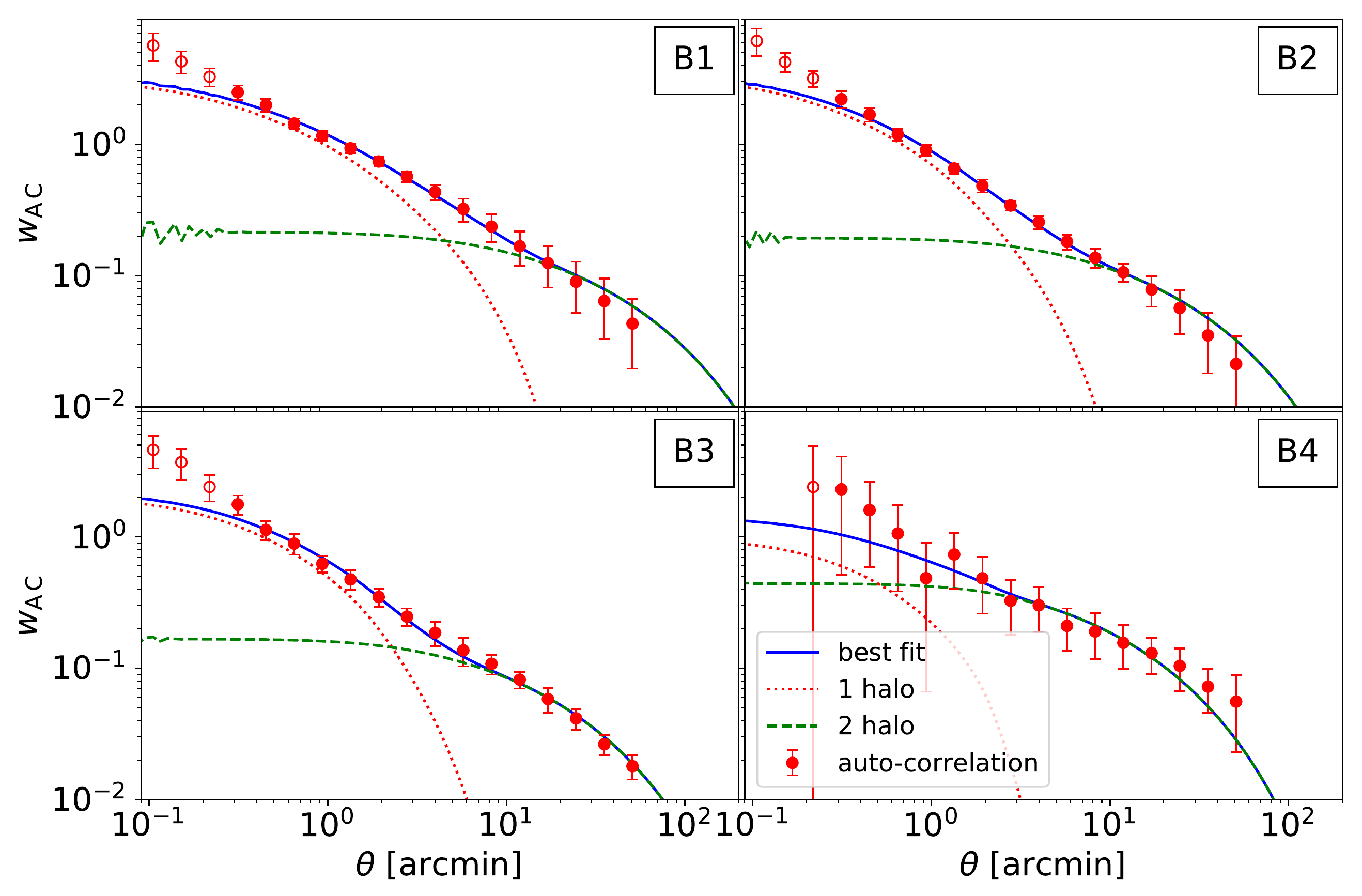}
\caption{Auto-correlation data points (in red) and the best fit to the model (blue solid line) for the four redshift bins. The red dotted line and the green dashed line show the one-halo and two-halo terms, respectively.}
\label{fig:ACF}
\end{figure}

The auto-correlation data and results are listed in this section. Figure \ref{fig:ACF} shows the auto-correlation data points (red dots) and the best fit obtained with the mean values of the posterior distributions (blue solid line) for BIN1 to BIN4 (from left to right, top to bottom). The one-halo (red dotted line) and two-halo (green dashed line) terms are also shown separately. Table \ref{Tab:auto-corr} lists the results in each redshift bin (first column, BIN1 to BIN4 from top to bottom) for the analysed parameters (second column, $logM_{min}$, $logM_1$, $\alpha$, and $A$). The priors, mean, median, 68\% C.I., and peak are shown in columns three to seven, respectively.

\begin{table*} 
\caption{Prior distributions and results in each redshift bin for the parameters studied in the auto-correlation run.} 
\label{Tab:auto-corr} 
\centering 
\begin{tabular}{c c c c c c c} 
\hline 
\hline 
Bin & Parameter & Prior & $\mu$ & median & $68\% \text{C.I.}$ & Peak \\ 
\hline 
BIN1 & $\log (M_{min}/M_{\odot})$ &  $\mathcal{U}$[  9.0, 16.0]   & $11.58$  & 11.57 & $ [11.52,11.61 ] $  & $ 11.55$ \\   
  & $\log (M_1/M_{\odot})$     &  $\mathcal{U}$[  9.0, 16.0]   & $12.68$  & 12.69 & $ [12.61,12.77 ] $ & $ 12.74$ \\   
  & $\alpha$       &  $\mathcal{U}$[  0.1,  2.5]   & $ 1.06$  & 1.06  & $ [ 1.01, 1.12 ] $ & $  1.07$ \\   
  & $A$            &  $\mathcal{U}$[ -5.0, -1.0]   & $-2.49$  & -2.48 & $ [-2.57,-2.36 ] $  & $ -2.41$ \\   
\hline 
BIN2 & $\log (M_{min}/M_{\odot})$ &  $\mathcal{U}$[  9.0, 16.0]   & $11.97$  & 11.97 & $ [11.94,11.99 ] $ & $ 11.95$ \\   
  & $\log (M_1/M_{\odot})$    &  $\mathcal{U}$[  9.0, 16.0]   & $13.05$  & 13.06 & $ [13.03,13.08 ] $ & $ 13.06$ \\   
  & $\alpha$       &  $\mathcal{U}$[  0.1,  2.5]   & $ 1.08$  & 1.08  & $ [ 1.01, 1.14 ] $ & $  1.08$ \\   
  & $A$            &  $\mathcal{U}$[ -5.0, -1.0]   & $-2.55$  & -2.54 & $ [-2.58,-2.48 ] $ & $ -2.53$ \\   
\hline 
BIN3 & $\log (M_{min}/M_{\odot})$ &  $\mathcal{U}$[  9.0, 16.0]   & $12.60$  & 12.60 & $ [12.59,12.60 ] $ & $ 12.60$ \\   
  & $\log (M_1/M_{\odot})$     &  $\mathcal{U}$[  9.0, 16.0]   & $13.81$  & 13.77 & $ [13.68,13.83 ] $ & $ 13.76$ \\   
  & $\alpha$       &  $\mathcal{U}$[  0.1,  2.5]   & $ 1.22$  & 1.25  & $ [ 1.06, 1.47 ] $ & $  1.28$ \\   
  & $A$            &  $\mathcal{U}$[ -5.0, -1.0]   & $-2.72$  & -2.72 & $ [-2.79,-2.66 ] $ & $ -2.72$ \\   
\hline 
BIN4 & $\log (M_{min}/M_{\odot})$ &  $\mathcal{U}$[  9.0, 16.0]   & $13.39$  & 13.38 & $ [13.38,13.39 ] $ & $ 13.38$ \\   
  & $\log (M_1/M_{\odot})$     &  $\mathcal{U}$[  9.0, 16.0]   & $15.29$  & 15.34 & $ [15.06,16.00 ] $ & $ 16.00$ \\   
  & $\alpha$       &  $\mathcal{U}$[  0.1,  2.5]   & $ 1.69$  & 1.71  & $ [ 1.42, 2.50 ] $ & $  1.84$ \\   
  & $A$            &  $\mathcal{U}$[ -5.0, -1.0]   & $-2.63$  & -2.61 & $ [-2.66,-2.54 ] $ & $ -2.60$ \\   
\hline 
\hline 
\end{tabular} 
\tablefoot{ The columns denote the bin, the parameter in question, and its prior distribution, and  the mean ($\mu$), median, $68\%$ C.I., and peak of its marginalised 1D posterior distribution.}
\end{table*}  

\newpage
\section{Corner plots and tables}
\label{App:corner_table}

\begin{figure*}[ht]
\centering
\includegraphics[width=0.25\textwidth]{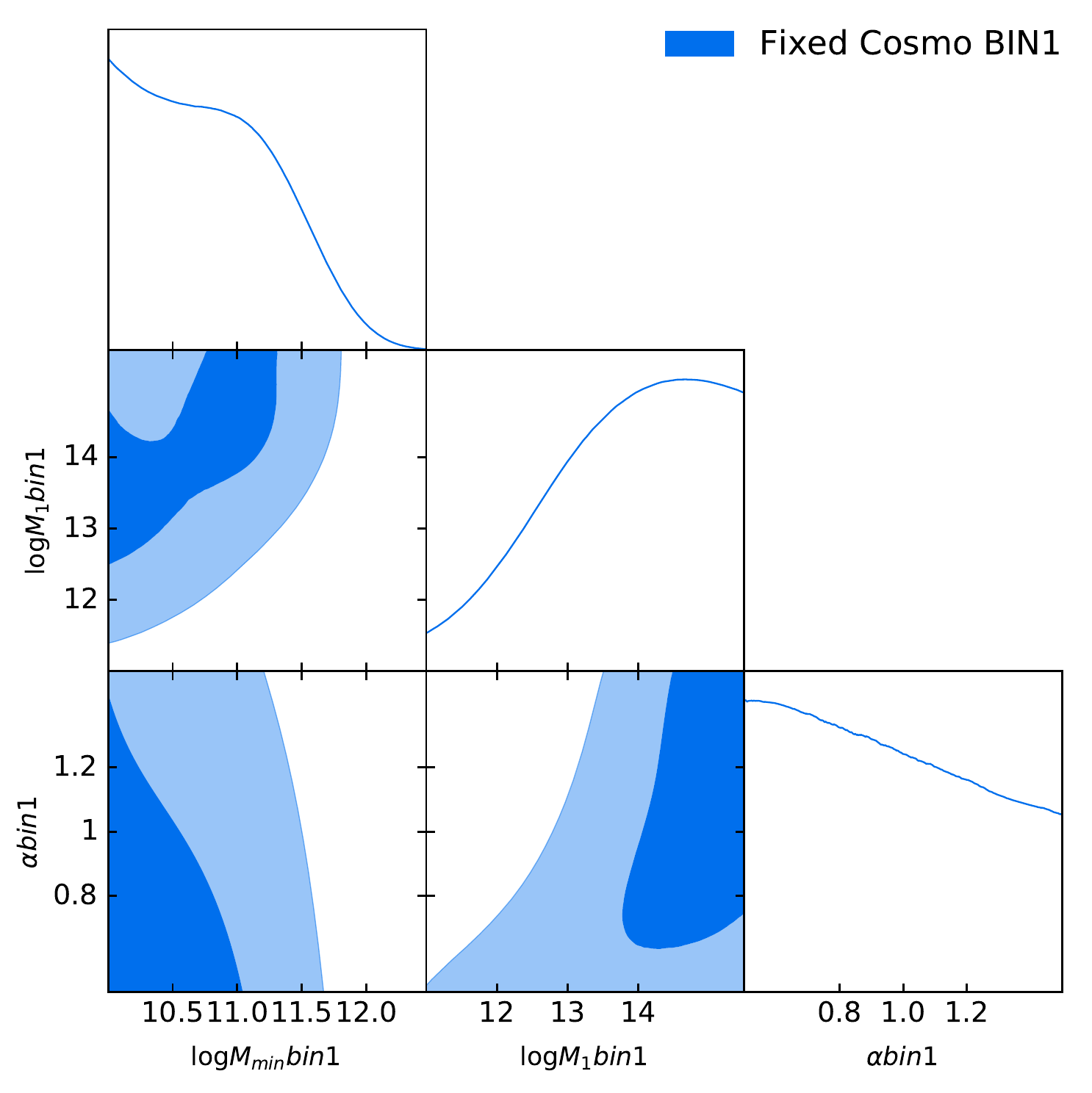}
\includegraphics[width=0.25\textwidth]{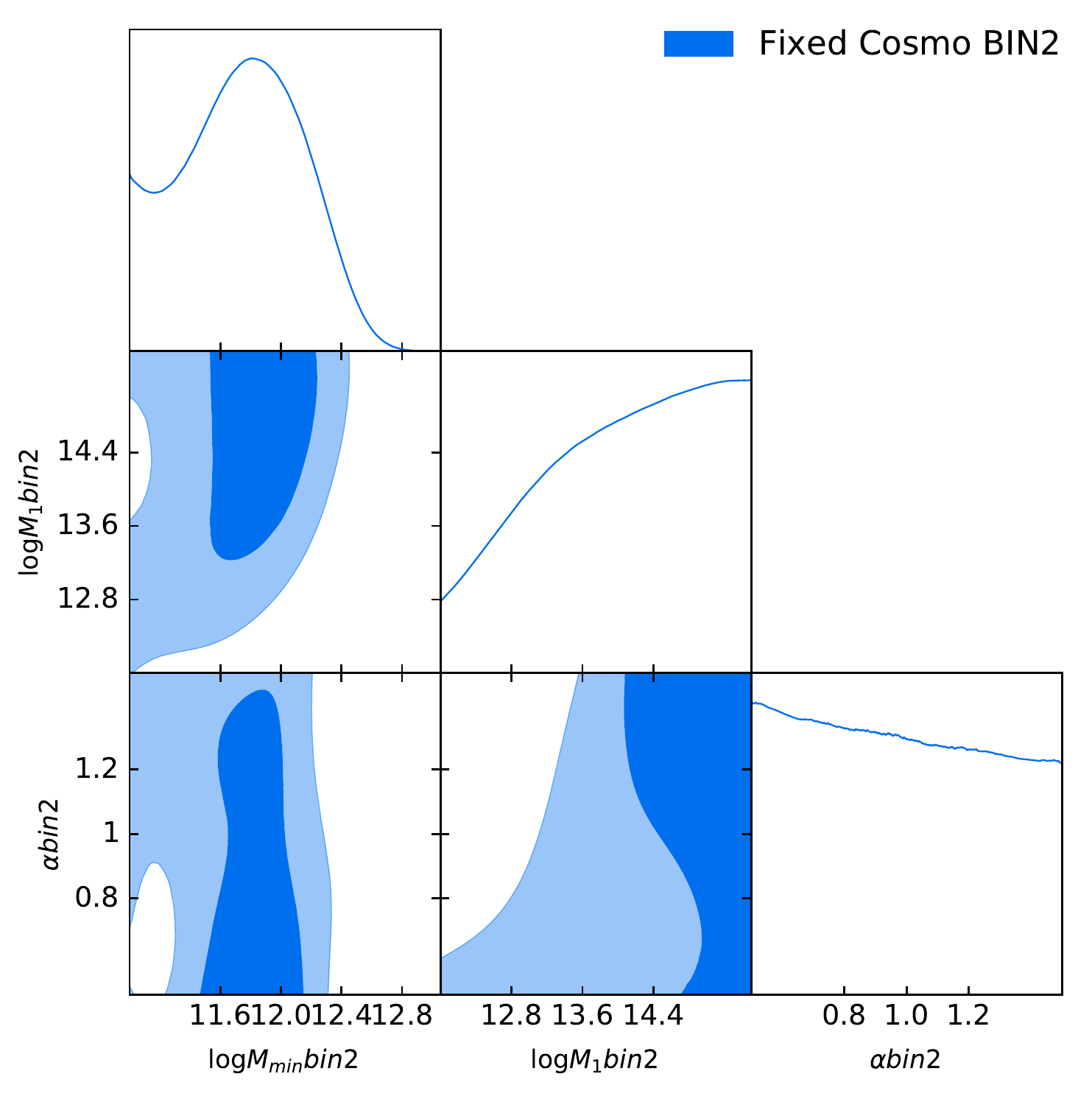}

\includegraphics[width=0.25\textwidth]{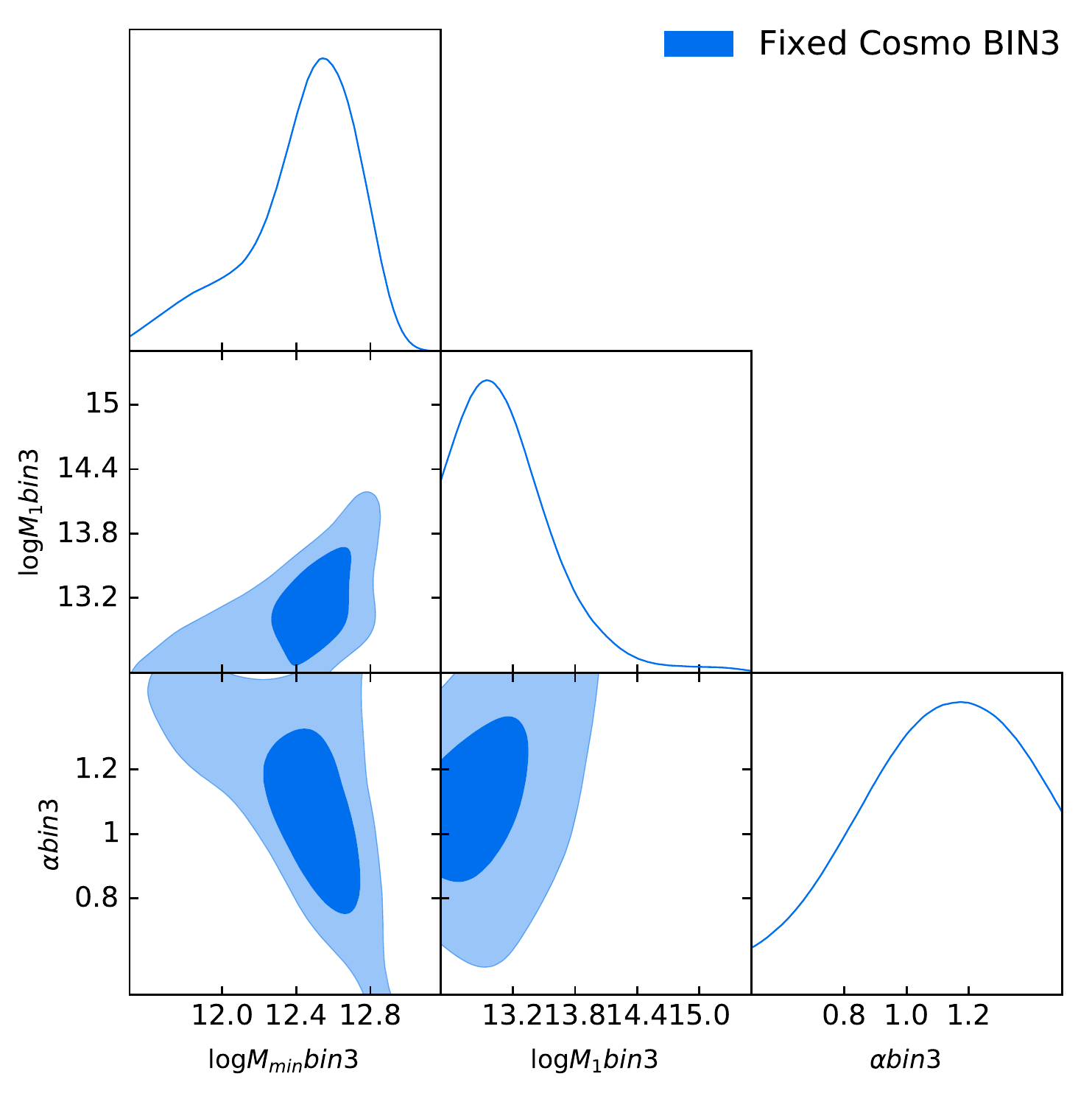}
\includegraphics[width=0.25\textwidth]{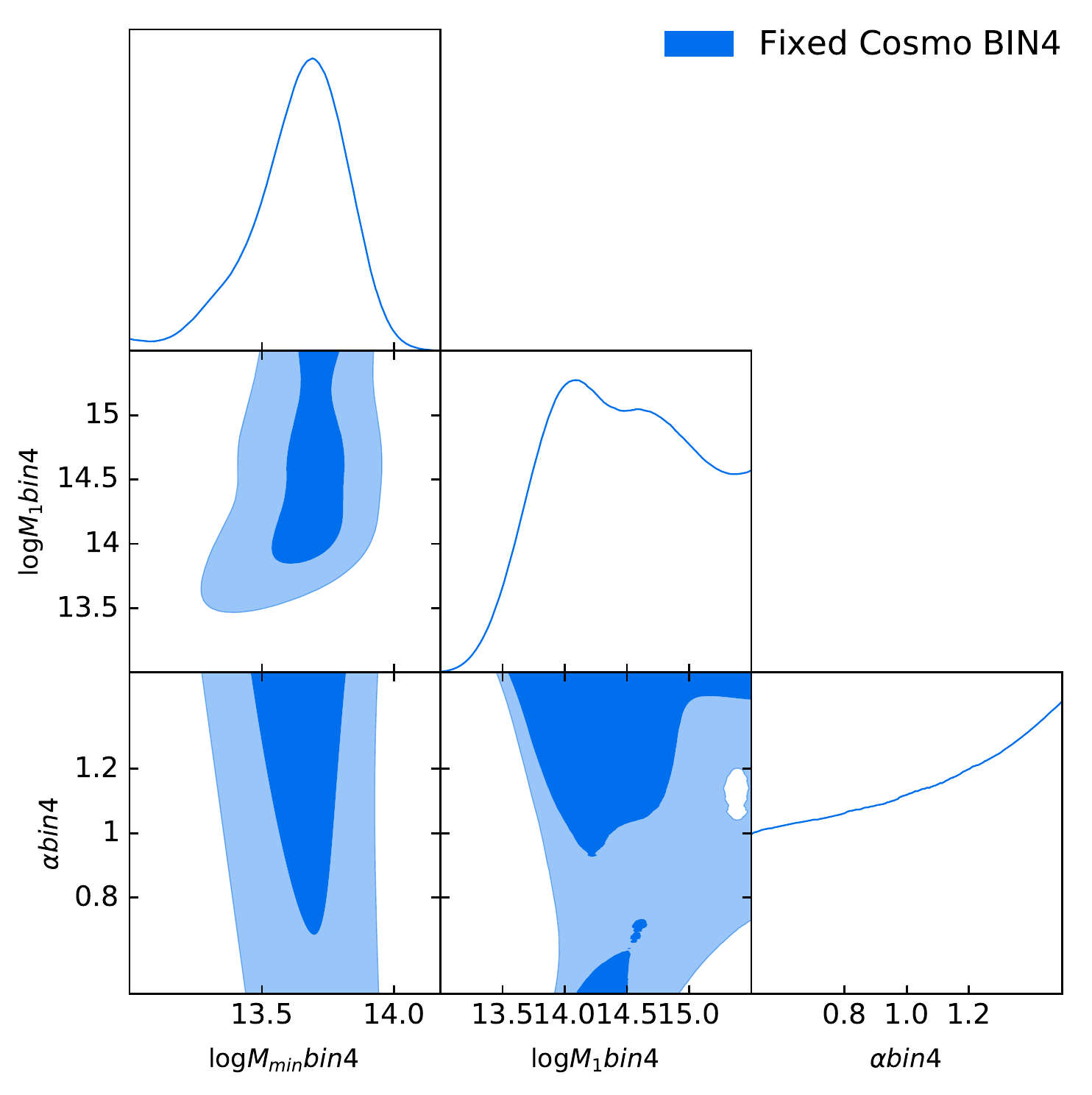}
\caption{Corner plots for the fixed cosmology case, the cosmology being fixed to the \textit{Planck} value. The posteriors distributions in Bin1, Bin2, Bin3, and Bin4 are shown from left to right, top to bottom. The contours are set to 39.3\% and 86.5\%}
\label{Fig:FPA}
\end{figure*}

\begin{figure*}[ht]
\centering
\includegraphics[width=0.4\textwidth]{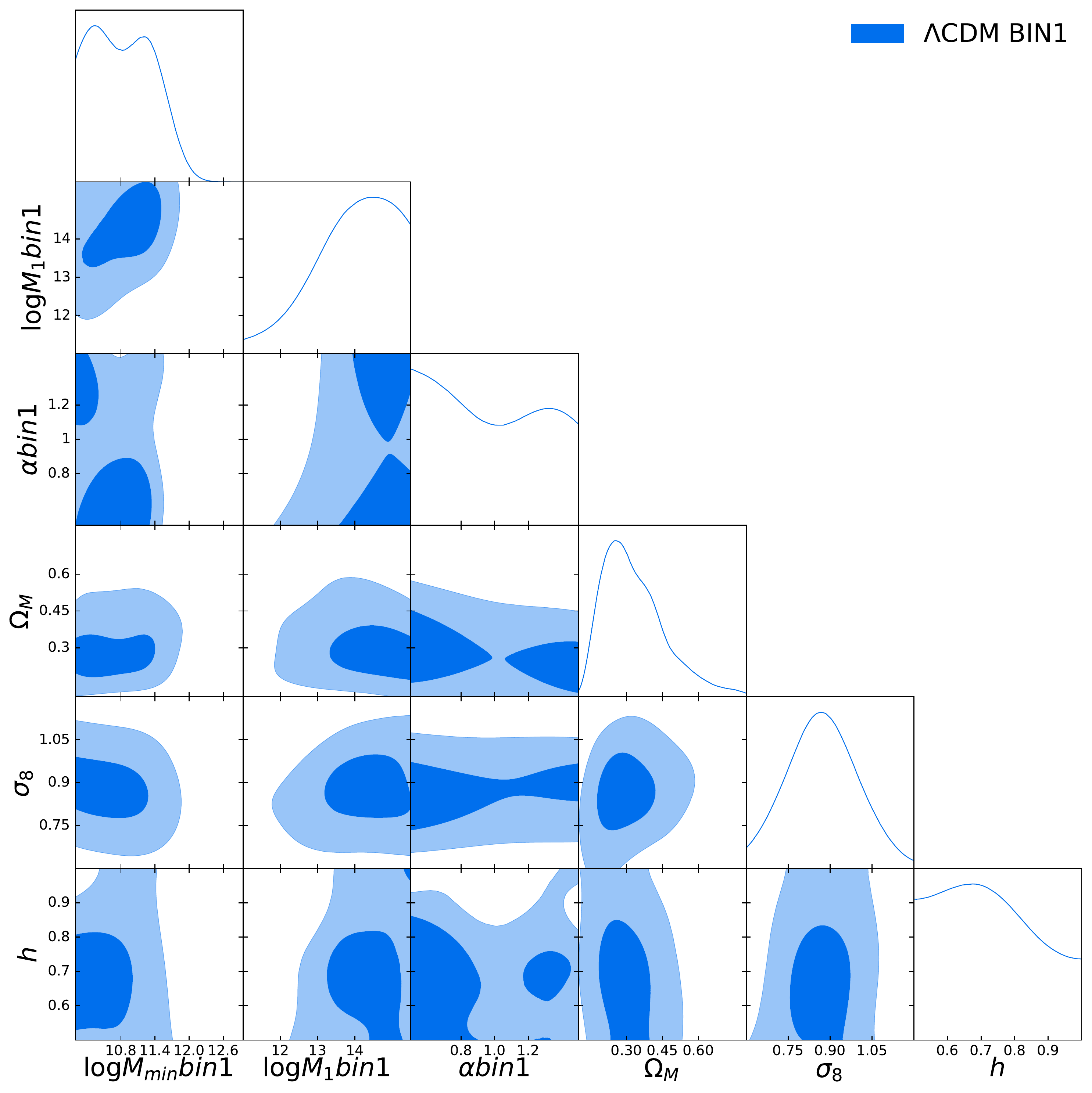}
\includegraphics[width=0.4\textwidth]{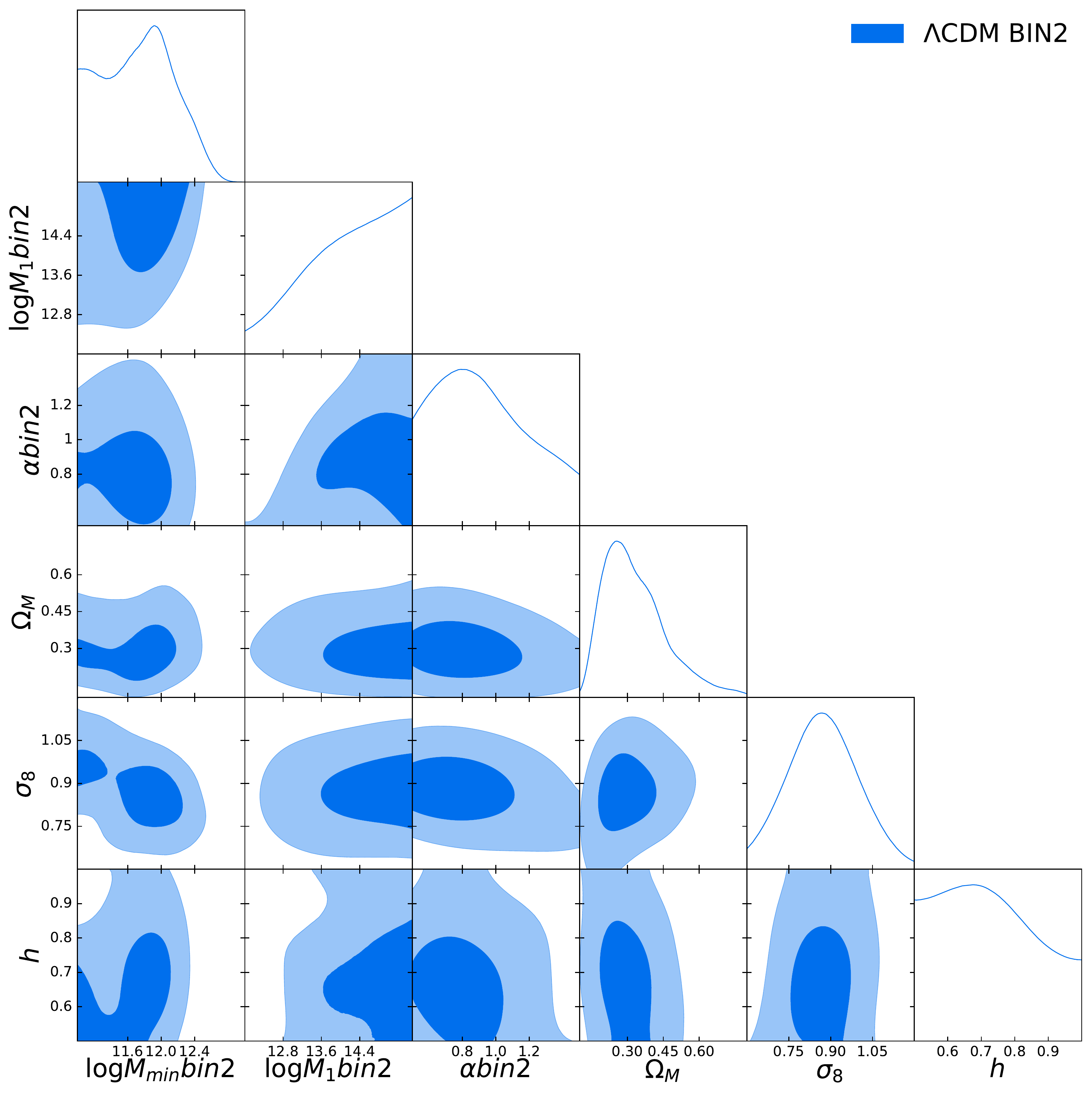}

\includegraphics[width=0.4\textwidth]{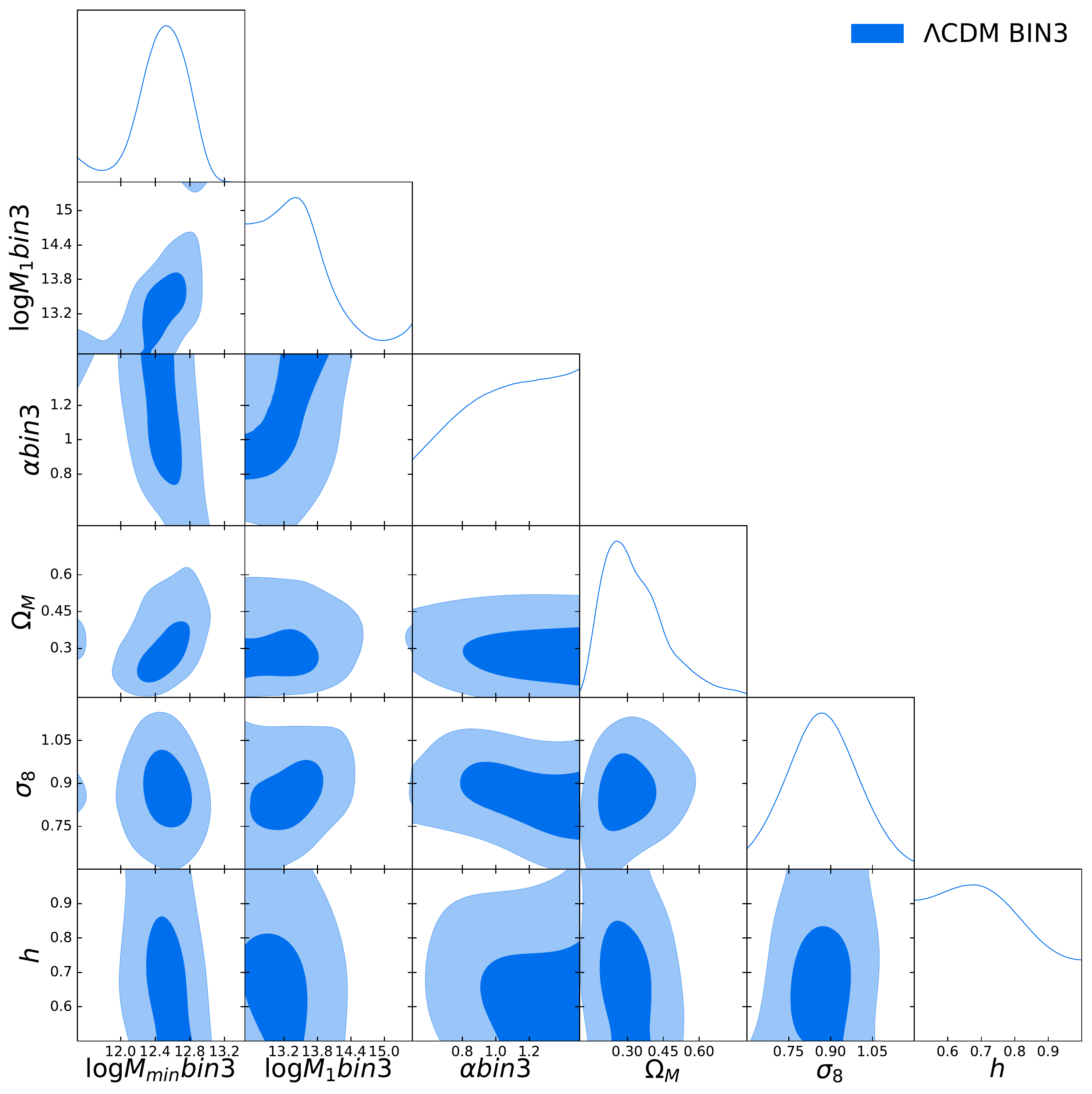}
\includegraphics[width=0.4\textwidth]{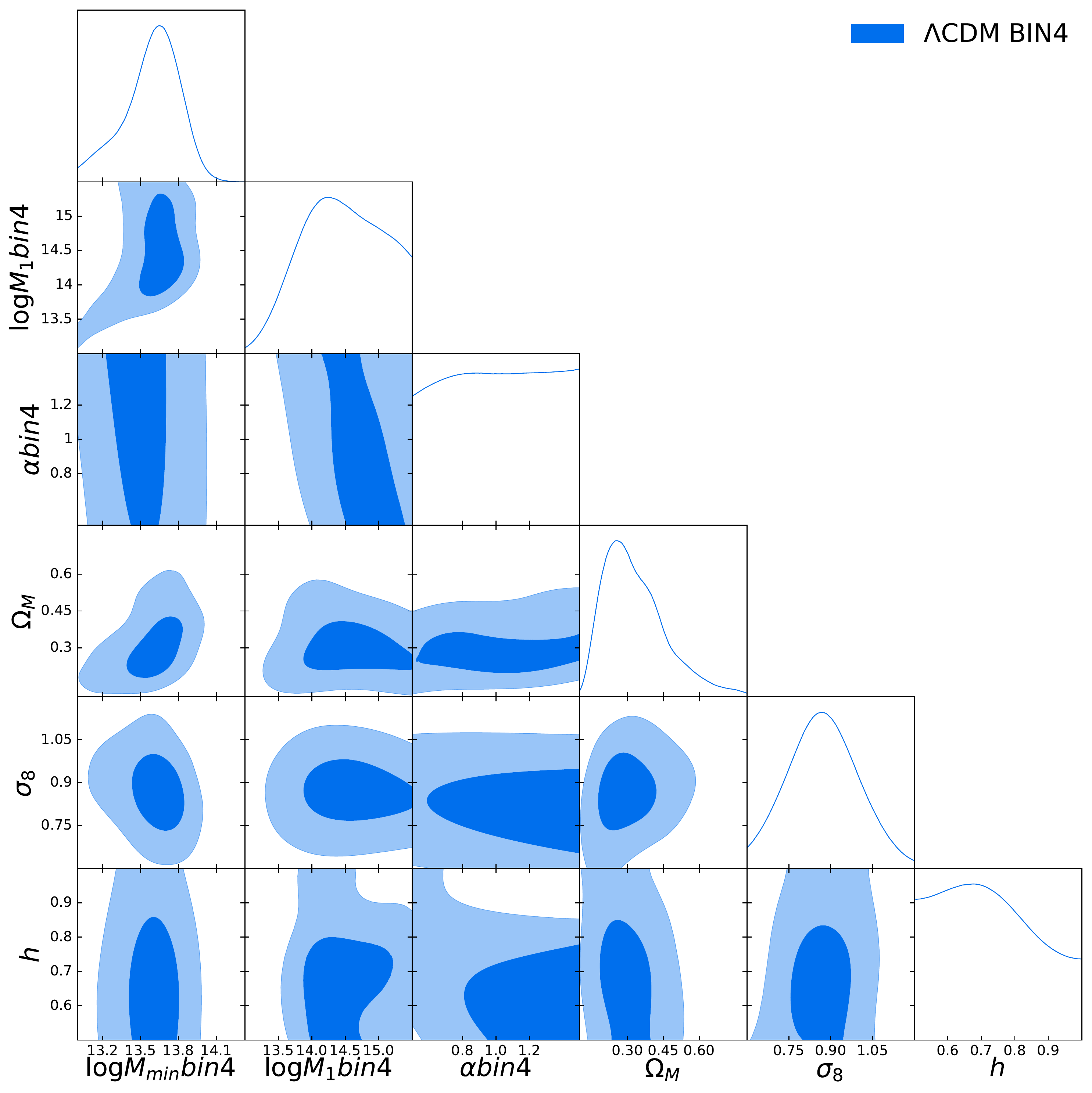}
\caption{Corner plots for the $\Lambda$CDM case, with the astrophysical parameters and $\Omega_M$, $\sigma_8$, and $h$ set as free parameters;  the rest of the cosmological parameters are fixed to the \textit{Planck} values. The posteriors distributions in Bin1, Bin2, Bin3, and Bin4 are shown from left to right, top to bottom. The contours are set to 39.3\% and 86.5\%.}
\label{Fig:FPA_FPC}
\end{figure*}

\begin{figure*}[ht]
\centering
\includegraphics[width=0.4\textwidth]{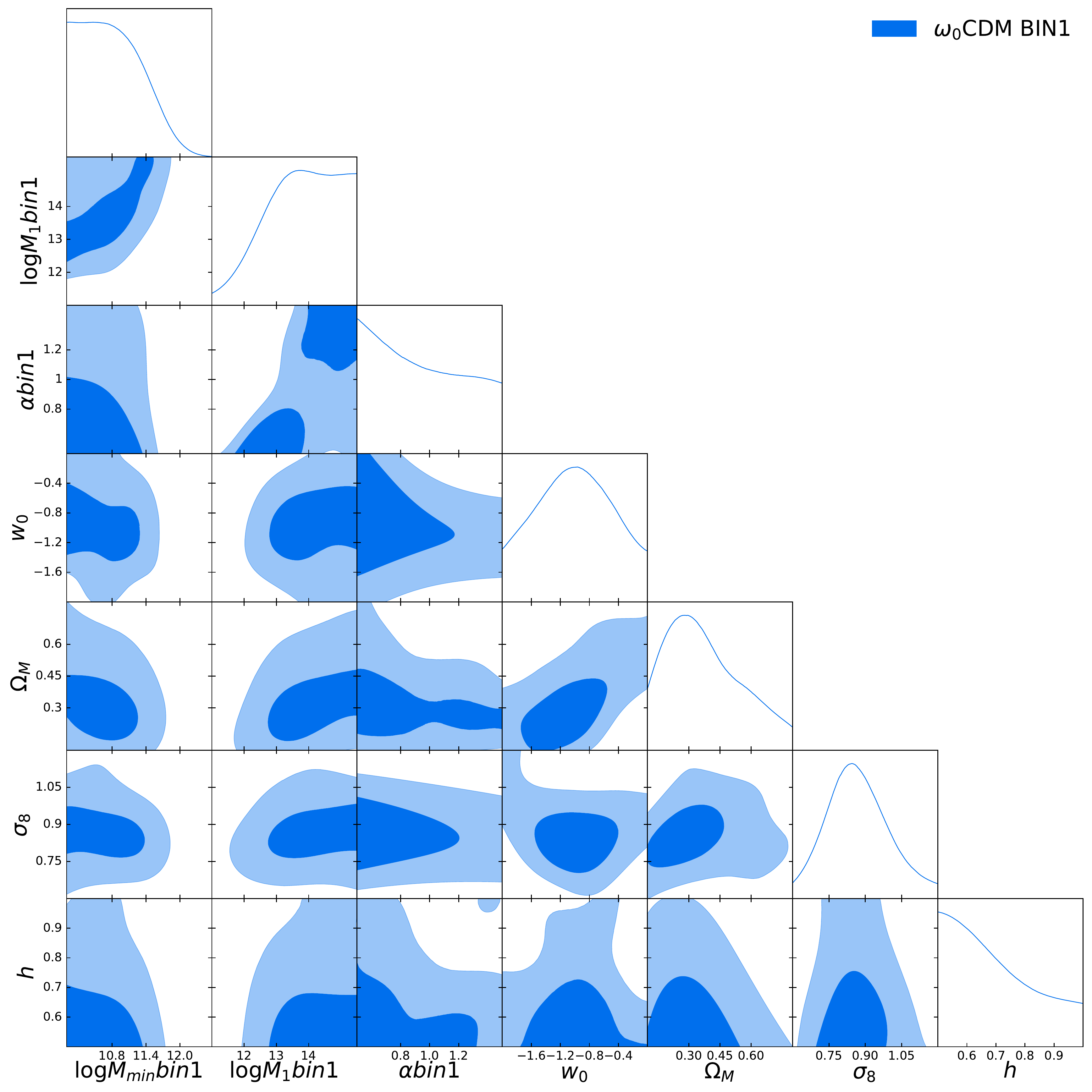}
\includegraphics[width=0.4\textwidth]{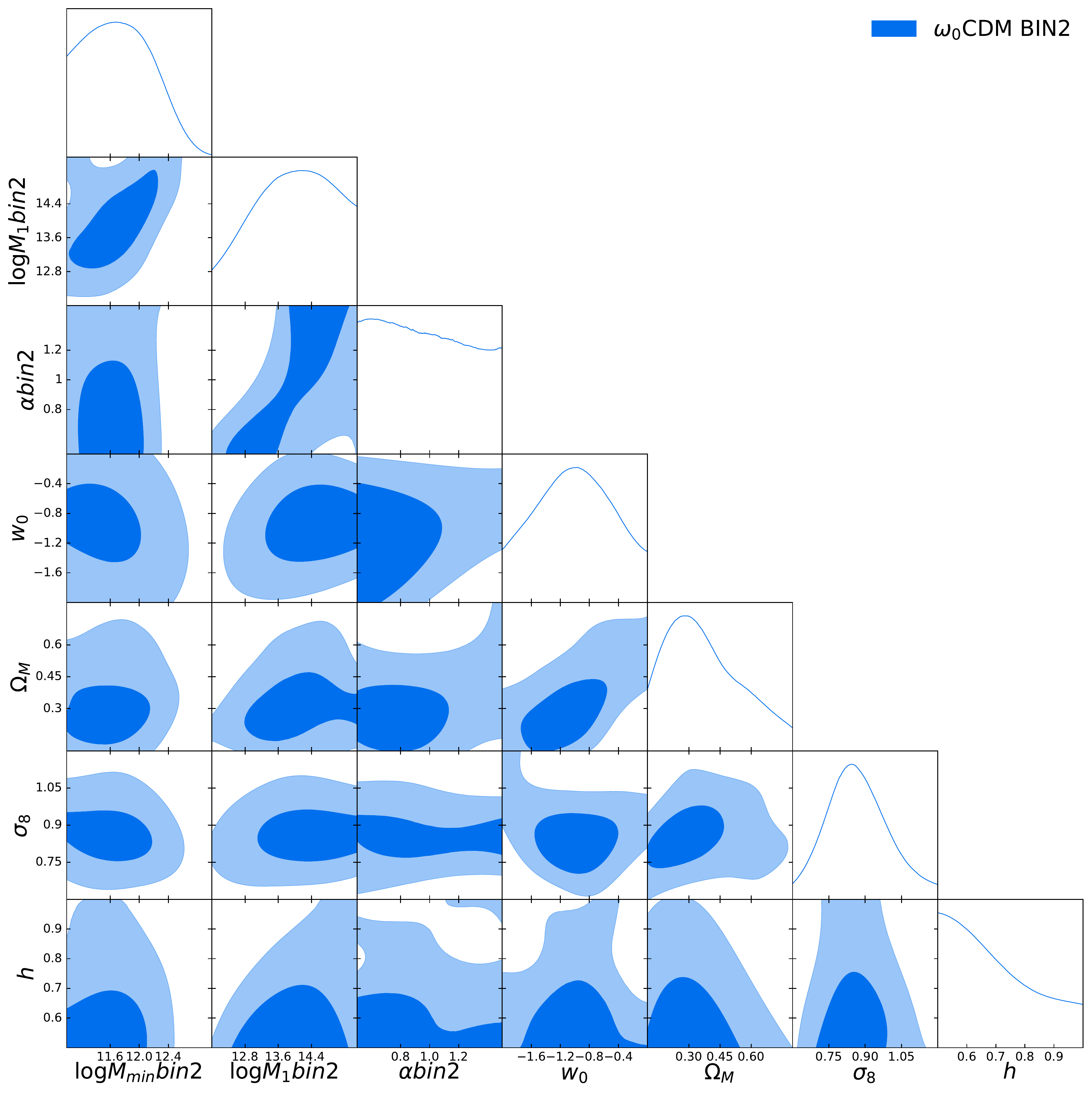}

\includegraphics[width=0.4\textwidth]{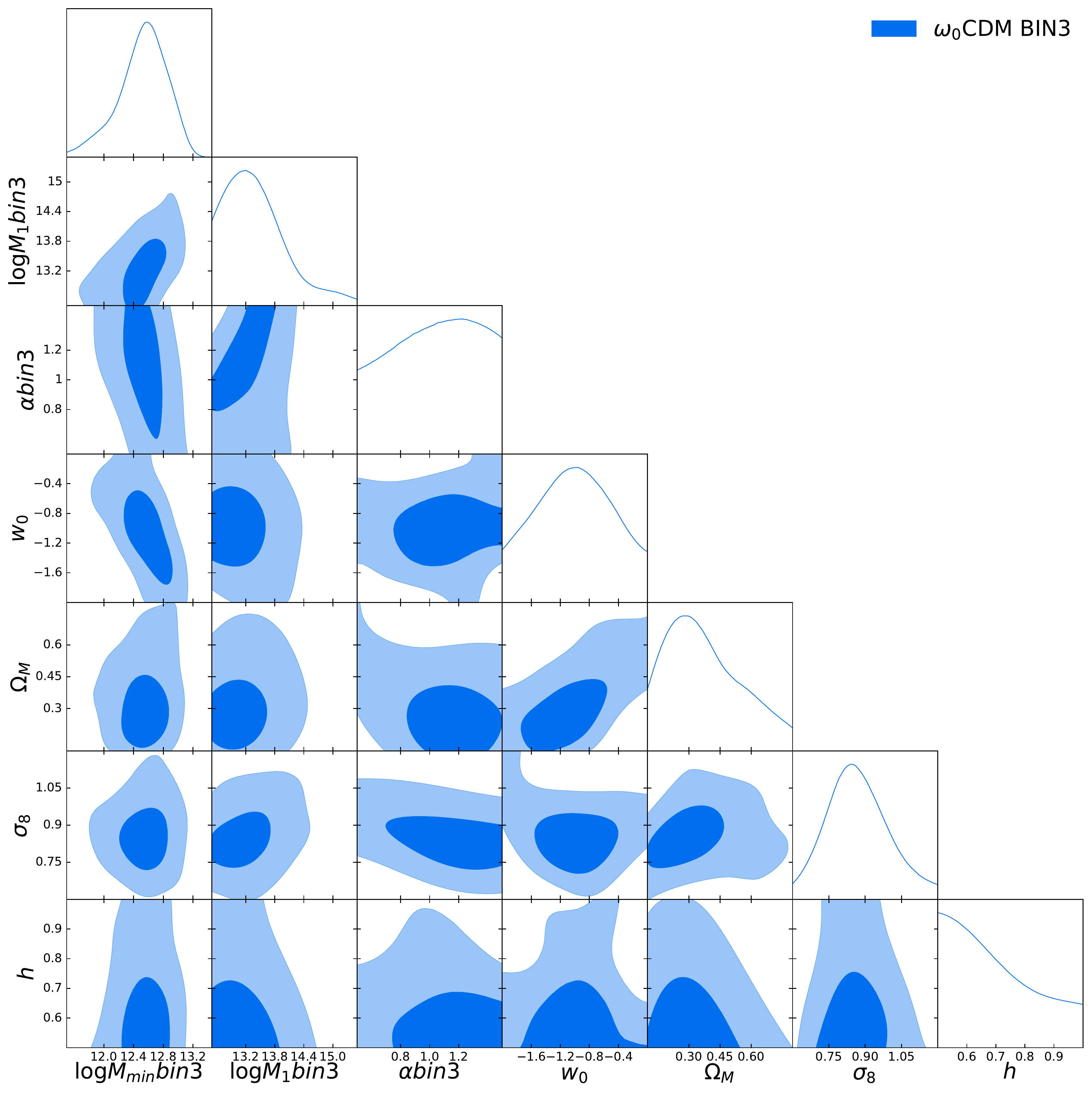}
\includegraphics[width=0.4\textwidth]{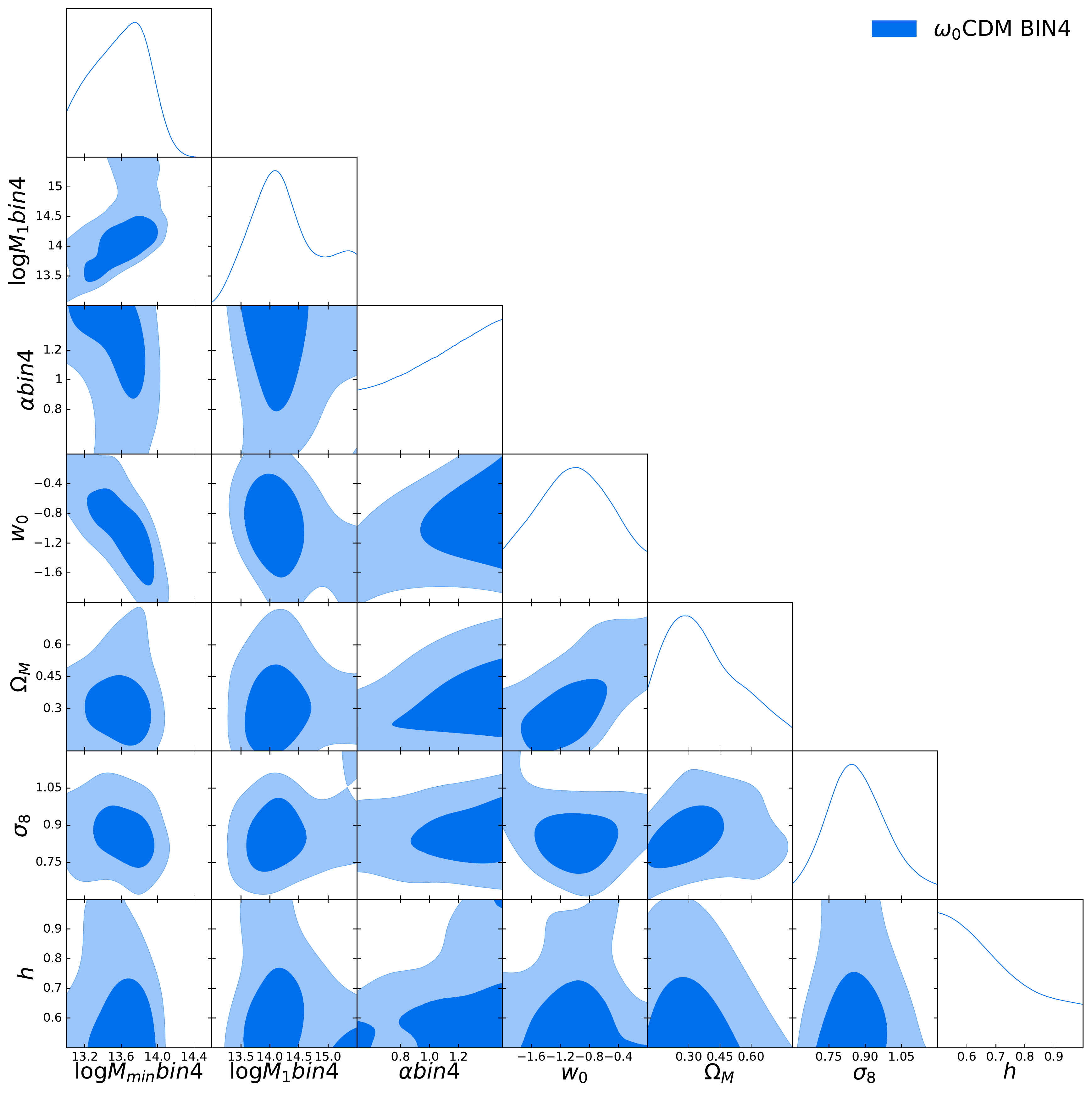}
\caption{Corner plots for the $\omega_0$CDM case, where the rest of the cosmological parameters are fixed to the \textit{Planck} values. The posteriors distributions in Bin1, Bin2, Bin3, and Bin4 are shown from left to right, top to bottom. The contours are set to 39.3\% and 86.5\%.}
\label{Fig:FPA_FPC_w0}
\end{figure*}

\begin{figure*}[h]
\centering
\includegraphics[width=0.35\textwidth]{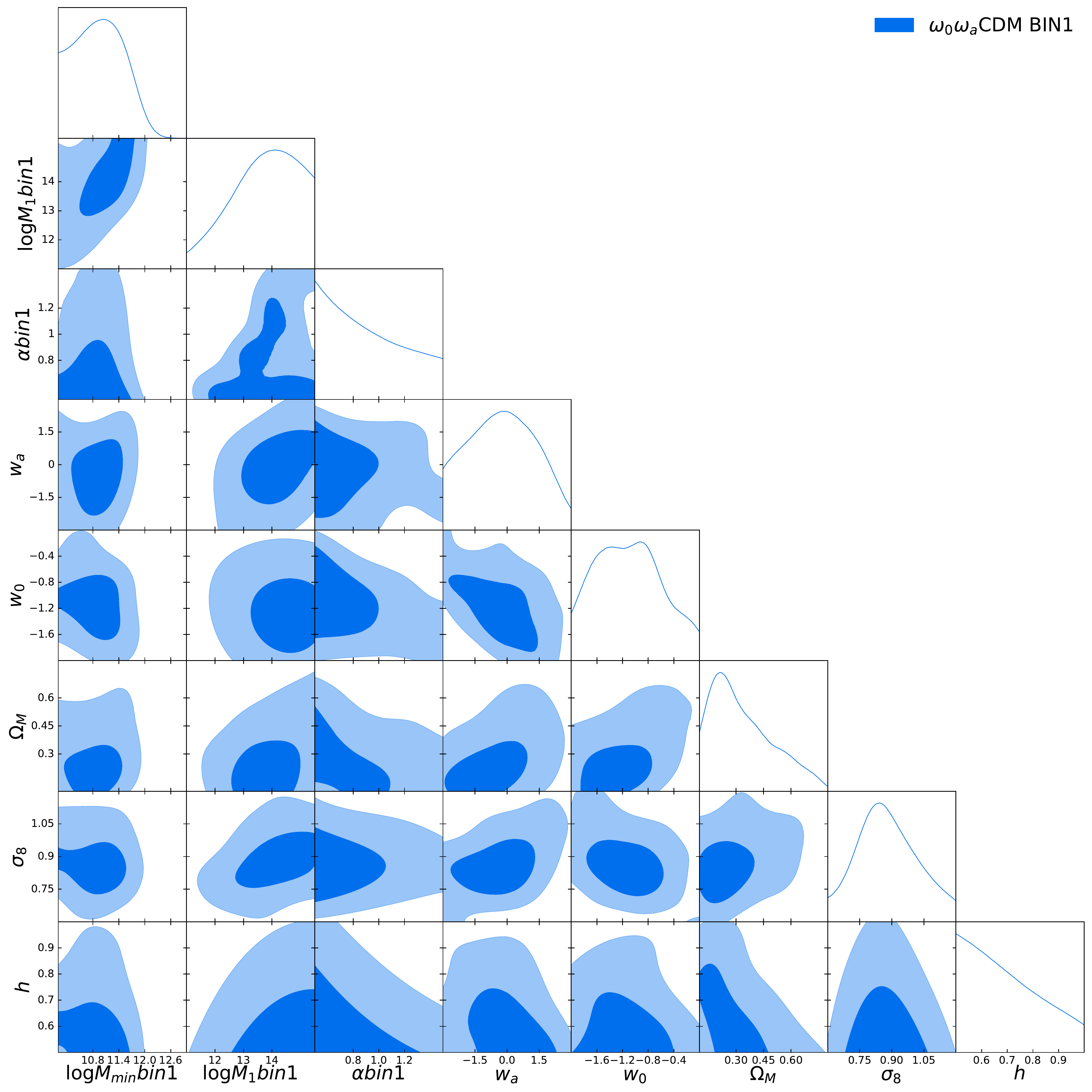}
\includegraphics[width=0.35\textwidth]{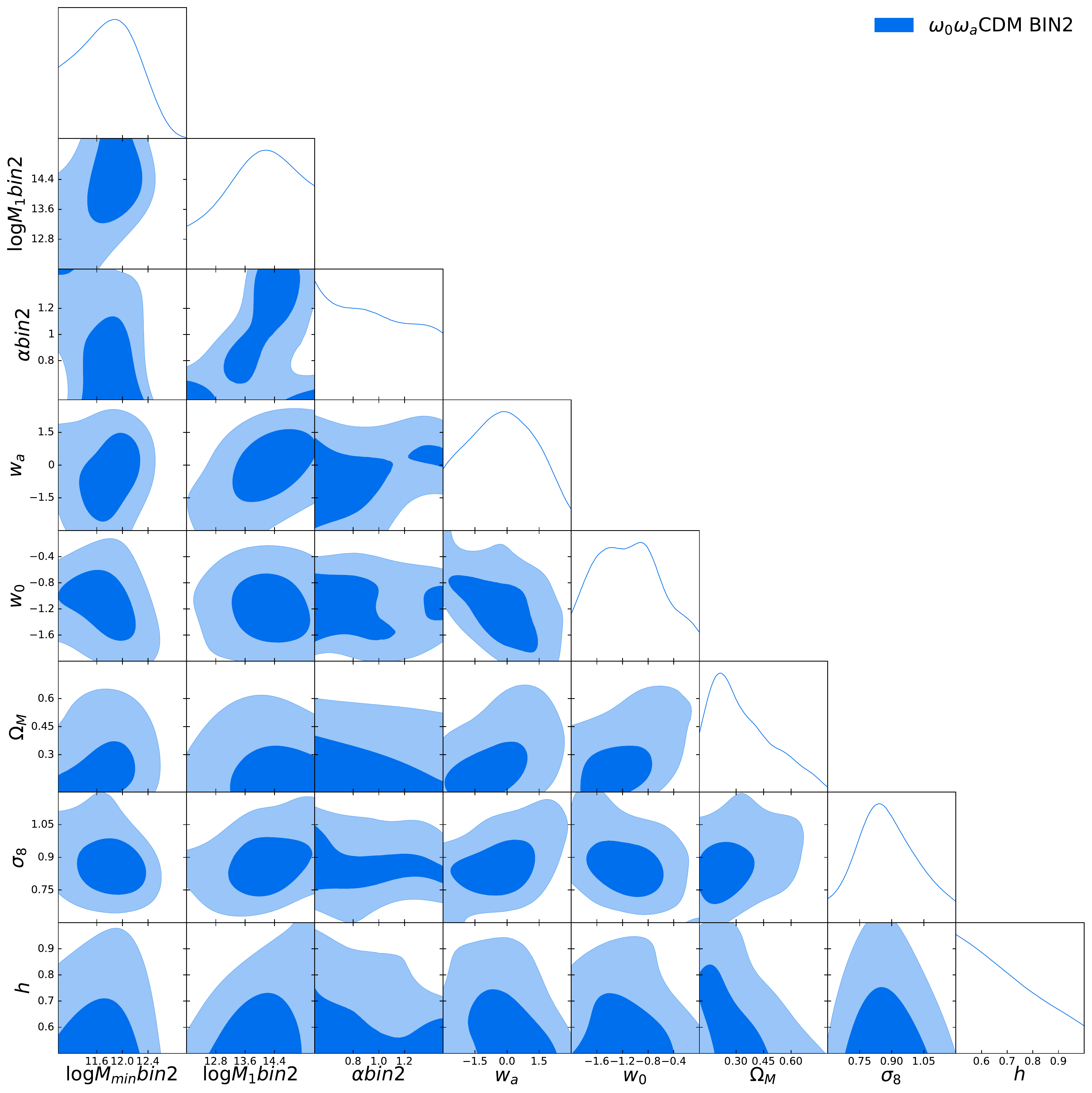}

\includegraphics[width=0.35\textwidth]{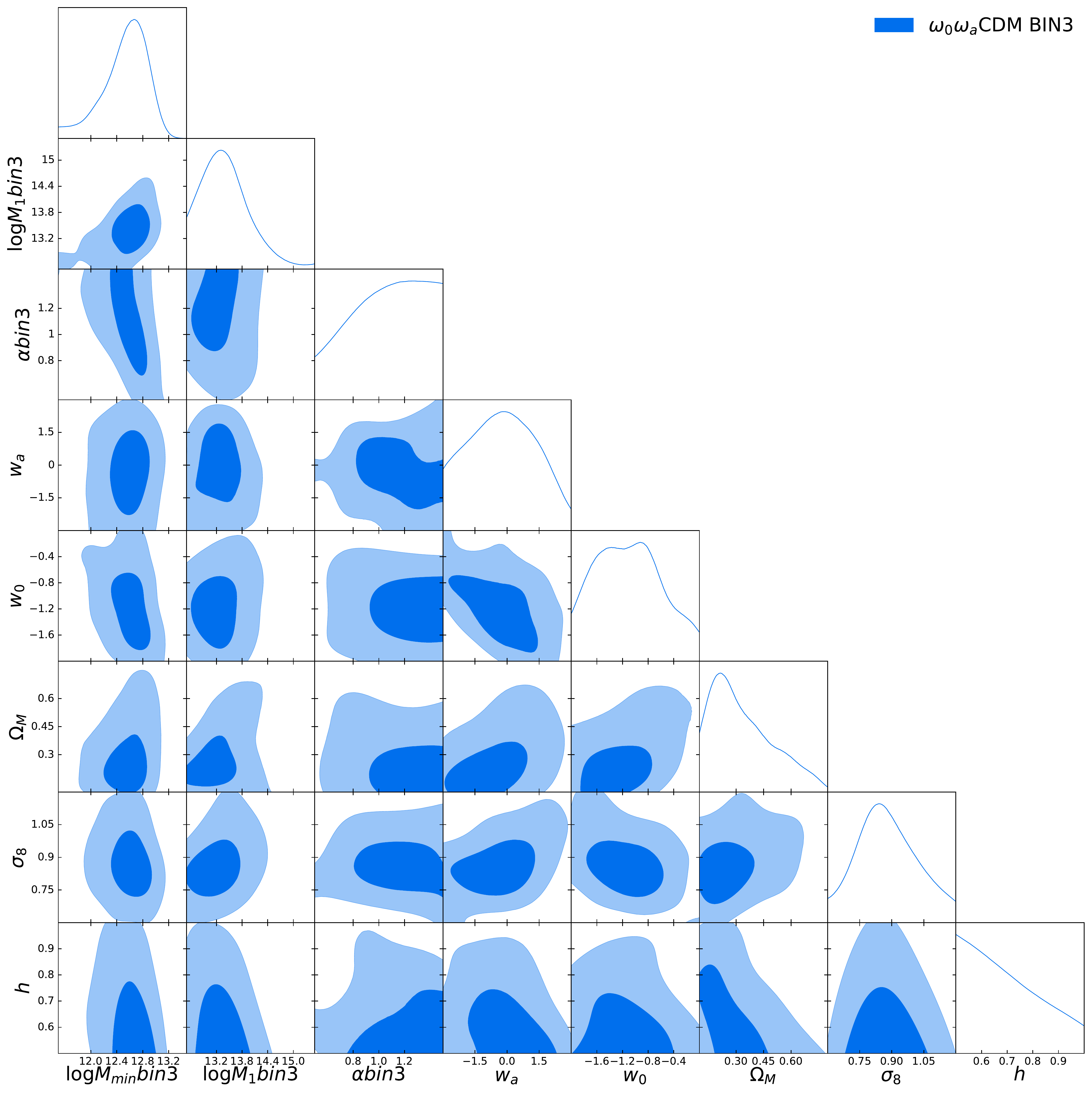}
\includegraphics[width=0.35\textwidth]{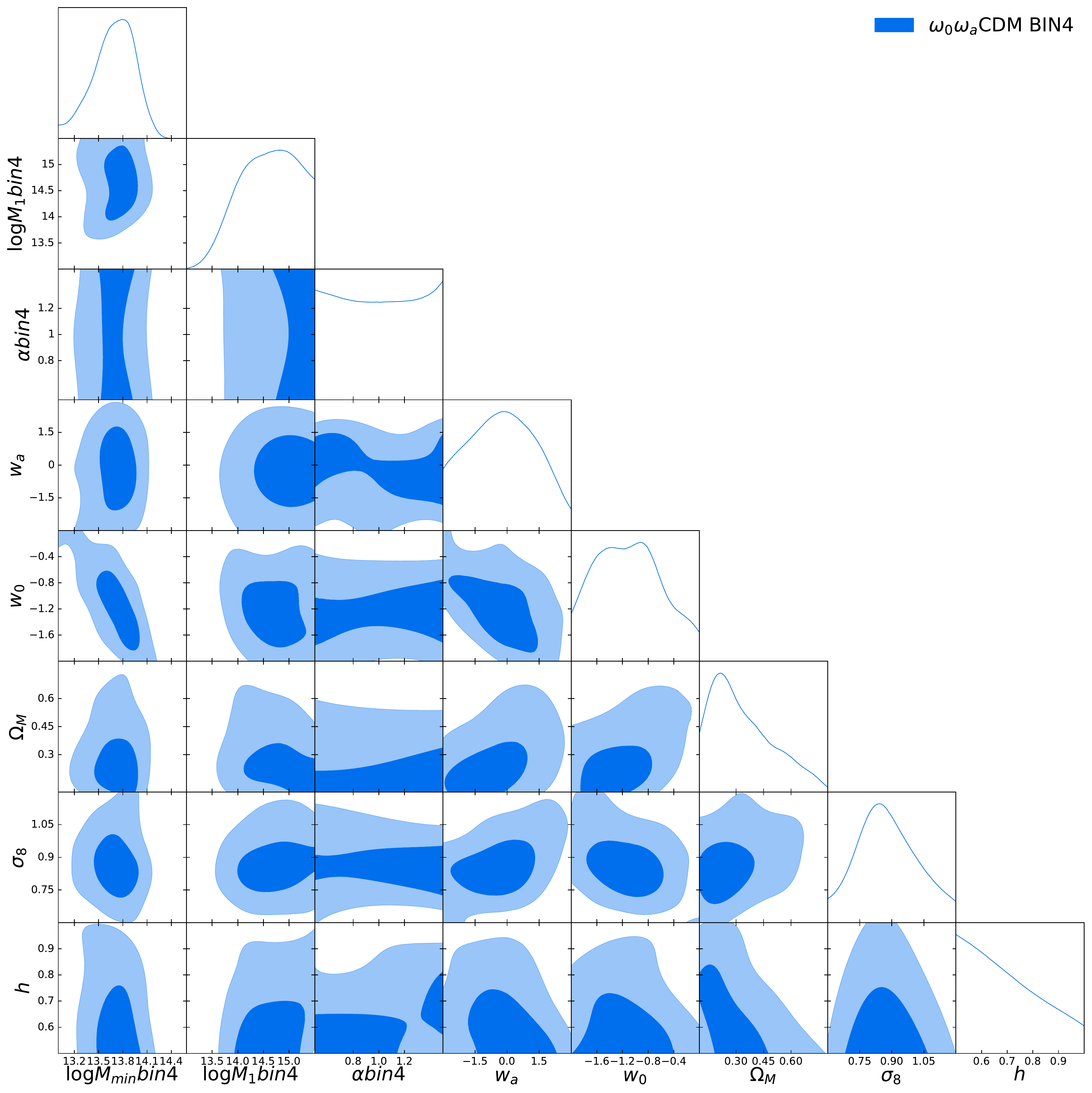}
\caption{Corner plots for the flat prior case on the astrophysical parameters and  $\Omega_M$, $\sigma_8$, $h$, $\omega_0$, and $\omega_a$ cosmological parameters. The posteriors distributions in Bin1, Bin2, Bin3, and Bin4 are shown from left to right, top to bottom. The contours are set to 0.393\%\ and 0.865\%.}
\label{Fig:FPA_FPC_wa}
\end{figure*}

\begin{figure*}[h]
\centering
\includegraphics[width=0.35\textwidth]{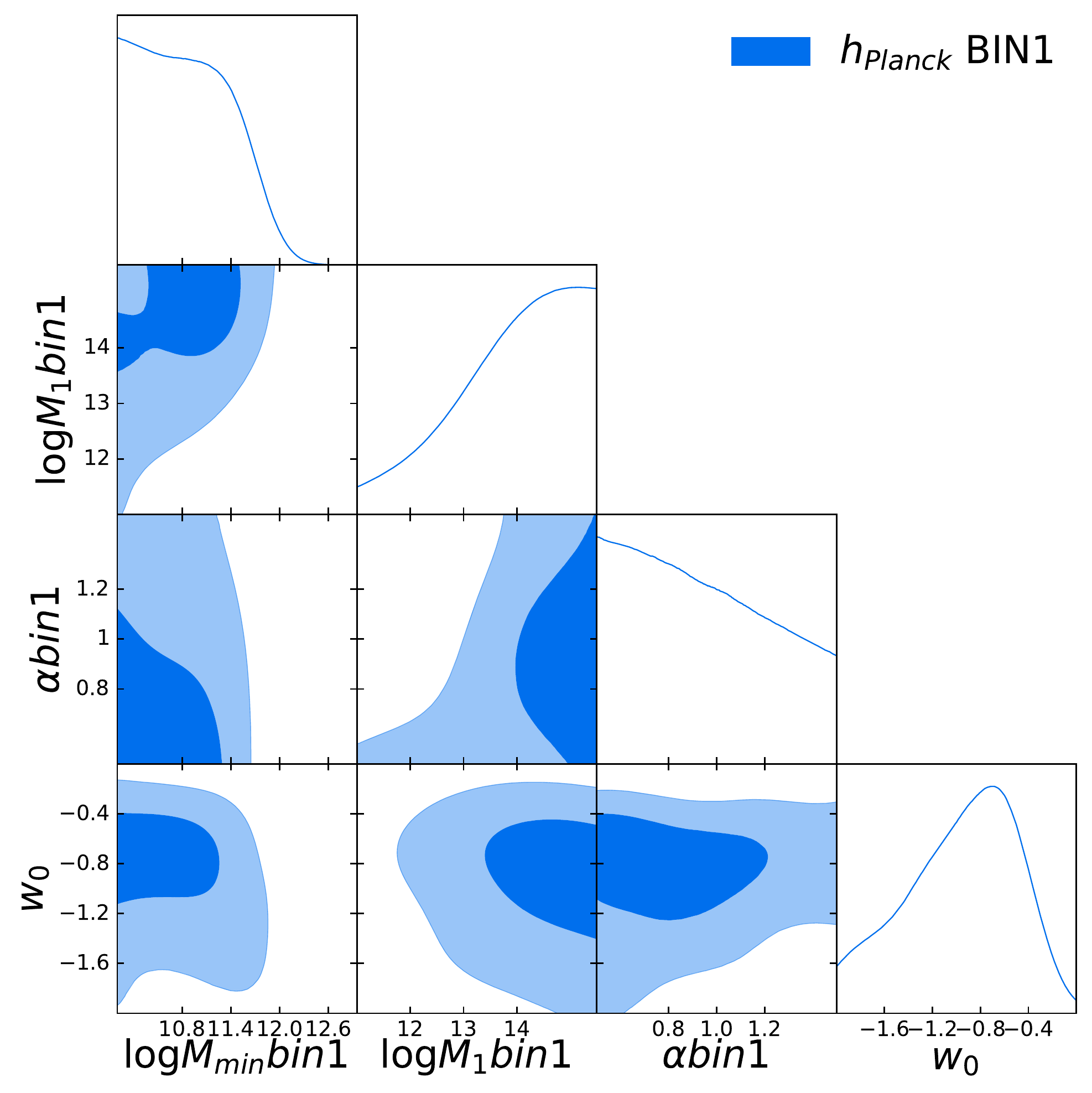}
\includegraphics[width=0.35\textwidth]{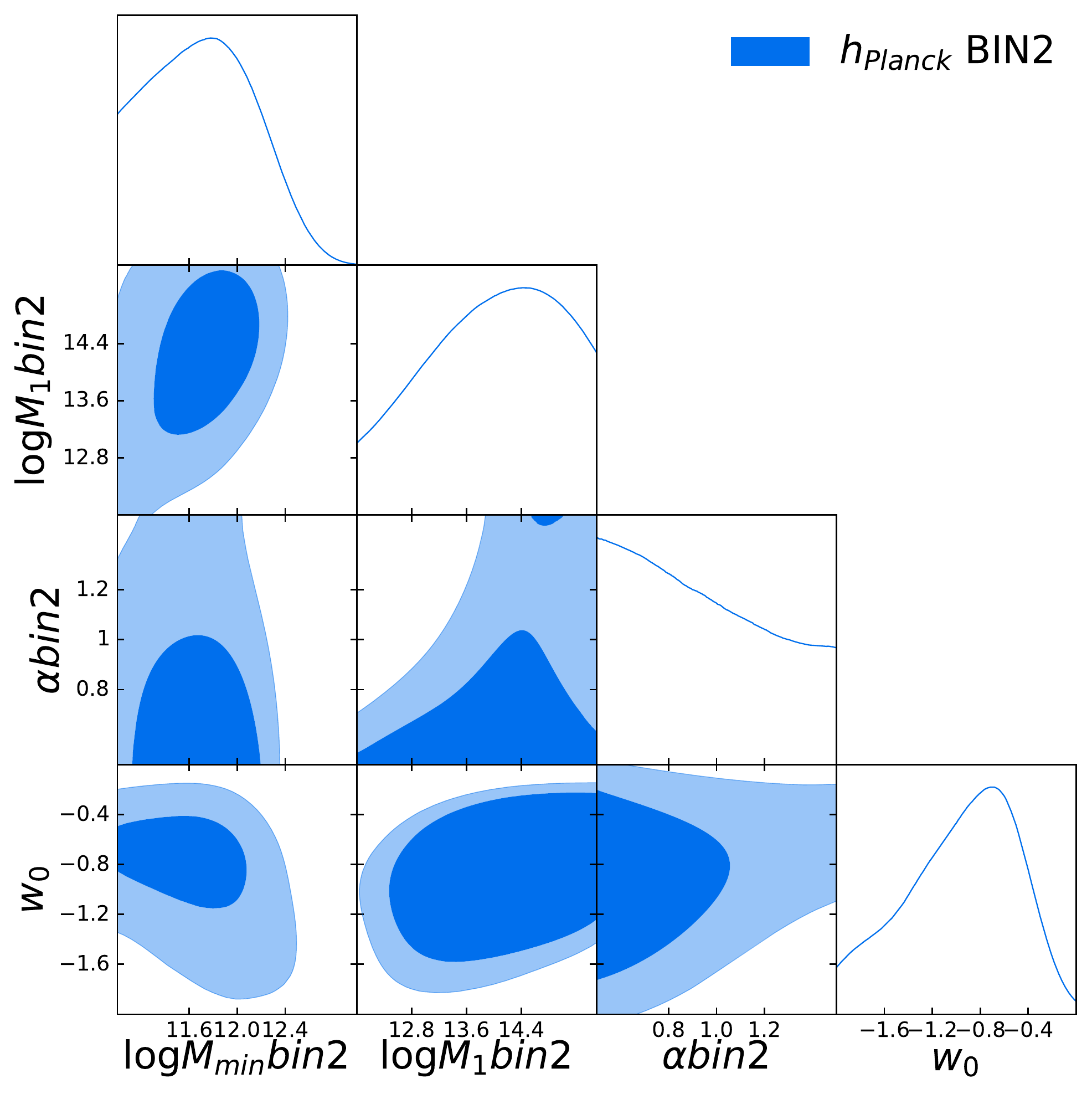}

\includegraphics[width=0.35\textwidth]{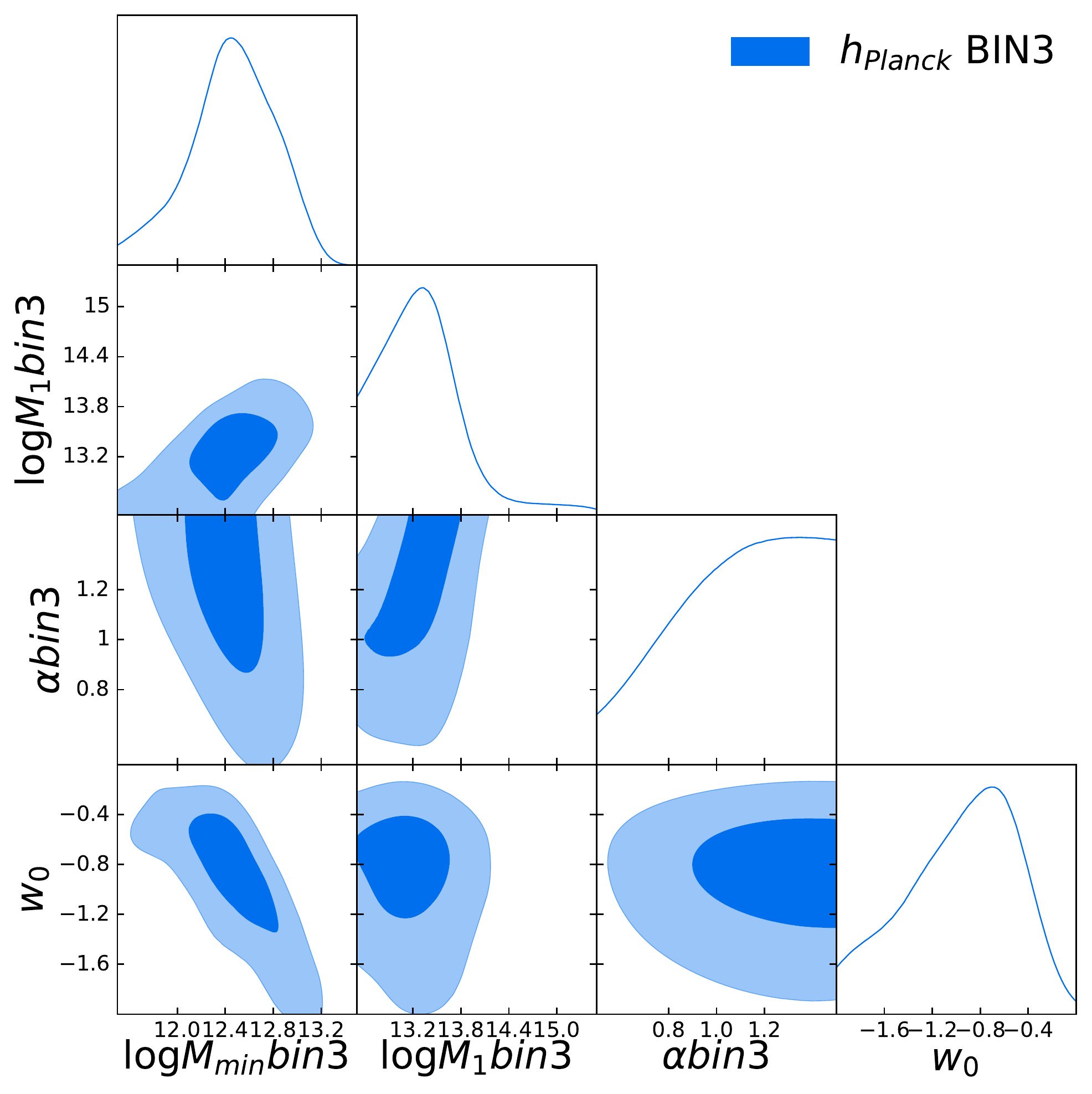}
\includegraphics[width=0.35\textwidth]{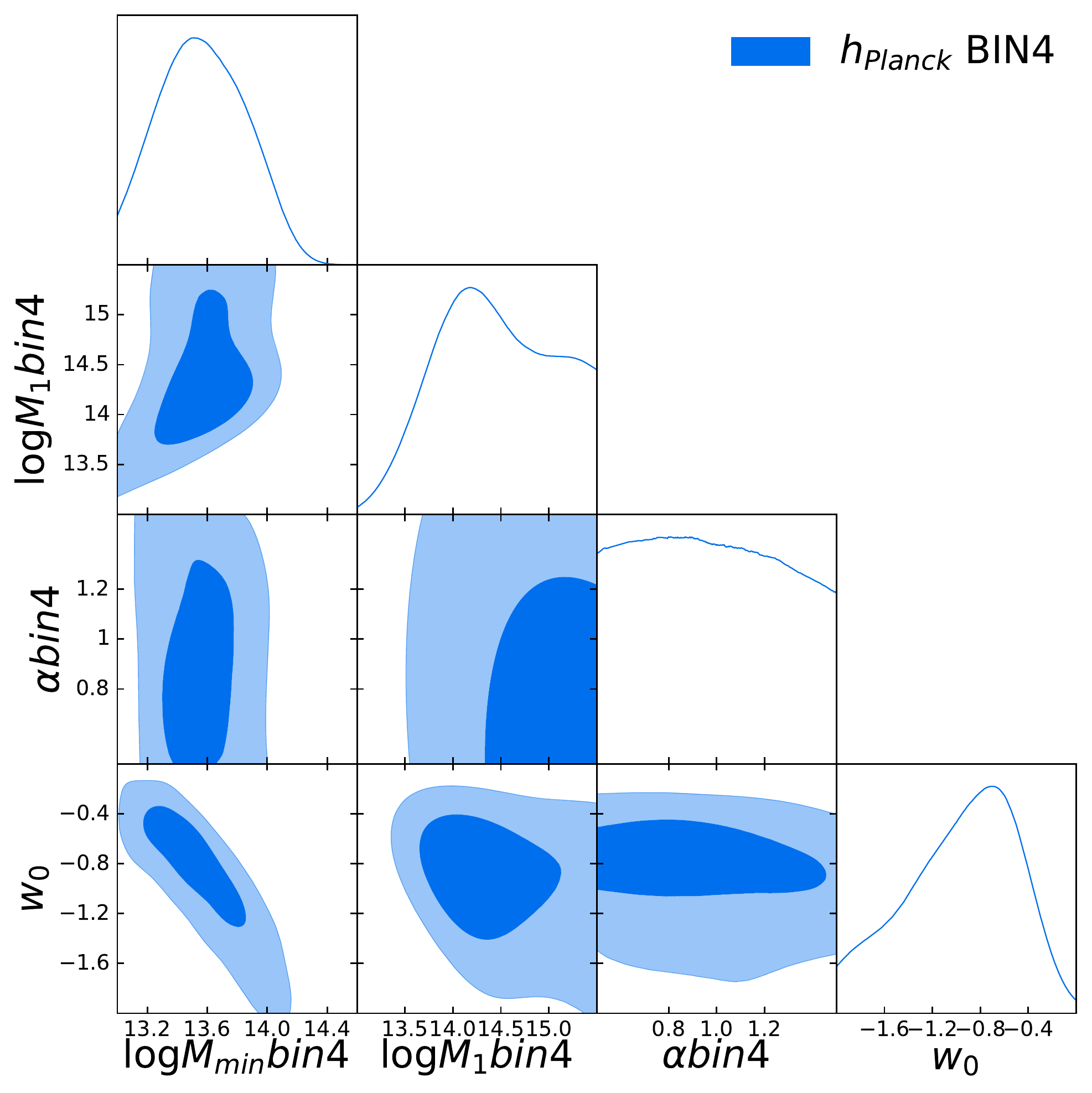}
\caption{Corner plots for the flat prior case on the astrophysical and $\omega_0$ parameters with h fixed to the  \textit{Planck} value. The posterior distributions in Bin1, Bin2, Bin3, and Bin4 are shown from left to right, top to bottom. The contours are set to 3.93\% and 86.5\%.}
\label{Fig:FPA_w0}
\end{figure*}

\begin{figure*}[h]
\centering
\includegraphics[width=0.35\textwidth]{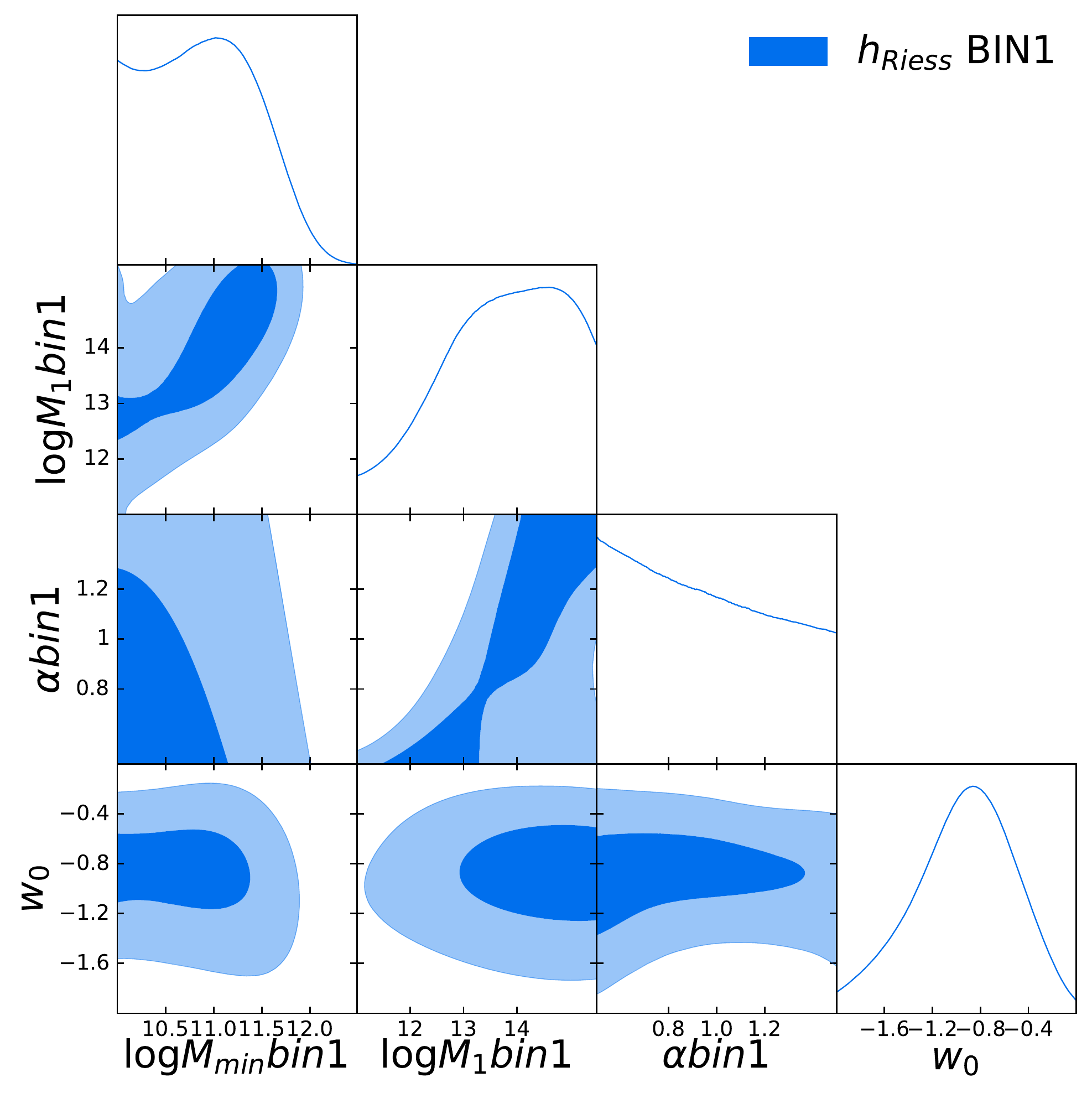}
\includegraphics[width=0.35\textwidth]{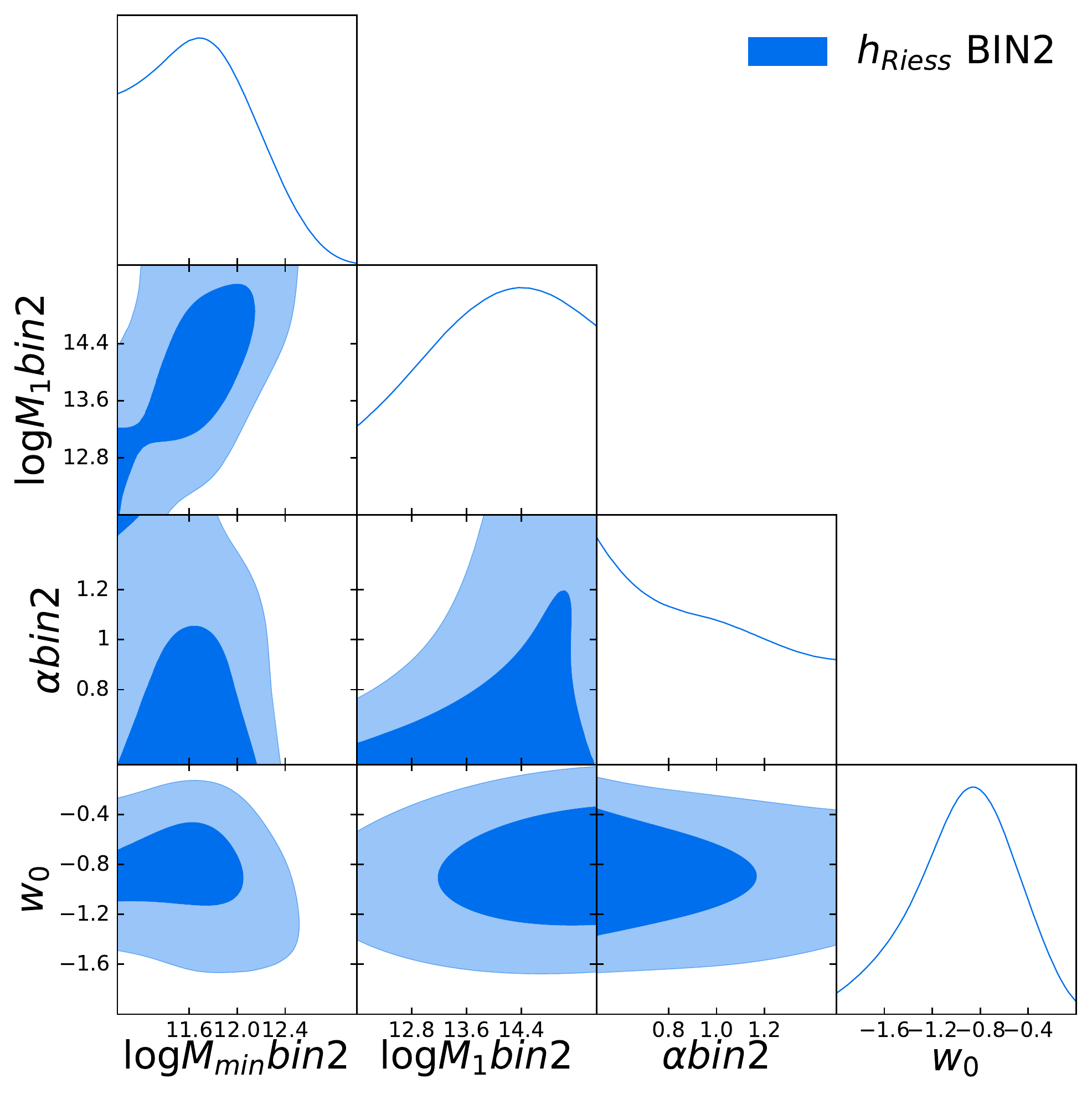}

\includegraphics[width=0.35\textwidth]{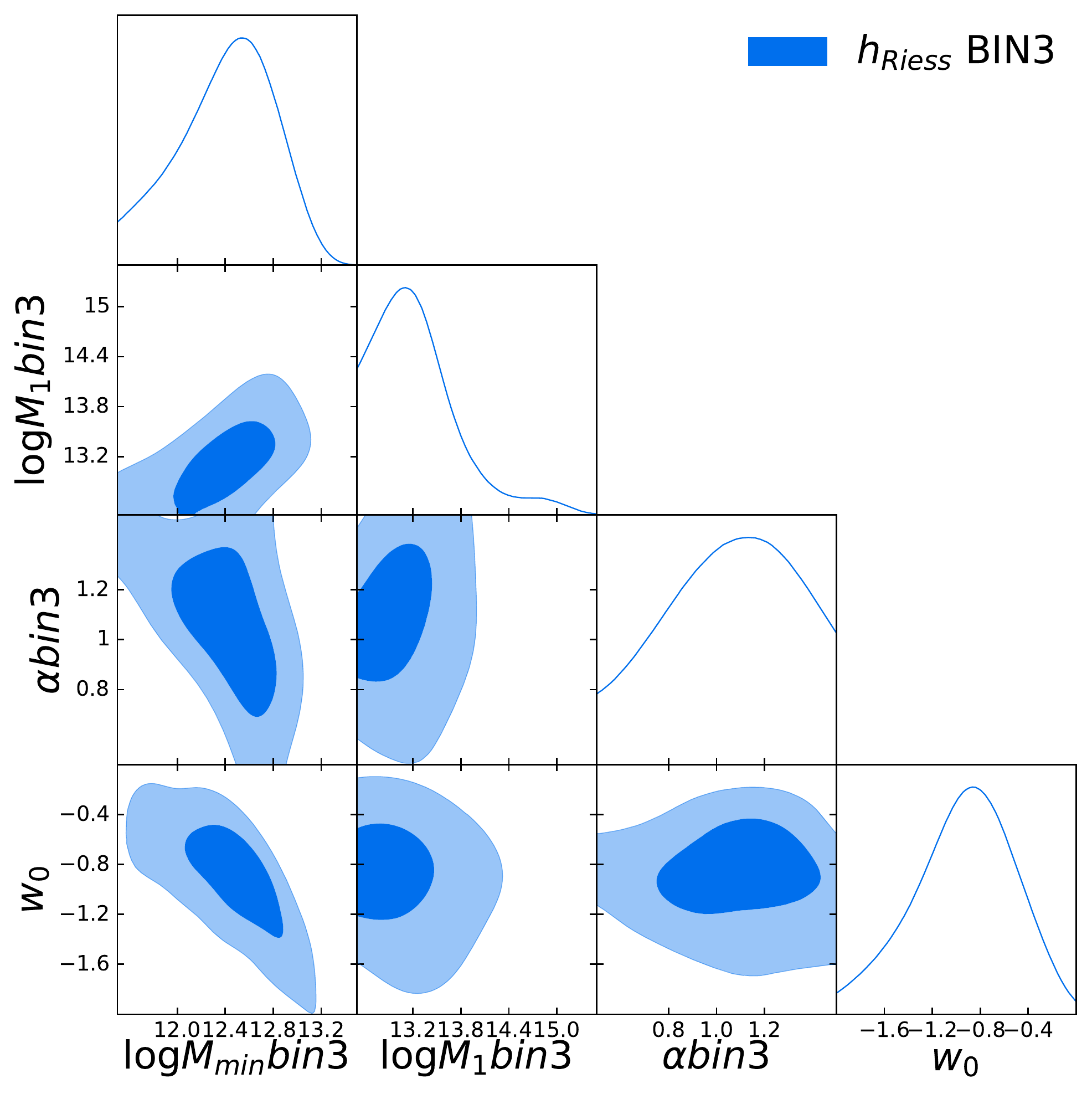}
\includegraphics[width=0.35\textwidth]{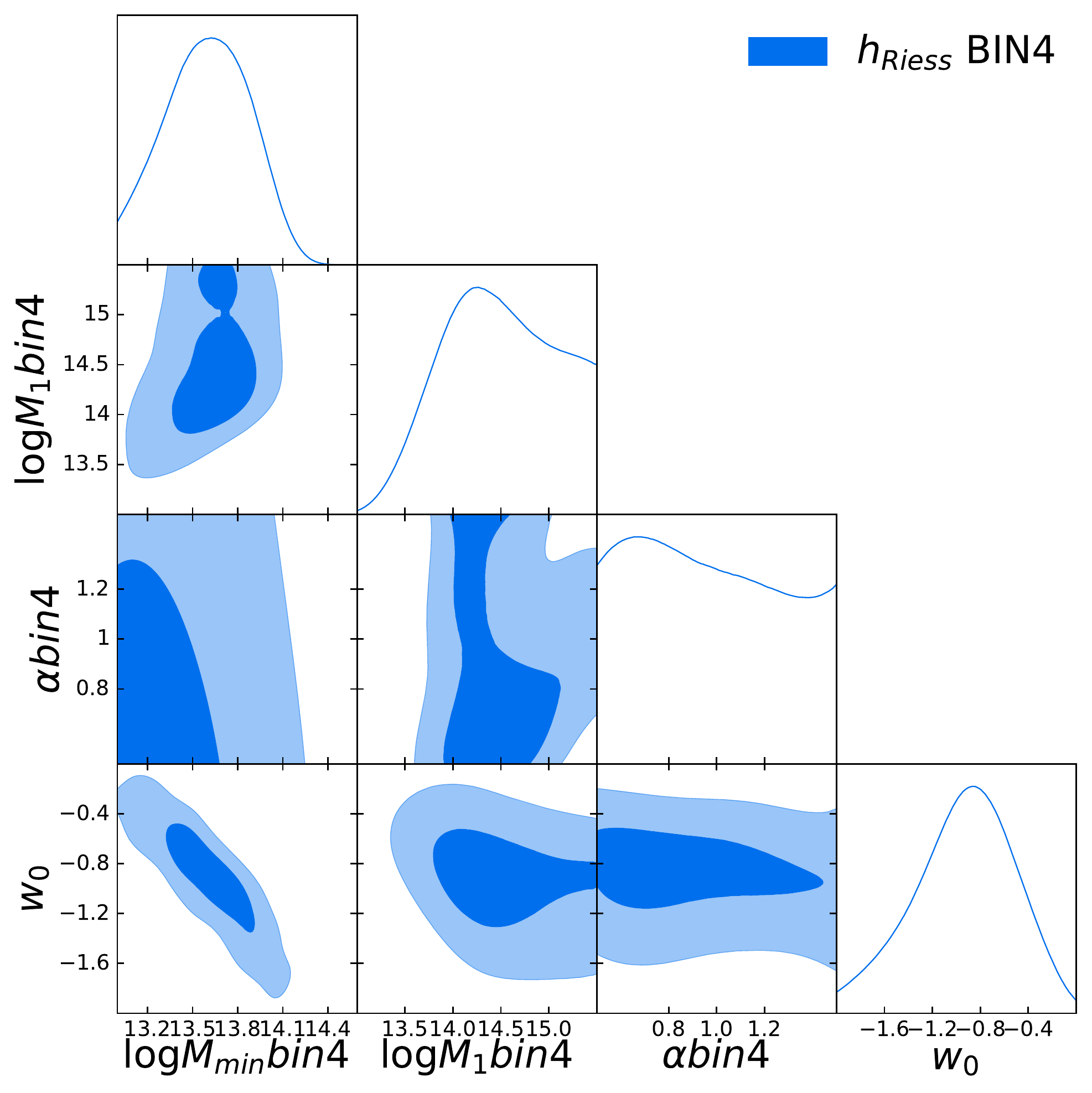}
\caption{Corner plots for the flat prior case on the astrophysical and $\omega_0$ parameters with h fixed to the Riess value. The posterior distributions in Bin1, Bin2, Bin3, and Bin4 are shown from left to right, top to bottom. The contours are set to 39.3\% and 8.65\%.}
\label{Fig:FPA_w0_hRiess}
\end{figure*}

In this section we present the 1D and 2D posterior distributions as well as the priors and the results for the MCMC tomographic runs discussed in this work for the CCF case: the astrophysical only case with fixed cosmology to \textit{Planck} (Fig. \ref{Fig:FPA} and Table \ref{Tab:FPA}) and the astrophysical and cosmological cases ($\Lambda$CDM in Fig. \ref{Fig:FPA_FPC} and Table \ref{Tab:FPA_FPC}, $\omega_0$CDM in Fig. \ref{Fig:FPA_FPC_w0} and Table \ref{Tab:FPA_FPC_w0} and $\omega_0\omega_a$CDM in Fig. \ref{Fig:FPA_FPC_wa} and Table \ref{Tab:FPA_FPC_wa}). Moreover, Fig. \ref{Fig:FPA_w0} and Table \ref{Tab:FPA_w0} show the case where the cosmology is fixed to the \textit{Planck} values, and Fig. \ref{Fig:FPA_w0_hRiess} and Table \ref{Tab:FPA_w0_hRiess} the case where the cosmology is fixed to the \textit{Planck} values, except for $h$ (which is the value provided by \citet{RIE19}).

The relevant 1$\sigma$ and 2$\sigma$ levels for a 2D histogram of samples is 39.3\% and 86.5\%, not 68\% and 95\%. Otherwise, there is no direct comparison with the 1D histograms above the contours. For visualisation purposes, each panel of Figs.~\ref{Fig:FPA_FPC}, \ref{Fig:FPA_FPC_w0}, and \ref{Fig:FPA_FPC_wa} shows the astrophysical parameters posterior distribution in each bin, together with the cosmological parameters. The cosmological parameters posterior distributions are  the same in all panels as the fit is performed jointly on the four redshift bins.

The columns in the tables are, from left to right: the bin to which the astrophysical parameters  refer and the cosmological case when present, the names of the parameters, the priors choice for the run and the mean, median, 68\% confidence intervals (C.I.), and the peak of the posterior distributions. The $>$ and $<$ signs in the C.I. column refer to lower and upper limits, respectively.

\begin{table*} 
\caption{Priors and results for the fixed cosmology case using  the \textit{Planck} values.} 
\label{Tab:FPA} 
\centering 
\begin{tabular}{c c c c c c c} 
\hline 
\hline 
\multicolumn{6}{c}{Fixed Cosmology} \\ 
Bin & Param & Priors & $\mu$ & median & $68\%C.I.$ & Peak \\ 
\hline 
BIN1 & $\log (M_{min}/M_{\odot})$ &  $\mathcal{U}$[ 10.0, 13.0]   & $10.82$ & 10.76 & $ <11.09 $  & $ 10.00$ \\   
     & $\log (M_1/M_{\odot})$     &  $\mathcal{U}$[ 11.0, 15.5]   & $13.79$ & 13.93 & $ > 13.28 $ & $ 14.63$ \\   
     & $\alpha$       &  $\mathcal{U}$[  0.5,  1.5]   & $ 0.96$ & 0.93 & $ < 1.12  $ & $  0.50$ \\   
\hline 
BIN2 & $\log (M_{min}/M_{\odot})$ &  $\mathcal{U}$[ 11.0, 13.0]   & $11.76$ & 11.76 & $ [11.37,12.21 ] $ & $ 11.80$ \\   
     & $\log (M_1/M_{\odot})$     &  $\mathcal{U}$[ 12.0, 15.5]   & $14.04$ & 14.13 & $ > 13.56 $        & $ 15.41$ \\   
     & $\alpha$       &  $\mathcal{U}$[  0.5,  1.5]   & $ 0.98$ & 0.97 & $ - $              & $  0.50$ \\   
\hline 
BIN3 & $\log (M_{min}/M_{\odot})$ &  $\mathcal{U}$[ 11.5, 13.5]   & $12.42$ & 12.48 & $ [12.25,12.80 ] $ & $ 12.54$ \\   
     & $\log (M_1/M_{\odot})$     &  $\mathcal{U}$[ 12.5, 15.5]   & $13.18$ & 13.10 & $ [12.54,13.36 ] $ & $ 12.95$ \\   
     & $\alpha$       &  $\mathcal{U}$[  0.5,  1.5]   & $ 1.09$ & 1.11 & $ [ 0.92, 1.43 ] $ & $  1.16$ \\   
\hline 
BIN4 & $\log (M_{min}/M_{\odot})$ &  $\mathcal{U}$[ 13.0, 15.5]   & $13.64$ & 13.66 & $ [13.51,13.85 ] $ & $ 13.69$ \\   
     & $\log (M_1/M_{\odot})$    &  $\mathcal{U}$[ 13.0, 15.5]   & $14.47$ & 14.46 & $ [13.79,15.03 ] $ & $ 14.09$ \\   
     & $\alpha$       &  $\mathcal{U}$[  0.5,  1.5]   & $ 1.05$ & 1.07 & $ > 0.88 $         & $  1.50$ \\   
\hline 
\hline 
\end{tabular} 
\tablefoot{ The columns denote the bin, the parameter in question, and its prior distribution, and  the mean ($\mu$), median, $68\%$ C.I., and peak of its marginalised 1D posterior distribution.}
\end{table*}  

%%%%%%%%%%%%%%%%%%%%

\begin{table*} 
\caption{Priors and results for the $\Lambda$CDM case, with the astrophysical parameters and $\Omega_M$, $\sigma_8$, and $h$ set as free parameters.}
%, being the rest of the cosmological parameters fixed to the \textit{Planck} values. }
\label{Tab:FPA_FPC} 
\centering 
\begin{tabular}{c c c c c c c} 
\hline 
\hline 
\multicolumn{6}{c}{$\Lambda$CDM} \\ 
Bin & Param & Priors & $\mu$ & median & $68\%C.I.$ & Peak \\ 
\hline 
BIN1 & $\log (M_{min}/M_{\odot})$ &  $\mathcal{U}$[ 10.0, 13.0]   & $10.87$ & 10.84 & $ [10.07,11.38 ] $ & $ 10.34$ \\   
     & $\log (M_1/M_{\odot})$     &  $\mathcal{U}$[ 11.0, 15.5]   & $13.91$ & 14.04 & $ > 13.49 $        & $ 14.43$ \\   
     & $\alpha$       &  $\mathcal{U}$[  0.5,  1.5]   & $ 0.97$ & 0.95 & $ - $              & $  0.50$ \\   
\hline 
BIN2 & $\log (M_{min}/M_{\odot})$ &  $\mathcal{U}$[ 11.0, 13.0]   & $11.74$ & 11.75 & $ < 11.96 $        & $ 11.92$ \\   
     & $\log (M_1/M_{\odot})$     &  $\mathcal{U}$[ 12.0, 15.5]   & $14.16$ & 14.27 & $ > 13.72 $        & $ 15.50$ \\   
     & $\alpha$       &  $\mathcal{U}$[  0.5,  1.5]   & $ 0.93$ & 0.90 & $ [ 0.52, 1.07 ] $ & $  0.80$ \\   
\hline 
BIN3 & $\log (M_{min}/M_{\odot})$ &  $\mathcal{U}$[ 11.5, 13.5]   & $12.47$ & 12.50 & $ [12.25,12.80 ] $ & $ 12.52$ \\   
     & $\log (M_1/M_{\odot})$     &  $\mathcal{U}$[ 12.5, 15.5]   & $13.46$ & 13.37 & $ < 13.64  $       & $ 13.41$ \\   
     & $\alpha$       &  $\mathcal{U}$[  0.5,  1.5]   & $ 1.05$ & 1.07 & $ > 0.91 $         & $  1.50$ \\   
\hline 
BIN4 & $\log (M_{min}/M_{\odot})$ &  $\mathcal{U}$[ 13.0, 15.5]   & $13.59$ & 13.61 & $ [13.43,13.85 ] $ & $ 13.65$ \\   
     & $\log (M_1/M_{\odot})$     &  $\mathcal{U}$[ 13.0, 15.5]   & $14.44$ & 14.44 & $ [13.83,15.12 ] $ & $ 14.23$ \\   
     & $\alpha$       &  $\mathcal{U}$[  0.5,  1.5]   & $ 1.01$ & 1.01 & $ - $              & $  1.50$ \\   
\hline 
COSMO & $\Omega_M$ &  $\mathcal{U}$[  0.1,  0.8]   & $ 0.33$ & 0.31 & $ [ 0.17, 0.41 ] $ & $  0.26$ \\   
      & $\sigma_8$ &  $\mathcal{U}$[  0.6,  1.2]   & $ 0.87$ & 0.87 & $ [ 0.75, 1.00 ] $ & $  0.87$ \\   
      & $h$        &  $\mathcal{U}$[  0.5,  1.0]   & $ 0.72$ & 0.71 & $ < 0.79 $         & $  0.67$ \\   
\hline 
\hline 
\end{tabular}
\tablefoot{The rest of the cosmological parameters are fixed to the \textit{Planck} values. The column information is the same as in Table \ref{Tab:FPA}.}
\end{table*}  

%%%%%%%%%%%%%%%%%%%

\begin{table*} 
\caption{Priors and results for the $\omega_0$CDM case and $\omega_a$ fixed to zero.}
%The rest of the cosmological parameters are fixed to the \textit{Planck} values.} 
\label{Tab:FPA_FPC_w0} 
\centering 
\begin{tabular}{c c c c c c c} 
\hline 
\hline 
\multicolumn{6}{c}{$\omega_0$CDM} \\ 
Bin & Param & Priors & $\mu$ & median & $68\%C.I.$ & Peak \\ 
\hline 
BIN1 & $\log (M_{min}/M_{\odot})$ &  $\mathcal{U}$[ 10.0, 13.0]   & $10.83$ & 10.77 & $ < 11.11 $ & $ 10.05$ \\   
     & $\log (M_1/M_{\odot})$     &  $\mathcal{U}$[ 11.0, 15.5]   & $13.77$ & 13.84 & $ > 13.23 $ & $ 13.74$ \\   
     & $\alpha$       &  $\mathcal{U}$[  0.5,  1.5]   & $ 0.95$ & 0.92 & $ < 1.12 $  & $  0.50$ \\   
\hline 
BIN2 & $\log (M_{min}/M_{\odot})$ &  $\mathcal{U}$[ 11.0, 13.0]   & $11.75$ & 11.72 & $ [11.16,12.10 ] $ & $ 11.67$ \\   
     & $\log (M_1/M_{\odot})$     &  $\mathcal{U}$[ 12.0, 15.5]   & $13.95$ & 13.99 & $ [13.19,15.18 ] $ & $ 14.20$ \\   
     & $\alpha$       &  $\mathcal{U}$[  0.5,  1.5]   & $ 0.98$ & 0.96 & $ - $              & $  0.59$ \\   
\hline 
BIN3 & $\log (M_{min}/M_{\odot})$ &  $\mathcal{U}$[ 11.5, 13.5]   & $12.53$ & 12.55 & $ [12.27,12.90 ] $ & $ 12.57$ \\   
     & $\log (M_1/M_{\odot})$     &  $\mathcal{U}$[ 12.5, 15.5]   & $13.43$ & 13.34 & $ [12.60,13.73 ] $ & $ 13.17$ \\   
     & $\alpha$       &  $\mathcal{U}$[  0.5,  1.5]   & $ 1.03$ & 1.05 & $ - $              & $  1.23$ \\   
\hline 
BIN4 & $\log (M_{min}/M_{\odot})$ &  $\mathcal{U}$[ 13.0, 15.5]   & $13.59$ & 13.60 & $ [13.31,13.94 ] $ & $ 13.75$ \\   
     & $\log (M_1/M_{\odot})$     &  $\mathcal{U}$[ 13.0, 15.5]   & $14.27$ & 14.19 & $ [13.49,14.67 ] $ & $ 14.09$ \\   
     & $\alpha$       &  $\mathcal{U}$[  0.5,  1.5]   & $ 1.07$ & 1.10 & $ > 0.92 $         & $  1.50$ \\   
\hline 
COSMO & $\omega_0$ &  $\mathcal{U}$[ -2.0,  0.0]   & $-1.00$ & -1.00 & $ [-1.56,-0.47 ] $ & $ -0.97$ \\   
      & $\Omega_M$ &  $\mathcal{U}$[  0.1,  0.8]   & $ 0.38$ & 0.35 & $ [ 0.13, 0.47 ] $ & $  0.26$ \\   
      & $\sigma_8$ &  $\mathcal{U}$[  0.6,  1.2]   & $ 0.87$ & 0.86 & $ [ 0.73, 0.98 ] $ & $  0.85$ \\   
      & $h$        &  $\mathcal{U}$[  0.5,  1.0]   & $ 0.70$ & 0.67 & $ < 0.75 $         & $  0.50$ \\   
\hline 
\hline 
\end{tabular} 
\tablefoot{The rest of the cosmological parameters are fixed to the \textit{Planck} values. The column information is the same as in Table \ref{Tab:FPA}.}
\end{table*}  

%%%%%%%%%%%%

\begin{table*} 
\caption{Priors and results for the MCMC run on the astrophysical parameters and the cosmological parameters: $\Omega_M$, $\sigma_8$, $h$, $\omega_0$, and $\omega_a$. }
%The rest of the cosmological parameters are fixed to the \textit{Planck} values.} 
\label{Tab:FPA_FPC_wa} 
\centering 
\begin{tabular}{c c c c c c c} 
\hline 
\hline 
\multicolumn{6}{c}{$\omega_0\omega_a$CDM} \\ 
Bin & Param & Priors & $\mu$ & median & $68\%C.I.$ & Peak \\ 
\hline 
BIN1 & $\log (M_{min}/M_{\odot})$ &  $\mathcal{U}$[ 10.0, 13.0]   & $10.98$ & 10.95 & $ [10.32,11.52 ] $ & $ 11.06$ \\   
     & $\log (M_1/M_{\odot})$     &  $\mathcal{U}$[ 11.0, 15.5]   & $13.71$ & 13.82 & $ [13.07,15.35 ] $ & $ 14.08$ \\   
     & $\alpha$       &  $\mathcal{U}$[  0.5,  1.5]   & $ 0.91$ & 0.87 & $ < 1.06  $        & $  0.50$ \\   
\hline 
BIN2 & $\log (M_{min}/M_{\odot})$ &  $\mathcal{U}$[ 11.0, 13.0]   & $11.80$ & 11.79 & $ [11.30,12.25 ] $ & $ 11.88$ \\   
     & $\log (M_1/M_{\odot})$     &  $\mathcal{U}$[ 12.0, 15.5]   & $13.94$ & 14.00 & $ [13.28,15.26 ] $ & $ 14.17$ \\   
     & $\alpha$       &  $\mathcal{U}$[  0.5,  1.5]   & $ 0.96$ & 0.94 & $ < 1.13  $        & $  0.50$ \\   
\hline 
BIN3 & $\log (M_{min}/M_{\odot})$ &  $\mathcal{U}$[ 11.5, 13.5]   & $12.53$ & 12.58 & $ [12.31,12.94 ] $ & $ 12.67$ \\   
     & $\log (M_1/M_{\odot})$     &  $\mathcal{U}$[ 12.5, 15.5]   & $13.44$ & 13.37 & $ [12.73,13.80 ] $ & $ 13.30$ \\   
     & $\alpha$       &  $\mathcal{U}$[  0.5,  1.5]   & $ 1.06$ & 1.09 & $ > 0.92 $         & $  1.27$ \\   
\hline 
BIN4 & $\log (M_{min}/M_{\odot})$ &  $\mathcal{U}$[ 13.0, 15.5]   & $13.69$ & 13.71 & $ [13.48,13.99 ] $ & $ 13.80$ \\   
     & $\log (M_1/M_{\odot})$     &  $\mathcal{U}$[ 13.0, 15.5]   & $14.58$ & 14.61 & $ [14.13,15.31 ] $ & $ 14.84$ \\   
     & $\alpha$       &  $\mathcal{U}$[  0.5,  1.5]   & $ 1.00$ & 1.00 & $ - $              & $  1.50$ \\   
\hline 
COSMO & $\omega_a$ &  $\mathcal{U}$[ -3.0,  3.0]   & $-0.19$ & -0.21 & $ [-1.88, 1.48 ] $ & $ -0.20$ \\   
      & $\omega_0$ &  $\mathcal{U}$[ -2.0,  0.0]   & $-1.09$ & -1.11 & $ [-1.72,-0.66 ] $ & $ -0.92$ \\   
      & $\Omega_M$ &  $\mathcal{U}$[  0.1,  0.8]   & $ 0.34$ & 0.31 & $ [ 0.11, 0.41 ] $ & $  0.21$ \\   
      & $\sigma_8$ &  $\mathcal{U}$[  0.6,  1.2]   & $ 0.88$ & 0.87 & $ [ 0.72, 1.01 ] $ & $  0.84$ \\   
      & $h$        &  $\mathcal{U}$[  0.5,  1.0]   & $ 0.70$ & 0.67 & $ < 0.76 $         & $  0.50$ \\   
\hline 
\hline 
\end{tabular} 
\tablefoot{The column information is the same as in Table \ref{Tab:FPA}. The rest of the cosmological parameters are fixed to the \textit{Planck} values.}
\end{table*}

%%%%%%%%%%%%%%%%

\begin{table*} 
\caption{Priors and results for the MCMC run on astrophysical parameters and the cosmological $\omega_0$.} 
%The rest of the cosmological parameters are fixed to the \textit{Planck}.} 
\label{Tab:FPA_w0} 
\centering 
\begin{tabular}{c c c c c c c} 
\hline 
\hline 
\multicolumn{6}{c}{Fixed Cosmology and $\omega_0$ with $h_{Planck}$} \\ 
Bin & Param & Priors & $\mu$ & median & $68\%C.I.$ & Peak \\ 
\hline 
BIN1 & $\log (M_{min}/M_{\odot})$ &  $\mathcal{U}$[ 10.0, 13.0]   & $10.88$ & 10.83 & $ < 11.19 $ & $ 10.01$ \\   
     & $\log (M_1/M_{\odot})$     &  $\mathcal{U}$[ 11.0, 15.5]   & $13.92$ & 14.10 & $ > 13.50 $ & $ 15.15$ \\   
     & $\alpha$       &  $\mathcal{U}$[  0.5,  1.5]   & $ 0.94$ & 0.91 & $ < 1.08 $  & $  0.50$ \\   
\hline 
BIN2 & $\log (M_{min}/M_{\odot})$ &  $\mathcal{U}$[ 11.0, 13.0]   & $11.74$ & 11.72 & $ [11.22,12.12 ] $ & $ 11.78$ \\   
     & $\log (M_1/M_{\odot})$     &  $\mathcal{U}$[ 12.0, 15.5]   & $13.96$ & 14.03 & $ [13.34,15.32 ] $ & $ 14.47$ \\   
     & $\alpha$       &  $\mathcal{U}$[  0.5,  1.5]   & $ 0.94$ & 0.90 & $ < 1.09 $         & $  0.50$ \\   
\hline 
BIN3 & $\log (M_{min}/M_{\odot})$ &  $\mathcal{U}$[ 11.5, 13.5]   & $12.46$ & 12.47 & $ [12.16,12.87 ] $ & $ 12.46$ \\   
     & $\log (M_1/M_{\odot})$     &  $\mathcal{U}$[ 12.5, 15.5]   & $13.33$ & 13.27 & $ [12.73,13.67 ] $ & $ 13.33$ \\   
     & $\alpha$       &  $\mathcal{U}$[  0.5,  1.5]   & $ 1.09$ & 1.12 & $ > 0.97 $         & $  1.34$ \\   
\hline 
BIN4 & $\log (M_{min}/M_{\odot})$ &  $\mathcal{U}$[ 13.0, 15.5]   & $13.58$ & 13.56 & $ [13.25,13.86 ] $ & $ 13.52$ \\   
     & $\log (M_1/M_{\odot})$     &  $\mathcal{U}$[ 13.0, 15.5]   & $14.43$ & 14.40 & $ [13.77,15.07 ] $ & $ 14.18$ \\   
     & $\alpha$       &  $\mathcal{U}$[  0.5,  1.5]   & $ 0.98$ & 0.98 & $ - $              & $  0.80$ \\   
\hline 
COSMO & $\omega_0$    &  $\mathcal{U}$[ -2.0,  0.0]   & $-0.95$ & -0.90 & $ [-1.31,-0.38 ] $ & $ -0.69$ \\   
\hline 
\hline 
\end{tabular} 
\tablefoot{The rest of the cosmological parameters are fixed to the \textit{Planck} values. The column information is the same as in Table \ref{Tab:FPA}.}
\end{table*}  

%%%%%%%%%%%

\begin{table*} 
\caption{Priors and results for the MCMC run on astrophysical parameters and the cosmological $\omega_0$, with the $h$ fixed to \cite{RIE19}.}
%The rest of the cosmological parameters are fixed to the \textit{Planck} values except for the $h$ that is fixed to the value by \cite{RIE19}.} 
\label{Tab:FPA_w0_hRiess} 
\centering 
\begin{tabular}{c c c c c c c} 
\hline 
\hline 
\multicolumn{6}{c}{Fixed Cosmo and $\omega_0$ with $h_{Riess}$} \\ 
Bin & Param & Priors & $\mu$ & median & $68\%C.I.$ & Peak \\ 
\hline 
BIN1 & $\log (M_{min}/M_{\odot})$ &  $\mathcal{U}$[ 10.0, 13.0]   & $10.92$ & 10.89 & $ > 11.23 $        & $ 11.02$ \\   
     & $\log (M_1/M_{\odot})$     &  $\mathcal{U}$[ 11.0, 15.5]   & $13.71$ & 13.81 & $ [13.02,15.32 ] $ & $ 14.63$ \\   
     & $\alpha$       &  $\mathcal{U}$[  0.5,  1.5]   & $ 0.95$ & 0.93 & $ < 1.11 $         & $  0.50$ \\   
\hline 
BIN2 & $\log (M_{min}/M_{\odot})$ &  $\mathcal{U}$[ 11.0, 13.0]   & $11.71$ & 11.68 & $ [11.13,12.03 ] $ & $ 11.68$ \\   
     & $\log (M_1/M_{\odot})$     &  $\mathcal{U}$[ 12.0, 15.5]   & $13.95$ & 14.02 & $ [13.42,15.46 ] $ & $ 14.36$ \\   
     & $\alpha$       &  $\mathcal{U}$[  0.5,  1.5]   & $ 0.94$ & 0.92 & $ < 1.10 $         & $  0.50$ \\   
\hline 
BIN3 & $\log (M_{min}/M_{\odot})$ &  $\mathcal{U}$[ 11.5, 13.5]   & $12.42$ & 12.45 & $ [12.09,12.89 ] $ & $ 12.53$ \\   
     & $\log (M_1/M_{\odot})$     &  $\mathcal{U}$[ 12.5, 15.5]   & $13.27$ & 13.18 & $ [12.58,13.49 ] $ & $ 13.10$ \\   
     & $\alpha$       &  $\mathcal{U}$[  0.5,  1.5]   & $ 1.05$ & 1.07 & $ [ 0.84, 1.40 ] $ & $  1.14$ \\   
\hline 
BIN4 & $\log (M_{min}/M_{\odot})$ &  $\mathcal{U}$[ 13.0, 15.5]   & $13.61$ & 13.61 & $ [13.32,13.90 ] $ & $ 13.62$ \\   
     & $\log (M_1/M_{\odot})$     &  $\mathcal{U}$[ 13.0, 15.5]   & $14.46$ & 14.46 & $ [13.84,15.10 ] $ & $ 14.27$ \\   
     & $\alpha$       &  $\mathcal{U}$[  0.5,  1.5]   & $ 0.97$ & 0.95 & $ - $              & $  0.66$ \\   
\hline 
COSMO & $\omega_0$ &  $\mathcal{U}$[ -2.0,  0.0]   & $-0.93$ & -0.91 & $ [-1.29,-0.45 ] $ & $ -0.86$ \\   
\hline 
\hline 
\end{tabular} 
\tablefoot{The rest of the cosmological parameters are fixed to the \textit{Planck} values. The column information is the same as in Table \ref{Tab:FPA}.}
\end{table*}  

\end{document}